%% file: main-cosmo.tex
\documentclass[aps,prd,reprint,11pt,amsmath,amssymb,nofootinbib,groupedaddress,superscriptaddress,floatfix,onecolumn,tightenlines,eqsecnum]{revtex4-2}

\usepackage{macros}

\graphicspath{{./plots/}}

\begin{document}

\title{Cosmology with varying fundamental constants from hyperlight, coupled scalars}

\author{Masha Baryakhtar}\email[]{mbaryakh@uw.edu}
\affiliation{Department of Physics, University of Washington, Seattle, Washington, 98195, USA}
\author{Olivier Simon}\email[]{osimon@princeton.edu}
\affiliation{Princeton Center for Theoretical Science, Princeton University, Princeton, New Jersey, 08544, USA}
\affiliation{Department of Physics, Princeton University, Princeton, New Jersey, 08544, USA}
\affiliation{Stanford Institute for Theoretical Physics, Stanford University, Stanford, California, 94305, USA}
\author{Zachary J. Weiner}\email[]{zweiner@uw.edu}
\affiliation{Department of Physics, University of Washington, Seattle, Washington, 98195, USA}

\date{\today}
\begin{abstract}
The fundamental constants at recombination can differ from their present-day values due to
degeneracies in cosmological parameters, raising the possibility of
yet-undiscovered physics coupled directly to the Standard Model.
We study the cosmology of theories in which a new, hyperlight scalar field modulates the electron
mass and fine-structure constant at early times.
We find new degeneracies in cosmologies that pair early recombination with a new contribution to the
matter density arising at late times, whose predictions can be simultaneously consistent with CMB
and low-redshift distance measurements.
Such ``late dark matter'' already exists in the Standard Model in the form of massive neutrinos but is necessarily realized by the scalar responsible for shifting the early-time
fundamental constants.
After detailing the physical effects of varying constants and hyperlight scalar
fields on cosmology, we show that variations of the electron mass and fine structure constant are
constrained at the percent and permille level, respectively, and a hyperlight scalar in the mass
range $10^{-32}~\eV \lesssim m_\phi \lesssim 10^{-28}~\eV$ can impact what variations are allowed
while composing up to a percent of the present dark matter density.
We comment on the potential for models with a varying electron mass to reconcile determinations
of the Hubble constant from cosmological observations and distance-ladder methods, and we show that
parameter inference varies significantly between recent baryon acoustic oscillation and type Ia
supernova datasets.
\end{abstract}

\maketitle
\makeatletter
\def\l@subsubsection#1#2{}
\makeatother
\tableofcontents

\section{Introduction}
\label{sec:introduction}

Cosmology is an essential frontier to study the dark sector and, at the same time,  is a unique probe of
the Standard Model and gravity at very early times.
The cosmic microwave background (CMB) is largely governed by the physics of photon recombination and acoustic plasma waves, which in turn depend sensitively on quantum electrodynamics and thus the electromagnetic coupling strength and the electron mass in the early Universe.
Some theories of physics beyond the Standard Model motivate variations of the fundamental physical constants over cosmological history; such microphysical models introduce new degrees of freedom which influence cosmological dynamics through gravity.
The CMB anisotropies are also sensitive to all of the Universe's content through gravitation, both via cosmic expansion and gravitational potentials, and both during and after recombination.

Many previous studies of the CMB have placed limits on phenomenological variations of the
fine-structure constant $\alpha$~\cite{Hannestad:1998xp, Kaplinghat:1998ry, Avelino:2000ea,
Battye:2000ds, Avelino:2001nr, Landau:2001st, Martins:2003pe, Rocha:2003gc, Stefanescu:2007aa,
Nakashima:2008cb, Menegoni:2009rg, Menegoni:2012tq} and of the mass of the electron
$m_e$~\cite{Kujat:1999rk, Ichikawa:2006nm, Landau:2008re, Scoccola:2008jw,
Nakashima:2009cs,Landau:2010zs, Scoccola:2012ny} at recombination under the assumption that the
gravitational effects of any new degrees of freedom are negligible.
The possibility that fundamental constants---in particular, the electron mass---differed from their
present-day values in the early Universe, triggering recombination earlier than standard, is also a
candidate proposal~\cite{Hart:2019dxi, Sekiguchi:2020teg, Schoneberg:2021qvd, Khalife:2023qbu} to
reconcile determinations of the present-day expansion rate $H_0$ as inferred by the CMB in
$\Lambda$--cold-dark-matter (\LCDM{}) cosmology with those based on local, distance-ladder
measurements~\cite{Riess:2016jrr,Riess:2019cxk,Riess:2021jrx}.
Early recombination alone is insufficient to explain observations of large-scale structure and type
Ia supernovae in addition to the CMB, however; existing proposals therefore further invoke spatial
curvature~\cite{Sekiguchi:2020teg, Schoneberg:2021qvd}.
Comparatively little literature has considered specific microphysical models responsible for the
variations and investigated whether added matter content qualitatively alters the predicted
cosmologies~\cite{Fung:2021fcj,Fung:2021wbz,Luu:2021yhl,Solomon:2022qqf,Hoshiya:2022ady}.
Implementing concrete theories is likewise imperative to understand the field evolution and the interplay of measurements
throughout cosmic time---whether nucleosynthesis at earlier times, astrophysical probes at later
times, or laboratory experiments today.

In this work and a companion article~\cite{particle-paper}, we test models of a new scalar field
that induces early-time shifts in fundamental constants against precision cosmological datasets.
We specifically consider a massive, hyperlight field with particle mass
$10^{-32}~\eV \lesssim m_\phi \lesssim 10^{-28}~\eV$ whose Standard Model (SM) couplings effectively
modulate the strength of electromagnetism and the mass of the electron.
The scalar is produced as a cold, spatially homogeneous field in the early Universe and makes
up a subcomponent of cosmological dark matter today.
In the mass range of interest, the amplitude of the scalar field is frozen until some point after
recombination, after which it oscillates and decays; 
the fundamental constants
thus take values before recombination that are constant in space and time but different from their
values today.
Our work complements prior study of models akin to quintessence ($m_\phi \lesssim 10^{-33}~\eV$) in
which $\alpha$ varies via a coupling to the photon~\cite{Tohfa:2023zip, Schoneberg:2023lun,
Vacher:2023gnp}.

We show on semianalytic grounds that models with contributions to the matter density at late times---``late dark matter'' scenarios---open up a novel degeneracy in the background cosmology that, taken at face value, has the potential
to reconcile late-time observations with scenarios of early recombination.
The scalar field conjectured to shift the values of constants naturally makes such a late contribution to dark matter, and so do massive neutrinos.
The hyperlight scalar fields we consider also suppress structure formation to an extent controlled
by their mass and abundance, yielding distinct signatures in the CMB that further test the scenario.
We assess the interplay of these effects on quantitative constraints with a detailed analysis of cosmological data and study the effect of various dataset combinations.
That the additional freedom of scalar field models allows for CMB observations and low-redshift
distance measurements to be simultaneously satisfied in varying-$m_e$ scenarios was observed in the
related analysis of Ref.~\cite{Luu:2021yhl}, though without identifying its physical origin.

The outline of this paper is as follows.
In \cref{sec:varying-constants}, we detail the physical impact of varying fundamental constants on
the CMB and other cosmological observables.
We devise a general set of parameter combinations that encode the main physical effects relevant to
the CMB anisotropies in \LCDM{}, including parameters that are typically fixed in standard
analyses---not just the electron mass and fine-structure constant, but also the present-day CMB
temperature, the helium yield, and the density of radiation beyond photons.
We then discuss the distinct effects arising from a scalar field hypothesized to be responsible for
the variations in the fundamental constants in \cref{sec:cosmology-hyperlight-scalars}, considering
the scalar's purely gravitational effects and their interplay with varying constants.
\Cref{sec:parameter-inference} presents our inference of parameters within this extension to the
standard cosmological model, comparing the results of scenarios with and without a scalar field.
We also critically assess the concordance of current cosmological datasets within the context of
cosmologies with a varying electron mass.
In particular, we survey the impact of recent low-redshift distance datasets from galaxy surveys,
including the Sloan Digital Sky Survey (SDSS)~\cite{eBOSS:2020lta, eBOSS:2020hur} and the Dark
Energy Spectroscopic Instrument (DESI)~\cite{DESI:2024mwx, DESI:2024lzq, DESI:2024uvr}, and type Ia
supernova samples, including Pantheon~\cite{Pan-STARRS1:2017jku}, Pantheon+~\cite{Brout:2022vxf,
Scolnic:2021amr}, the Dark Energy Survey (DES)~\cite{DES:2024tys}, and Union3~\cite{Rubin:2023ovl}.
We conclude in \cref{sec:conclusions}, discussing the potential impact of near-term and future
datasets and possible model extensions motivated by Ref.~\cite{particle-paper}.

We use natural units in which $\hbar = c = 1$ and define the reduced Planck mass
$\Mpl \equiv 1 / \sqrt{8 \pi G} = 2.435 \times 10^{18}~\GeV$.
Repeated spatial indices (latin characters) are implicitly summed regardless of their placement.
We use upright boldface to denote spatial vectors.
Unless otherwise specified, we fix a homogeneous, conformal-time
Friedmann-Lema\^itre-Robertson-Walker (FLRW) spacetime with metric
\begin{align}\label{eqn:flrw-metric}
    g_{\mu \nu}
    \equiv a(\tau)^2 \eta_{\mu \nu}
\end{align}
where $\eta_{\mu \nu}$ is the ``mostly positive'' Minkowski metric and $a(\tau)$ the scale factor.
We use primes to denote derivatives with respect to conformal time $\tau$.
We occasionally work in terms of cosmic time $t$ (defined by $\ud t = a \ud \tau$), and use dots to
denote $t$ derivatives; the Hubble rate is $H \equiv \dot{a} / a$.

\input{cosmo-varconst}

\input{cosmo-scalars}

\input{cosmo-results}

\section{Conclusions}
\label{sec:conclusions}

The cosmic microwave background, long lauded as a precision probe of the early
Universe, depends sensitively on the physics of quantum electrodynamics.
In this work we consider microphysical realizations of scenarios in which the fundamental
parameters of QED---the electromagnetic fine-structure constant and electron mass---take on
different values during recombination than at the present day.
Companion work~\cite{particle-paper} discusses theories of new scalar fields whose couplings vary
fundamental constants in spacetime, emphasizing that consistency with theoretical considerations and
independent probes restricts the viable space of models.
Crucially, Ref.~\cite{particle-paper} shows that mechanizing shifted early-time fundamental
constants that only start evolving toward their present-day values after recombination
\emph{requires} the scalar field to have a nonnegligible abundance.
Here we investigate the cosmological impact of a new scalar through both its direct coupling and
its gravitational coupling to SM matter.

Variations in the fine-structure constant and the electron mass, regardless of their origin,
primarily affect the release of the CMB via their well-studied impacts on recombination, shifting
the temperature and duration of last scattering, and on Thomson scattering.
These effects are not independently constrained by data in \LCDM{} cosmology, however; in
\cref{sec:varying-constants} we clarify the physical origin of degeneracies between the
fundamental constants and \LCDM{} parameters.
In particular, we enumerate how CMB anisotropies largely probe the state of the Universe at last
scattering, whenever it may have occurred, and how the rate of recombination---namely, the redshift
interval over which visible CMB photons last scattered---impacts the generation of polarization and
the suppression of small-scale anisotropies.
\Cref{sec:early-universe-signatures,sec:scalar-impact-on-cosmological-background} describe the
combinations of standard parameters that directly encode the most important physical effects that
imprint in CMB anisotropies, generalizing those of prior work~\cite{Efstathiou:1998xx,
Howlett:2012mh, Kosowsky:2002zt}.
In doing so, we uncover the connection between modified recombination scenarios and the CMB
monopole temperature, if treated as an otherwise unmeasured parameter.
In particular, when either of the present-day CMB temperature and the early-time electron mass is
not fixed, the Hubble constant is no longer uniquely constrained to fix the angular extent of the
sound horizon precisely measured by CMB anisotropies---the same ``geometric'' degeneracy
characteristic to cosmologies with nonzero spatial curvature~\cite{Efstathiou:1998xx}.

The correlation of fundamental constants with \LCDM{} parameters inevitably affects cosmological
dynamics at late times, to which the CMB anisotropies are only weakly sensitive.
Low-redshift distance measurements, whether from type Ia supernovae or baryon acoustic oscillations,
therefore play a crucial role in breaking degeneracies by directly measuring the shape of the late-time expansion
history, even without calibrating their absolute distances.
When including low-redshift distance datasets, cosmological data constrain electron mass variations
at the percent level and fine-structure constant variations at nearly the permille level.
On the other hand, \cref{sec:scalar-impact-on-cosmological-background} identifies a novel solution
that combines early recombination with ``late dark matter,'' i.e., an enhancement of the matter
abundance after recombination.
The SM already offers a source of late dark matter: massive neutrinos, which become nonrelativistic
well before the present but after recombination.
Moreover, \cref{sec:cosmology-hyperlight-scalars} shows that a hyperlight scalar field, invoked to
mechanize variation of fundamental constants between recombination and the present day, must begin
oscillating and boost the matter abundance after recombination as well.
The positive correlation between $h$ and the abundance of late dark matter (measured by the sum of
the neutrino masses or by $\fphi$ for a scalar) along this degeneracy is especially notable in
contrast to the \emph{anti}correlation between the two in \LCDM{}.

Though early recombination with late dark matter realizes a degeneracy in the cosmological
background, the scenarios are distinguished (and therefore testable) through the behavior of the new
component's perturbations.
\Cref{sec:parameter-inference} carefully studies the complementary impact of individual datasets in
constraining these scenarios and provides quantitative evidence for the physical effects outlined in
\cref{sec:varying-constants,sec:cosmology-hyperlight-scalars}.
\Cref{sec:results-scalars} further illustrates how the perturbations of a hyperlight scalar suppress
the growth of structure and source an integrated Sachs-Wolfe effect, signatures which ultimately
restrict the scalar abundance allowed by \Planck{} data to no more than $\mathcal{O}(1\%)$ of the
dark matter's.

A heavier electron at early times, which triggers recombination at a higher temperature, has been
touted as an extension to \LCDM{} for which cosmological data infer a Hubble constant $H_0$ in
better agreement with distance-ladder measurements.
In \cref{sec:concordance} we critically assess this claim, arguing that because the physical effects
of \cref{sec:varying-constants} lead \Planck{} data to prefer late recombination to early
recombination, a superficial reduction in the $H_0$ tension comes at the cost of degrading the fit
to individual cosmological datasets.
Moreover, though the degeneracy introduced by the scalar's contribution to the late-time matter
abundance (identified in \cref{sec:cosmology-hyperlight-scalars}) offers the potential to
simultaneously satisfy CMB and low-redshift observations at arbitrarily large $H_0$, \Planck{} data
too strongly disfavors the impact of a hyperlight scalar's perturbations to fully realize this
effect.
To the extent that \Planck{} data \emph{do} allow for hyperlight scalars, however, slightly larger
electron masses and Hubble constants are allowed compared to scenarios without a scalar.
We show that similar conclusions apply to massive neutrinos as an alternative realization of late
dark matter.
An interesting question for further study is whether any motivated models of late dark matter affect
the dynamics of perturbations in a manner the CMB is more amenable to.

Because low-redshift datasets play such a crucial role in breaking degeneracies in varying electron
mass scenarios, any discord between them is put on full display.
\Cref{sec:results-varying-constants-low-z,sec:results-scalars-low-z} show that the diverging
preferences in the uncalibrated late-time expansion history (i.e., its shape as encoded by the scale
factor of matter--dark-energy equality $\amL$) propagate directly to the electron mass and $h$,
regardless of whether a hyperlight scalar is included in the model.
In particular, \cref{sec:concordance} highlights an increasing discrepancy in the cosmologies
preferred by recent BAO datasets and recent SNe datasets that, combined with \Planck{}, vary in
their inferred $h$ between $0.704 \pm 0.011$ and $0.632 \pm 0.018$.
The former result derives from the recent DESI DR1 BAO measurements~\cite{DESI:2024mwx,
DESI:2024lzq, DESI:2024uvr} and is larger than that from prior BAO data, a shift driven in part by
an increase its inferred the uncalibrated amplitude $h r_\mathrm{d}$.
\Cref{sec:concordance} also emphasizes the importance of a holistic interpretation of the agreement
between posteriors derived from, e.g., CMB and BAO data with those from the SH0ES-calibrated
distance ladder: even with the additional freedom afforded to early-recombination (and
varying-$T_0$) cosmologies, no independent dataset combination prefers a late-time expansion history
that agrees with the SH0ES-calibrated distance ladder in both calibration ($h$) and shape ($\amL$).

Future high-resolution CMB observations~\cite{CMB-S4:2016ple, SimonsObservatory:2018koc} will
greatly improve measurements of temperature and polarization anisotropies deep into the damping
tail, promising substantially stronger constraints on fine-structure constant variations.
Even current and near-term datasets from ground-based experiments~\cite{ACT:2020gnv, SPT-3G:2021eoc,
SPT-3G:2022hvq} offer increased constraining potential.
While no physical effect positions high-resolution observations to better constrain electron-mass
variations, alternative CMB datasets can weigh in on the influence of features in \Planck{} data
suspected to be systematic in origin.
The lensing anomaly in particular is absent in other datasets, including recent reanalyses of
\Planck{} data~\cite{Rosenberg:2022sdy, Tristram:2023haj}.
Bounds on the abundance of hyperlight scalars (coupled to the SM or not) could too be alleviated
because the same systematic effects effectively disfavor the their impact on structure growth, as
has already been observed for bounds on the neutrino masses~\cite{Tristram:2023haj}.
Future BAO and SNe datasets will also provide substantially more precise measurements of the
late-time expansion history, better breaking degeneracies in CMB constraints on early recombination.
We will study constraints from these more recent datasets and forecast bounds from upcoming
observations in future work.

In a companion article~\cite{particle-paper}, we apply the results of \cref{sec:results-scalars} to
constrain concrete extensions of the SM with new scalar fields coupled to the electron and photon.
Reference~\cite{particle-paper} discusses numerous challenges in constructing models that are both
theoretically consistent and allowed by independent constraints from other astrophysical and
laboratory measurements.
In viable theories, the cosmological constraints of \cref{sec:results-scalars} in particular yield
leading bounds on hyperlight, quadratically coupled scalars that become matterlike in the matter
dominated era.

This work restricts its analysis to regimes that realize the phenomenological scenarios of prior
study, in which the change to the fundamental constants remains fixed through recombination.
The theoretical developments of Ref.~\cite{particle-paper}, however, motivate scenarios in which the
scalar (and so the fundamental constants) evolve before recombination under the influence of the
cosmological abundance of SM matter.
Such a regime resembles attempts to address coincidence problems in early dark energy via couplings
to dark matter~\cite{Karwal:2021vpk, Lin:2022phm} or neutrinos~\cite{Sakstein:2019fmf,
CarrilloGonzalez:2020oac} to explain why the early dark energy field becomes dynamical around the
time of recombination.
Moreover, a complete treatment of coupled scalars requires an extension to heavier masses for
which the scalar begins oscillating before recombination of its own volition.
Time-dependent fundamental constants around recombination may yield qualitatively
different CMB signatures via, for instance, the interplay of changes to the shape of the visibility
function and diffusion damping rate at earlier times.

In addition, scalars may equally well couple to other SM content like the Higgs, quarks, and gluons,
which would affect the masses of baryons and mediate fifth forces between them.
Fifth forces between electrons are present in the scenarios we considered here, but their
effect is negligible because inhomogeneities in the the electron fluid are themselves negligible in
linear perturbation theory (see \cref{sec:cosmology-hyperlight-scalars}).
We leave a general study of the cosmological signatures of coupled scalars, including the effect of evolution prior to recombination and scalar-mediated forces on the CMB and large-scale structure, to future work.

\begin{acknowledgments}
We thank Nikita Blinov, David Cyncynates, Junwu Huang, Mikhail Ivanov, Hongwan Liu, Cristina Mondino, Caio Nascimento, Maxim Pospelov, Murali Saravanan, Sergey Sibiryakov, Neal Weiner, and Tien-Tien Yu for helpful conversations and especially Marilena Loverde for many useful and extended discussions.
M.B.\ is supported by the U.S. Department of Energy Office of Science under Award No. DE-SC0024375.
M.B.\ and Z.J.W.\ are supported by the Department of Physics and College of Arts and Science at the University of Washington.
O.S.\ was supported by the Department of Physics and a DARE Fellowship from the Office of the Vice Provost for Graduate Education at Stanford University.
This work made use of the software packages
\textsf{emcee}~\cite{Foreman-Mackey:2012any,Hogg:2017akh,Foreman-Mackey:2019},
\textsf{corner.py}~\cite{corner}, \textsf{NumPy}~\cite{Harris:2020xlr},
\textsf{SciPy}~\cite{Virtanen:2019joe}, \textsf{matplotlib}~\cite{Hunter:2007ouj},
\textsf{xarray}~\cite{hoyer2017xarray}, \textsf{ArviZ}~\cite{arviz_2019},
\textsf{SymPy}~\cite{Meurer:2017yhf}, and \textsf{CMasher}~\cite{cmasher}.
\end{acknowledgments}

\appendix

\input{scalar-implementation}

\input{parameter-inference}

\bibliography{bib,manual}

\end{document}

%% file: cosmo-varconst.tex
\section{Cosmology with varying constants}\label{sec:varying-constants}

We begin by grounding our discussion with a detailed review of the impact of variations in
fundamental constants on cosmology, seeking to quantify the parameter dependence of physical effects
measured by cosmological data in the extended \LCDM{} parameter space.
We focus on the cases where the electron mass $m_e$ and the fine-structure constant $\alpha$ are
different prior to recombination than at the present day; in this section we do so without
considering additional effects from any specific microphysical model thereof.
We denote the constants' early-time values as $\alpha_i$ and $m_{e, i}$.
We first discuss early-Universe signatures, summarizing key parameter combinations that are well
constrained by data and the resulting degeneracies in the physics up to and including recombination.
We then consider complementary probes of the late Universe, after the recombination epoch.

As detailed in this section, the usual flat \LCDM{} model has sufficient parameter freedom to
allow for the scale factor of recombination to differ substantially from the standard prediction without affecting the primary CMB anisotropies.
In standard cosmology, the scale factor of recombination is fixed by the ratio of the CMB monopole
temperature at the present time to that at photon last scattering, $a_\star / a_0 = T_0 / T_\star$.
The former is directly and precisely measured~\cite{Fixsen:1996nj,Fixsen:2009ug}, while the latter
is determined by the standard atomic and thermal physics that govern recombination.
This degeneracy direction has been explored in the context of the Hubble tension by changing
$T_\star$ through variations of the early-time electron mass~\cite{Sekiguchi:2020teg}.
Separately, Ref.~\cite{Ivanov:2020mfr} reframed the Hubble tension as a (counterfactual, if illustrative) tension in the present-day CMB temperature $T_0$.
Our discussion captures the arguments of Refs.~\cite{Ivanov:2020mfr,Sekiguchi:2020teg} in terms of the common physics from which they derive; we also comment on connections to the rescaling invariance of the Boltzmann and Einstein equations as explored in detail by Refs.~\cite{Cyr-Racine:2021oal,Ge:2022qws,Greene:2023cro,Greene:2024qis}.

Given that the CMB, baryon acoustic oscillations, and supernovae datasets we consider here are
individually fit well by the \LCDM{} model compared to any known extension thereof, our discussion
centers on what degeneracies can be realized in extended parameter spaces that identically reproduce
\LCDM{} predictions.
That is, we take the best-fit \LCDM{} model as a proxy for the data itself and investigate
correlated changes to parameters that leave its predictions invariant.
Such a description is sufficient to anticipate whether an extension of \LCDM{} might contain
directions in parameter space that should be poorly constrained by data.
However, only a quantitative analysis can assess whether a model's additional freedom in fact
improves upon the fit to individual datasets by explaining additional (possibly unanticipated)
features in those data.
We present the results of parameter inference in \cref{sec:parameter-inference}.

Throughout, we refer to the contributions of various species to the present-day, average energy density with
the abundance parameters $\omega_X = \bar{\rho}_{X, 0} / 3 H_{100}^2 \Mpl^2$, where
$H_0 = h \cdot 100~\mathrm{Mpc}^{-1} \mathrm{km} / \mathrm{s} \equiv h H_{100}$.
Specifically, $\omega_b$, $\omega_c$, $\omega_\Lambda$, $\omega_\gamma$, and $\omega_\nu$ are the
abundances of baryons, cold dark matter, dark energy, photons, and neutrinos, respectively.
We also use $\omega_m = \omega_b + \omega_c$ and $\omega_r = \omega_\gamma + \omega_\nu$
to denote the total matter and radiation abundances, including contributions only from species that are
nonrelativistic and relativistic at early times, respectively.
The early-time energy density of neutrinos is typically quantified in terms of the effective number
of degrees of freedom $N_\mathrm{eff}$ as
\begin{align}
    \omega_\nu
    &= \frac{7}{8} \left( \frac{4}{11} \right)^{4/3} N_\mathrm{eff} \cdot \omega_\gamma,
    \label{eqn:omega-nu-omega-gamma}
\end{align}
which accounts for the increase in the photon temperature relative to the neutrino temperature due
to the entropy transferred from electrons and positrons when they annihilate.
In standard cosmology, the neutrinos' temperature is slightly enhanced by their residual coupling to
the plasma during electron-positron annihilation; numerical solutions of the SM kinetic equations
yield $N_\mathrm{eff} = 3.044$~\cite{Akita:2020szl,Froustey:2020mcq,Bennett:2020zkv}.\footnote{Calculations
of the final digit continue to be refined~\cite{Cielo:2023bqp}, including finite-temperature QED
corrections and neutrino mass mixing~\cite{Akita:2020szl,Froustey:2020mcq,Bennett:2020zkv}.}
If electron-positron annihilation occurs earlier (later) due to a heavier (lighter) electron, then the neutrino temperature would be larger (smaller) than the SM prediction.
However, this effect is small: over the range of electron-mass variations relevant to our main results (at most $20\%$ in either direction), the relative change in $N_\mathrm{eff}$ is a few tenths of a percent~\cite{Birrell:2014uka, Grohs:2017iit}.
As differences of this size are smaller than the anticipated $1 \sigma$ sensitivity of future CMB experiments like
CMB-S4~\cite{CMB-S4:2016ple,Abazajian:2019eic,Dvorkin:2022jyg}, we neglect this effect and fix $N_\mathrm{eff} = 3.044$.

\subsection{Early-Universe signatures}\label{sec:early-universe-signatures}

At linear order in perturbation theory, the effect of varying constants is entirely captured by
accounting for their time dependence in the Thomson cross section and in the recombination equations
whose solution gives the ionization history. These dynamics (implemented in \textsf{CLASS}~\cite{Blas:2011rf, Lesgourgues:2011re} and \textsf{HyRec}~\cite{Ali-Haimoud:2010hou,Lee:2020obi}) constitute the focus of our analysis.
Subleading effects include changes to the electrons' and baryons' energy densities through a varying electron mass and fine-structure constant, respectively.
As electrons make up a small fraction $\propto m_e / m_p \approx 1/1836$ of the total energy density in ``baryonic'' matter and the electromagnetic contribution to the nucleon masses is at the permille level~\cite{particle-paper}, we neglect these corrections.

In \cref{sec:recombination}, we study the effect of the fundamental constants on both when
and how quickly recombination proceeds.
The earlier recombination occurs, the less abundant dark matter and baryons are relative to radiation;
\cref{sec:photon-baryon-plasma} explains the resulting effect on the dynamics of the photon-baryon
plasma.
Photon diffusion (\cref{sec:damping}) is sensitive to the fundamental constants not only implicitly
via the time recombination occurs but also explicitly via the Thomson cross section.
In addition, the faster recombination occurs, the narrower the interval over which photons last
scatter; \Cref{sec:damping,sec:polarization} discuss the resulting impact on small-scale damping and
the generation of polarization.
Finally, the earlier hydrogen recombines, the smaller the sound horizon is at last scattering.
\Cref{sec:angular-power-spectra} shows how the distance to last scattering must compensate to
preserve its manifestation in angular correlations on the sky.
Drawing from these derivations, in \cref{sec:degeneracies} we discuss numerous degeneracies in
extended \LCDM{} parameter space for which the primary CMB anisotropies are completely (or
partially) unchanged.

\subsubsection{Recombination}\label{sec:recombination}

The CMB observed today comprises photons that last scattered off electrons with a distribution in
time determined by the physics of electrons recombining into neutral atoms and Thomson scattering.
This distribution---the so-called visibility function---encodes both when (i.e., at what
temperature) and over what redshift interval most CMB photons last scattered; both features directly
impact the visible anisotropies in CMB temperature and polarization.
The Thomson scattering rate is
\begin{align}
    \dd{\kappa}{\tau}
    &\equiv a \bar{n}_e \sigma_T,
\end{align}
where the cross section $\sigma_T = 8 \pi \alpha_i^2 / 3 m_{e, i}^2$ and $\bar{n}_e$ is the number density
of free electrons.
The visibility function is then defined by
\begin{align}
    g(\tau)
    &= \dd{\kappa}{\tau}
        \exp\left[
            - \int_{\tau}^{\tau_0} \ud \tilde{\tau} \,
                \left. \dd{\kappa}{\tau} \right\vert_{\tau = \tilde{\tau}}
        \right]
    \equiv \kappa' e^{ - \kappa }
    \label{eqn:def-visibility}
\end{align}
[the latter equality defining $\kappa(\tau)$ as the optical depth from today, $\tau_0$].
We define the moment of recombination as the time of maximum visibility, with the scale factor of
recombination $a_\star$ given by $g_\star \equiv g(\tau(a_\star)) = \max g(\tau(a))$.
Quantities with a subscript star generally indicate those evaluated at recombination.

The dependence on the fundamental constants of the Thomson scattering cross section is manifest in
its definition, but that of the electron ionization history is slightly more complex.
The visibility function peaks roughly when the photons decouple from the plasma, which occurs when
the Universe becomes nearly neutral.
Defining the ionization fraction $X_e \equiv \bar{n}_e / n_\mathrm{H}$, where $n_\mathrm{H}$ is the number
density of hydrogen nuclei, ionized and not, and neglecting the presence of helium (which
effectively
finishes recombining before hydrogen recombination begins),
\begin{align}
    \bar{n}_e
    &= X_e \frac{(1 - Y_\mathrm{He}) \bar{\rho}_b}{m_\mathrm{H}}.
    \label{eqn:free-electron-number-def}
\end{align}
Here $Y_\mathrm{He} \equiv \bar{\rho}_\mathrm{He} / \bar{\rho}_b$ is the helium fraction by mass and
$m_\mathrm{H}$ the mass of hydrogen.
In spite of the complexities of the recombination
process~\cite{Peebles:1968ja,Peebles:1970ag,Seager:1999bc,Ali-Haimoud:2010hou,Ali-Haimoud:2010tlj},
its dynamics largely depend on temperature relative to the ionization energy of hydrogen,
$E_I = \alpha_i^2 m_{e, i} / 2$.
The temperature at recombination therefore scales as $T_\star \propto \alpha_i^2 m_{e, i}$, so
\begin{align}
    a_\star
    &= \frac{T_0}{T_\star}
    \propto \frac{\sqrt[4]{\omega_\gamma}}{\alpha_i^2 m_{e, i}}.
    \label{eqn:a-star-proportionality}
\end{align}

\begin{figure}[th!]
    \centering
    \includegraphics[width=\textwidth]{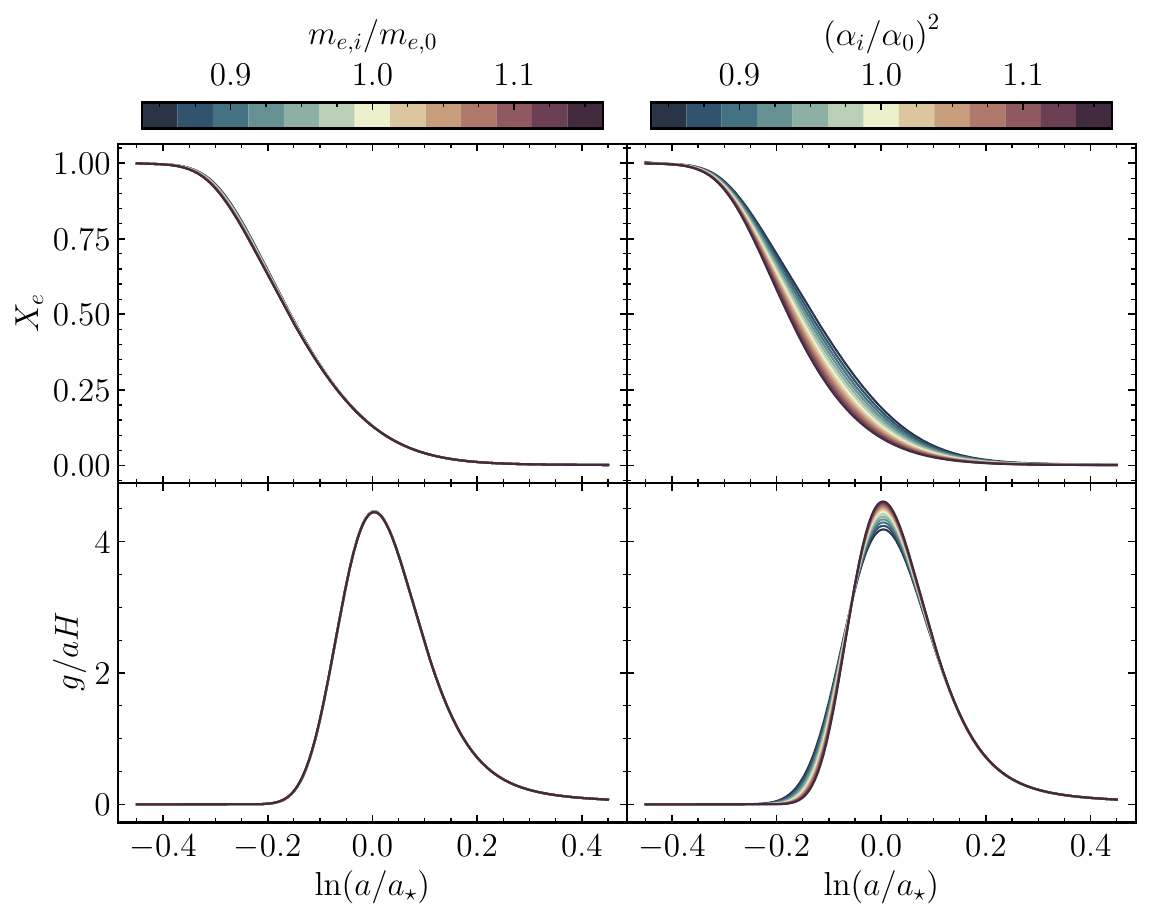}
    \caption{
        Impact on the ionization fraction [\cref{eqn:free-electron-number-def}, top panels] and the dimensionless visibility function [\cref{eqn:g-star-dependence}, bottom panels] in
        cosmologies varying the electron mass $m_{e, i}$ (left panels) and the fine-structure
        constant $\alpha_i$ (right panels) by color.
        The colors on the right-hand panels label values of $\alpha_i^2$ so that curves of the same
        color in the left-hand panels depict the same shift in the time of recombination.
        Results are plotted versus the logarithm of the scale factor relative to that when the
        visibility function peaks, $a_\star$, to emphasize changes in dimensionless shape; these are negligible for varying $m_{e, i}$ but significant for varying $\alpha_i$.
        Each curve varies $\omega_b$ and $\omega_c$ in proportion to $1/a_\star$ to keep baryon to
        photon and matter to radiation densities fixed at recombination
        (\cref{sec:photon-baryon-plasma}) and fixes the helium mass fraction $Y_\mathrm{He}$ (whose
        effect is discussed in \cref{sec:damping}).
        All results also fix the angular size of the sound horizon $\theta_s$
        (\cref{sec:angular-power-spectra}), a choice which has a negligible effect on these results.
    }
    \label{fig:ionization-visibility-varying-constants}
\end{figure}

The scale factor relative to that at recombination, $x \equiv a / a_\star$, proves to be the most
convenient time coordinate with which to characterize the CMB anisotropies~\cite{Sekiguchi:2020teg}.
Moreover, writing the photon visibility as a dimensionless distribution over $\ln x$, i.e., as
$g / a H = e^{- \kappa} \ud \kappa / \ud \ln x$, isolates superficial changes to its shape that only
arise in a dimensionful form [e.g., over conformal time as in \cref{eqn:def-visibility}].
\Cref{fig:ionization-visibility-varying-constants} depicts the effect of varying constants on the
ionization history $X_e$ and on the dimensionless visibility function; plotted against the scale
factor relative to that of peak visibility, both are nearly indistinguishable when varying the
electron mass by $15\%$ in either direction.
The dimensionless rate of recombination (i.e., the change in the ionization fraction with $\ln a$)
increases more noticeably with the fine-structure constant, evident in the change to $X_e(x)$ in
\cref{fig:ionization-visibility-varying-constants}, due to the more sensitive dependence of the
effective rates of recombination, photoionization, etc., on $\alpha_i$~\cite{Planck:2014ylh,
Hart:2017ndk}.
As a consequence, the visibility function narrows and broadens substantially more when varying the
fine-structure constant than the electron mass (i.e., at a commensurate variation in $a_\star$).
We discuss the physical impacts of these changes to the visibility and ionization history and
empirically determine their dependence on $\alpha_i$ in \cref{sec:damping}.
In addition, $a_\star$ in \cref{fig:ionization-visibility-varying-constants} increases slightly
faster than estimated from the ionization energy in \cref{eqn:a-star-proportionality} $\alpha_i$; a
numerical fit estimates $a_\star \propto \alpha_i^{-2.08}$.

\subsubsection{Acoustic oscillations and matter effects}\label{sec:photon-baryon-plasma}

The dynamics of the photon-baryon plasma are primarily driven by three effects: the competition
between radiation pressure and gravitational collapse, the increasing importance of matter to the
expansion rate and gravitational potentials, and the damping of plasma oscillations on scales below
the photons' mean free path.
The semianalytic solutions of Ref.~\cite{Weinberg:2008zzc} (see also Ref.~\cite{Baumann:2022mni})
suggest that the Sachs-Wolfe, Doppler, and damping effects depend on cosmological parameters only
through particular dimensionless combinations and depend on wave number only relative to the sound
scale at recombination $1/r_{s, \star}$, to the comoving Hubble scale at matter-radiation equality
$k_\mathrm{eq}$, and to the effective damping scale $k_{D, \star}$.
We discuss (and define) the role of each in turn.

The importance of baryons to the dynamics of the plasma is measured by the baryon-to-photon ratio
\begin{align}
    R(x)
    \equiv \frac{3 \bar{\rho}_b}{4 \bar{\rho}_\gamma}
    &= R_\star x,
    \label{eqn:baryon-to-photon-ratio-ito-x}
\end{align}
written in terms of that at recombination, $R_\star \equiv 3 \omega_b a_\star / 4 \omega_\gamma$.
Acoustic waves in the photon-baryon fluid travel with an effective sound speed
\begin{align}
    c_s(x)
    &= \frac{1}{\sqrt{3} \sqrt{1 + R(x)}},
\end{align}
with a corresponding sound horizon of
\begin{align}
    r_{s}(a)
    &\equiv \int_{0}^{a} \frac{\ud \tilde{a}}{\tilde{a}} \,
        \frac{c_s(\tilde{a})}{\tilde{a} H(\tilde{a})}
    = \frac{a_\star}{\sqrt{3}}
        \int^{a / a_\star}_0 \ud x \, \frac{1}{\sqrt{1 + R_\star x}}
        \frac{1}{(a_\star x)^2 H(x)}.
    \label{eqn:sound-horizon-ito-x}
\end{align}
Independently of the physical extent of the sound horizon itself, the measured, relative heights of
the acoustic peaks tightly constrain $R_\star$.
At recombination, dark energy is negligible and the Universe well described by matter and radiation
alone.
The latter term in the integrand of \cref{eqn:sound-horizon-ito-x} thus takes the form
\begin{align}
    a^2 H
    = (a_\star x)^2 H(x)
    &= H_{100} \sqrt{\omega_r}
        \sqrt{1 + x / x_\mathrm{eq}},
    \label{eqn:a2H-matter-radiation-ito-x}
\end{align}
where
\begin{align}
    x_\mathrm{eq}
    &= a_\mathrm{eq} / a_\star
    \label{eqn:def-x-eq}
\end{align}
is the scale factor at equality relative to that at recombination.
Since $a_\mathrm{eq} / a_0 = \omega_r / \omega_m$,
$x_\mathrm{eq} = \bar{\rho}_{r, \star} / \bar{\rho}_{m, \star}$
measures the ratio of the radiation and matter densities at recombination.
Because \cref{eqn:a2H-matter-radiation-ito-x} is independent of $a_\star$, the sound horizon at
recombination [\cref{eqn:sound-horizon-ito-x}] is simply proportional to
$a_\star / \sqrt{\omega_r}$ if $x_\mathrm{eq}$ and $R_\star$ are held fixed; indeed, in a
matter-radiation Universe \cref{eqn:sound-horizon-ito-x} takes the closed
form~\cite{Eisenstein:1997ik}
\begin{align}
    r_{s, \star}
    &= \frac{2 a_\star \sqrt{ x_\mathrm{eq} / R_\star } }{H_{100} \sqrt{3 \omega_r}}
        \ln \left(
            \frac{
                \sqrt{R_\star} \sqrt{1 + x_\mathrm{eq}}
                + \sqrt{1 + R_\star}
            }{
                1 + \sqrt{R_\star x_\mathrm{eq}}
            }
        \right)
    ,
    \label{eqn:rs-in-matter-radiation-ito-wr-R_star}
\end{align}
which is exact in \LCDM{} up to the $\mathcal{O}(10^{-9})$ relative contribution from dark energy at
recombination.

The relative importance of matter to gravity (at the background and perturbed level) at early times
is fully characterized by when matter-radiation equality occurs relative to recombination
($x_\mathrm{eq}$).
The heights of the peaks are sensitive to the onset of matter domination via the radiation driving
effect~\cite{Hu:1996mn,Hu:1995en}: oscillations of the plasma are driven by the decay of the
gravitational potentials at horizon crossing, but to a decreasing extent as the Universe becomes
increasingly dominated by matter.
The early integrated Sachs-Wolfe (ISW) effect similarly depends on the time evolution of
metric potentials as the Universe transitions from radiation domination, i.e., on $x_\mathrm{eq}$
only~\cite{Vagnozzi:2021gjh}.
Moreover, in Universes that are well described by the matter-radiation solution, the full
scale-dependent shape of these effects is identical for cosmologies with the same
$x_\mathrm{eq}$---a feature Ref.~\cite{Adil:2022hkj} emphasized as important beyond simply fixing
the relative size of the sound horizon and the comoving horizon at equality.
In particular, the comoving Hubble rate in a matter-radiation Universe is
\begin{align}
    a H
    &= \frac{\sqrt{\omega_r}}{a_\star} \frac{H_{100}}{x_\mathrm{eq}}
        \sqrt{(x / x_\mathrm{eq})^{-2} + (x / x_\mathrm{eq})^{-1}}
    \equiv \frac{k_\mathrm{eq}}{\sqrt{2}}
        \sqrt{(x / x_\mathrm{eq})^{-2} + (x / x_\mathrm{eq})^{-1}},
    \label{eqn:comoving-hubble-mr}
\end{align}
defining $k_\mathrm{eq}$ as the comoving Hubble scale at equality ($x = x_\mathrm{eq}$).
Just like the sound horizon, the comoving horizon at equality ($1 / k_\mathrm{eq}$) scales with
$a_\star / \sqrt{\omega_r}$ if $x_\mathrm{eq}$ is held fixed, situating all associated features at
the same relative scales.

In sum, the shape of the CMB power spectrum, as a dimensionless observable, by itself measures
relative energy densities at recombination.
The balance of photon pressure and the baryons' weight determines the amplitude of acoustic
oscillations, constraining the parameter combination
$R_\star \propto \omega_b a_\star / \omega_\gamma$, while the radiation driving effect depends on
when (relative to recombination) matter overtakes radiation in the Universe's energy budget,
constraining $x_\mathrm{eq} \propto \omega_r / \omega_m a_\star$.
These combinations may be held fixed trivially because $\omega_b$ and
$\omega_c = \omega_m - \omega_b$ are free parameters in standard cosmology, in which case
$r_{s, \star} k_\mathrm{eq}$ is constant.
These phenomena alone are therefore unable to discern whether the photon temperature today or at
recombination differ from the SM prediction [\cref{eqn:a-star-proportionality}], or even if the
radiation density $\omega_r$ is different than that from photons and neutrinos as predicted in the
SM.

\subsubsection{Small-scale damping}\label{sec:damping}

Fluctuations in the plasma are damped on small scales by two distinct effects, one due to photon
diffusion and the other due to the finite duration of recombination.
We consider each of their parameter dependence in turn.
The damping scale associated to photon diffusion is~\cite{Hu:1995em,Zaldarriaga:1995gi}
\begin{align}
    \frac{1}{k_{D, \star}^2}
    &\equiv \int_{0}^{1}
        \frac{\ud x / x}{(a_\star x H)^2}
        \frac{R^2 + 16 (1 + R) / 15}{6 (1 + R)^2}
        \left( \dd{\kappa}{\ln x} \right)^{-1}
    \label{eqn:inverse-damping-scale-squared}
    .
\end{align}
The baryon-to-photon ratio $R$ is invariant as a function of $x$ in cosmologies with the same value
for $R_\star$.
Because the Thomson scattering rate per $e$-fold, $\ud \kappa / \ud \ln x = \bar{n}_e \sigma_T / H$,
is dimensionless, the comoving diffusion scale $k_{D, \star}$ is proportional to $k_\mathrm{eq}$
[\cref{eqn:comoving-hubble-mr}] at fixed $x_\mathrm{eq}$.
Whether or not $k_{D, \star} / k_\mathrm{eq}$ varies with parameters is therefore determined by the
dimensionless Thomson scattering rate.

Inserting the definition of free electron number density $\bar{n}_e$,
\cref{eqn:free-electron-number-def}, and rewriting the baryon abundance in terms of $R_\star$ gives
\begin{align}
    \dd{\kappa}{\ln x}
    = \frac{n_e(x) \sigma_T}{H}
    &= \frac{R_\star K_\star}{x_\mathrm{eq}}
        \frac{X_e(x)}{\sqrt{(x / x_\mathrm{eq})^{2} + (x / x_\mathrm{eq})^{3}}}
    \label{eqn:scattering-rate-combo-for-damping-scale}
\end{align}
where
\begin{align}
    K_\star
    &\equiv 4 H_{100} \Mpl^2
        \frac{\sqrt{\omega_\gamma} \sigma_T}{a_\star^2}
        \sqrt{\frac{\omega_\gamma}{\omega_r}}
        \frac{1 - Y_\mathrm{He}}{m_\mathrm{H}}
    \label{eqn:K-star}
\end{align}
is a rescaling of the Thomson scattering rate at recombination that encodes its parameter dependence
complementary to that of $R_\star$, $x_\mathrm{eq}$, and the ionization history.
In more detail, the acoustic dynamics of the plasma (\cref{sec:photon-baryon-plasma}) motivate
expressing the density of electron scatterers relative to the photon energy density at
recombination, i.e., in terms of the baryon-to-photon ratio $R_\star$ via
$n_e \propto R_\star x \rho_\gamma = R_\star \rho_{\gamma, \star} / x^4$.
The impact of matter on plasma dynamics suggests expressing the Hubble rate in terms of the
(square root of) the total radiation density and encoding the matter contribution with
$x_\mathrm{eq}$ as
$H \propto \sqrt{\rho_{r, \star} / x^4} \sqrt{1 + x / x_\mathrm{eq}}$.
Their ratio yields an overall dependence on
$\sqrt{\rho_{\gamma, \star}} \propto \sqrt{\omega_\gamma} / a_\star^2 \propto T_\star^2$ times the
square root of the fraction of the total radiation density in photons.
Finally, the factor of $(1 - Y_\mathrm{He}) / m_\mathrm{H}$ in \cref{eqn:K-star} simply translates the total
number density of baryons to that of electrons that do not recombine into helium.

If the ionization history $X_e$ is unchanged as a function of $x$ and both $R_\star$ and
$x_\mathrm{eq}$ are fixed, then the full $x$ dependence of the integrand of
\cref{eqn:inverse-damping-scale-squared} is unchanged.
The diffusion damping scale then scales in tandem with $k_\mathrm{eq}$ and $1/r_{s, \star}$ if
$K_\star$ is fixed.
Inserting \cref{eqn:a-star-proportionality} for $a_\star$,
\begin{align}
    K_\star
    &\propto
        \alpha_i^6
        \sqrt{\frac{\omega_\gamma}{\omega_r}}
        \frac{1 - Y_\mathrm{He}}{m_\mathrm{H} / \Mpl}
    \label{eqn:damping-combo-ito-constants}
\end{align}
At fixed $K_\star$, $x_\mathrm{eq}$, and ionization history, the impact of diffusion damping on CMB
anisotropies is independent of both the early-time electron mass $m_{e, i}$ and the present photon
temperature $T_0$.
Namely, the increase of the electron density relative to the Hubble rate with $T_\star^2 \propto
m_{e, i}^2 \alpha_i^4$ compensates the Thomson cross section's inverse dependence on $m_{e, i}^2$ but compounds
with its proportionality to $\alpha_i^2$.
The irrelevance of $T_0$ merely reflects that diffusion damping only depends on the properties of
the Universe during recombination, not on the expansion that has elapsed since then.
\Cref{eqn:damping-combo-ito-constants} further encodes the salient dependence of the diffusion scale
on the helium yield and the effective number of neutrino species
[\cref{eqn:omega-nu-omega-gamma}].\footnote{
    The dynamics of the plasma are also sensitive to what fraction of radiation is collisionless
    (i.e., freely streams) or fluidlike; in particular, that SM neutrinos free stream sources
    distinctive signatures in the CMB by, for instance, shifting the location of the acoustic
    peaks~\cite{Bashinsky:2003tk, Hou:2011ec, Baumann:2015rya, Pan:2016zla, Ge:2022qws}.
    Therefore, our treatment tacitly takes the fraction of radiation that is free-streaming to
    be fixed---a rather artificial choice, but our purpose is only to highlight the connection of
    the net radiation density (independent of its characteristics) to other cosmological parameters
    and not to specifically explore scenarios featuring new light degrees of freedom.
}

The diffusion parameter $K_\star$ captures the relevant parameter dependence of diffusion damping,
except for any incurred impact to the ionization history;
\cref{fig:ionization-visibility-varying-constants} shows that this effect is negligible when varying
$m_{e, i}$ but not $\alpha_i$.
The damping scale involves an integral over $X_e$, and to the extent that $\alpha_i$ simply
rescales $X_e$, we could estimate $k_{D, \star} \propto \sqrt{X_{e}}$ (on top of its other parameter
dependence).
However, in the short interval before recombination ($\Delta \ln a \sim 0.4$), $X_e / X_{e, \LCDM}$
varies rapidly in proportionality between $\alpha_i^0$ and $\alpha_i^{-5}$ (with $R_\star$ and
$x_\mathrm{eq}$ fixed); the net effect on the damping scale is an integral over this variation.
Changes to the ionization history therefore cancel some of the direct dependence of the damping
scale on $\alpha_i$ in $K_\star$.
We determine empirically that $a_\star k_{D, \star} \sim \alpha_i^{1.63}$ at fixed $Y_\mathrm{He}$.
Similarly, the helium mass fraction $Y_\mathrm{He}$ enters both directly in the Thomson scattering
rate and through its effect on the ionization history, which does shift appreciably.
Leading up to recombination, $X_e$ varies rapidly in proportionality between
$(1 - Y_\mathrm{He})^0$ and $(1 - Y_\mathrm{He})^{-0.8}$, reducing the damping scale's overall
dependence on $Y_\mathrm{He}$ just as for $\alpha_i$.
We compute that $a_\star k_{D, \star} \propto (1 - Y_\mathrm{He})^{0.24}$ in total.
Moreover, the helium mass fraction is also itself sensitive to $\alpha_i$ and $m_{e, i}$ through the
dynamics of nucleosynthesis.
We derive the dependence of $Y_\mathrm{He}$ on parameters in Ref.~\cite{particle-paper}, leading to
$a_\star k_{D, \star} \sim \alpha_i^{1.42}$.
The helium mass fraction scales much more weakly with the electron mass, leading
to $a_\star k_{D, \star} \propto m_{e, i}^{-0.05}$.

The second effect that damps small-scale CMB power is due to the nonzero width of the visibility
function, which effectively averages rapidly oscillating modes over multiple
oscillations~\cite{Zaldarriaga:1995gi,Weinberg:2008zzc}.
This effect is referred to as Landau damping for its analogy to the damping of oscillations
comprising a spread of frequencies.
Approximating the visibility function as a Gaussian about its peak as
$g(\tau) \approx \exp( - [\tau - \tau_\star]^2 / 2 \sigma_g^2) / \sqrt{2 \pi} \sigma_g$, the
comoving width of the last-scattering surface may be approximated by
$\sigma_g = 1 / \sqrt{2 \pi} g_\star$.
However, $\sigma_g$ (as an interval in conformal time) also depends explicitly on the expansion rate
at recombination, which, for example, is larger for earlier recombination.
The effect solely due to the shape of the visibility function is captured by the width of the
visibility function relative to $a H$ (i.e., as a dimensionless distribution over
$\ln x$), which is approximately $a_\star H_\star / \sqrt{2 \pi} g_\star$.
One may define an effective Landau damping scale in analogy to the diffusion scale by combining a
Gaussian approximation to the visibility function with analytic solutions for the photon
perturbations in the tight-coupling approximation~\cite{Weinberg:2008zzc}, yielding
\begin{align}
    k_L
    &= \frac{\sqrt{6 (1 + R_\star)}}{\sigma_g}
    = \frac{\sqrt{\omega_r}}{a_\star} H_{100}
        \frac{g_\star}{a_\star H_\star}
        \sqrt{12 \pi (1 + R_\star) \left( 1 + x_\mathrm{eq}^{-1} \right)}
    \label{eqn:landau-damping-scale}
    .
\end{align}
To the extent that $X_e$ is a fixed function of $x$ (such that $g_\star / a_\star H_\star$ is
unchanged) the Landau damping scale is proportional to the aforementioned physical scales at fixed
$R_\star$ and $x_\mathrm{eq}$.

\Cref{fig:ionization-visibility-varying-constants} shows that $g_\star / a_\star H_\star$ is
effectively independent of $m_{e, i}$ (scaling only $\sim m_{e, i}^{-0.01}$); we empirically find that varying
$\alpha_i$ and $Y_\mathrm{He}$ independently yields
\begin{align}
    \frac{g_\star}{a_\star H_\star}
    &\approx 4.45 \left( \frac{\alpha_i}{\alpha_0} \right)^{0.61}
        \left( \frac{1 - Y_\mathrm{He}}{1-0.245} \right)^{0.11}
    \label{eqn:g-star-dependence}
\end{align}
at fixed $R_\star$ and $x_\mathrm{eq}$.
Therefore, $r_{s, \star} k_L$ scales with roughly one fewer power of $\alpha_i$ than does
$r_{s, \star} k_{D, \star}$.
Since the Landau and diffusion damping scales are of similar size (and combine in quadrature),
measurements of the damping tail do not cleanly translate to measurements of either of the
individual parameter combinations in \cref{eqn:damping-combo-ito-constants,eqn:g-star-dependence};
however, CMB polarization is independently sensitive to $g_\star / a_\star H_\star$ as
discussed next.

\subsubsection{Polarization}\label{sec:polarization}

The amplitude of the CMB polarization spectrum relative to the temperature spectrum is also
sensitive to the duration of recombination~\cite{Zaldarriaga:1995gi}.
Polarization is generated by Thomson scattering when the temperature distribution has a nonnegligible
quadrupole.
The quadrupole is sourced by the dipole of the distribution, which is itself proportional to the time
derivative of the temperature fluctuation (the monopole)~\cite{Zaldarriaga:1995gi,Hu:1997hv}.
Efficient Thomson scattering, however, impedes the growth of anisotropy (like the quadrupole).
The observed CMB polarization is thus generated as the photons decouple, when scattering still
occurs and generates polarization but not so frequently that the quadrupole is suppressed.

The average interval between scatterings around decoupling is comparable to the width of the
visibility function and sets the relative amplitude of the polarization and temperature
perturbations proportional to $c_s k \sigma_g$---i.e., truly instantaneous recombination generates
no polarization.
In other words, the longer recombination takes, the more the quadrupole of the temperature
distribution is able to grow, generating greater polarization at last scattering.
(Incidentally, the amplitude of the polarization spectrum is also sensitive to the baryon-to-photon
ratio $R_\star$ via the factor of the sound speed $c_s$.)
Thus, the polarization amplitude uniquely probes the width of the visibility function, with
parameter dependence given by \cref{eqn:g-star-dependence}.
Reference~\cite{Hadzhiyska:2018mwh} provides subpercent-level constraints directly on the width of the
visibility function, introduced as a phenomenological extension of the \LCDM{} model; the
constraints are driven to a comparable extent by the effects on Landau damping and on the amplitude of
the polarization spectrum.
We therefore expect CMB polarization data to provide complementary constraining power on variations
in the fine-structure constant.
Because the shape of the visibility function varies negligibly with the electron mass
(per \cref{fig:ionization-visibility-varying-constants}), we further anticipate that polarization
offers no unique information for electron-mass variations compared to temperature data.

\subsubsection{Angular power spectra}\label{sec:angular-power-spectra}

The angular power spectrum of the CMB measures the scale of the aforementioned features---the sound
horizon, the horizon at equality, and the diffusion and Landau damping scales---only relative to the
transverse comoving distance to last scattering.
In cosmologies with nonzero curvature $\Omega_k$, the transverse comoving distance
is~\cite{Hogg:1999ad}
\begin{align}
    D_M(a)
    &\equiv \chi(a) \sinc \left[ \sqrt{-\Omega_k} \frac{\chi(a)}{1 / H_0} \right],
    \label{eqn:transverse-distance}
\end{align}
where the line-of-sight comoving distance is
\begin{align}
    \chi(a)
    &= \int_{a}^{a_0} \frac{\ud \tilde{a}}{\tilde{a}} \,
        \frac{1}{\tilde{a} H(\tilde{a})}.
    \label{eqn:comoving-distance}
\end{align}
The transverse distance reduces to the line-of-sight one [$D_M(a) = \chi(a)$] for zero curvature, which we assume for the remainder of the paper.
In a Universe with only matter and dark energy, characterized by a scale factor of matter-$\Lambda$
equality $\amL = \sqrt[3]{\omega_m / \omega_\Lambda}$ and a Hubble constant $h = \sqrt{\omega_m + \omega_\Lambda}$, the comoving distance
\cref{eqn:comoving-distance} integrates to~\cite{Baes:2017rfj}
\begin{align}
    \chi(a)
    &= \frac{2}{\sqrt{\omega_m} H_{100}}
        \left[ F_M(1; \amL) - F_M(a; \amL) \right],
    \label{eqn:comoving-distance-matter-Lambda}
\end{align}
where
\begin{align}
    F_M(a; \amL)
    &\equiv \sqrt{a} \cdot {}_2{F}_1(1/6, 1/2; 7/6, - [a / \amL]^3)
    \label{eqn:matter-lambda-distance-function}
\end{align}
and ${}_2{F}_1$ is a hypergeometric function.

The precisely measured angular extent of the sound horizon is a ratio of two scales,
$\theta_s = r_{s, \star} / D_{M, \star}$, that each depend only on early- or late-time physics.
The support of the comoving distance integral in \cref{eqn:comoving-distance} is dominated at late
times around matter-$\Lambda$ equality, since $a H$ decreases in the matter era but increases in
dark-energy domination; the sound horizon \cref{eqn:sound-horizon-ito-x} is of course only sensitive
to early times, dominated by the period shortly before recombination.
Attempts to explain the discrepancy in $H_0$ inferences are thus typically categorized as early-time
or late-time solutions~\cite{Knox:2019rjx}.

The sound horizon at recombination \cref{eqn:rs-in-matter-radiation-ito-wr-R_star} may be written as
\begin{align}
    r_{s, \star}
    &= \frac{2 a_\star}{H_{100} \sqrt{3 \omega_r}} F_{r_s}(R_\star, x_\mathrm{eq}),
\end{align}
where the dependence on $x_\mathrm{eq}$ and $R_\star$, which are constrained by the shape of the CMB
power spectrum (\cref{sec:photon-baryon-plasma}), scales as
$F_{r_s}(R_\star, x_\mathrm{eq}) \propto x_\mathrm{eq}^{0.247} R_\star^{-0.0964}$ when numerically
approximated about the best-fit \LCDM{} parameters~\cite{Planck:2018vyg}.
The comoving distance to last scattering does not depend explicitly upon either $a_\star$
or $\omega_r$ but does so implicitly via the matter density when holding $x_\mathrm{eq}$ fixed.
Replacing $\omega_m$ in \cref{eqn:comoving-distance-matter-Lambda} with
$\omega_r / a_\star x_\mathrm{eq}$ and taking the limit of a matter Universe, for which
$F_M(a; \amL \to \infty) \to \sqrt{a}$,
$\chi_\star \approx \chi(0) = 2 \sqrt{a_\star x_\mathrm{eq}} / \sqrt{\omega_r} H_{100}$.
In this limit,
$\theta_s = \sqrt{a_\star} F_{r_s}(R_\star, x_\mathrm{eq}) / \sqrt{3 x_\mathrm{eq}} \propto \sqrt{a_\star} / x_\mathrm{eq}^{0.253} R_\star^{0.0964}$.
In reality, $F_M(1; \amL) \propto (\sqrt{\omega_m} / h)^{0.193}$ about the best-fit \LCDM{}
cosmology, which encodes the correction to the distance to last scattering due to dark energy.
Substituting for $\omega_m$ as above [and accounting for the small difference between $\chi_\star$
and $\chi(0)$],
\begin{align}
    \label{eqn:matter-radiation-rs-power-law}
    r_{s, \star}
    &\propto \frac{a_\star}{\sqrt{\omega_r}}
        x_\mathrm{eq}^{0.247} R_\star^{-0.0964}
\intertext{and}
    \chi_{\star}
    &\propto \omega_m^{-0.403} h^{-0.193} a_\star^{-0.0173}
        = \frac{a_\star}{\sqrt{\omega_r}}
        \left(
            \frac{\sqrt{\omega_r}}{a_\star^{3.18} h}
            x_\mathrm{eq}^{2.09}
        \right)^{1/5.173}.
    \label{eqn:matter-radiation-chi-power-law}
\end{align}
Combining the two generalizes the well-known dependence of $\theta_s$ on cosmological
parameters~\cite{Planck:2018vyg,2dFGRSTeam:2002tzq,Howlett:2012mh} to
\begin{align}
    \theta_s
    &\propto \left(
            \frac{a_\star^{3.18} h}{\sqrt{\omega_r}}
            x_\mathrm{eq}^{-0.810} R_\star^{-0.499}
        \right)^{1/5.173}
    \propto \left(
            \frac{T_0^{1.18} h}{\alpha_i^{6.36} m_{e, i}^{3.18}}
            \sqrt{\frac{\omega_\gamma}{\omega_r}}
            x_\mathrm{eq}^{-0.810} R_\star^{-0.499}
        \right)^{1/5.173},
    \label{eqn:theta-s-degeneracy}
\end{align}
the second proportionality substituting for $a_\star$ with \cref{eqn:a-star-proportionality}.
Within the \LCDM{} model, $a_\star$ is effectively fixed and \Planck{}~\cite{Planck:2018vyg}
measures $x_\mathrm{eq}$ and $R_\star$ with roughly $0.9\%$ and $0.7\%$ uncertainty, respectively,
while $\theta_s$ is measured with a comparatively negligible $0.03\%$ uncertainty.
Holding $T_0$ and $\omega_r / \omega_\gamma$ fixed, the uncertainty on $x_\mathrm{eq}$ therefore
translates to a $\sim 0.7\%$ uncertainty on $h$, an estimate reproduced in full parameter inference.

Any given parametrization of the primordial power spectrum must be rescaled to account for
the shift in correspondence between wave number (in arbitrary units) and the scales that
characterize the physical effects discussed above.
The standard power-law model,
\begin{align}
    \Delta_\mathcal{R}^2(k)
    &= A_s \left( \frac{k}{k_\mathrm{p}} \right)^{n_s - 1},
    \label{eqn:def-scalar-power-spectrum}
\end{align}
is overparametrized with a normalization $A_s$ and pivot scale
$k_\mathrm{p}$, the latter of which is conventionally fixed.
Thus, instead of taking $k_\mathrm{p}$ to scale in tandem with other scales in the problem
(i.e., in proportion to $\sqrt{\omega_r} / a_\star$), we choose $A_s$ to fix the value of
$A_s \left( \sqrt{\omega_r} / a_\star \right)^{n_s - 1}$.
Note that, because the Universe reionizes at late times, the observed amplitude of CMB anisotropies
is suppressed by a further factor of $e^{- 2 \tau_\mathrm{reion}}$, where $\tau_\mathrm{reion}$ is
the optical depth to reionization.
CMB observations on small scales therefore only constrain the combination $A_s e^{- 2
\tau_\mathrm{reion}}$, but reionization also enhances CMB polarization on large scales (multipoles
$\ell \lesssim 30$), breaking this degeneracy.

\subsubsection{Degeneracies of the early Universe}\label{sec:degeneracies}

We now summarize what parameter directions leave the primary CMB anisotropies invariant per the
preceding discussion.
Namely, we identify degeneracies that preserve the relative abundance of baryons and photons at
recombination,
\begin{subequations}\label{eqn:constrained-parameter-combos}
\begin{align}
    R_\star
    &= \frac{3 \rho_{b, \star}}{4 \rho_{\gamma, \star}}
    \propto \frac{3 \omega_b}{4 \omega_\gamma} \frac{T_0}{\alpha_i^2 m_{e, i}}
    \propto \frac{3 \omega_b}{4 \alpha_i^2 m_{e, i} T_0^3}
\intertext{the relative abundance of matter and radiation at recombination,}
    x_\mathrm{eq}
    &= \frac{\rho_{r, \star}}{\rho_{m, \star}}
    \propto \frac{\omega_r}{\omega_m} \frac{\alpha_i^2 m_{e, i}}{T_0}
    \propto \frac{\alpha_i^2 m_{e, i} T_0^3}{\omega_m}
        \frac{\omega_r}{\omega_\gamma}
\intertext{the Thomson scattering rate per Hubble time,}
    K_\star
    &\propto \alpha_i^6
        \sqrt{\frac{\omega_\gamma}{\omega_r}}
        \frac{1 - Y_\mathrm{He}}{m_\mathrm{H} / \Mpl},
    \label{eqn:constrained-parameter-combos-K-star}
\intertext{the width of the dimensionless visibility function,}
    \frac{g_\star}{a_\star H_\star}
    &\propto \alpha_i^{0.61} \left( 1 - Y_\mathrm{He} \right)^{0.11}
\intertext{and the angular size of the sound horizon,}
    \theta_s
    &\propto \left(
            \frac{T_0^{1.18} h}{\alpha_i^{6.36} m_{e, i}^{3.18}}
            \sqrt{\frac{\omega_\gamma}{\omega_r}}
            x_\mathrm{eq}^{-0.810} R_\star^{-0.499}
        \right)^{1/5.173}.
\end{align}
\end{subequations}
\Cref{eqn:constrained-parameter-combos} takes the scale factor of recombination to be proportional
to $T_0 / \alpha_i^2 m_{e, i}$, per \cref{eqn:a-star-proportionality}.
When all of these parameter combinations and the ionization history are fixed, all the scales that
encode early-Universe physics---the inverse sound horizon $1/r_{s, \star}$, the horizon at equality
$k_\mathrm{eq}$, the diffusion damping scale $k_{D, \star}$, and the Landau damping scale
$k_L$---scale in proportion to
$\sqrt{\omega_r} / a_\star \propto \alpha_i^2 m_{e, i} T_0 \sqrt{\omega_r / \omega_\gamma}$.

\begin{figure}[th!]
    \centering
    \includegraphics[width=\textwidth]{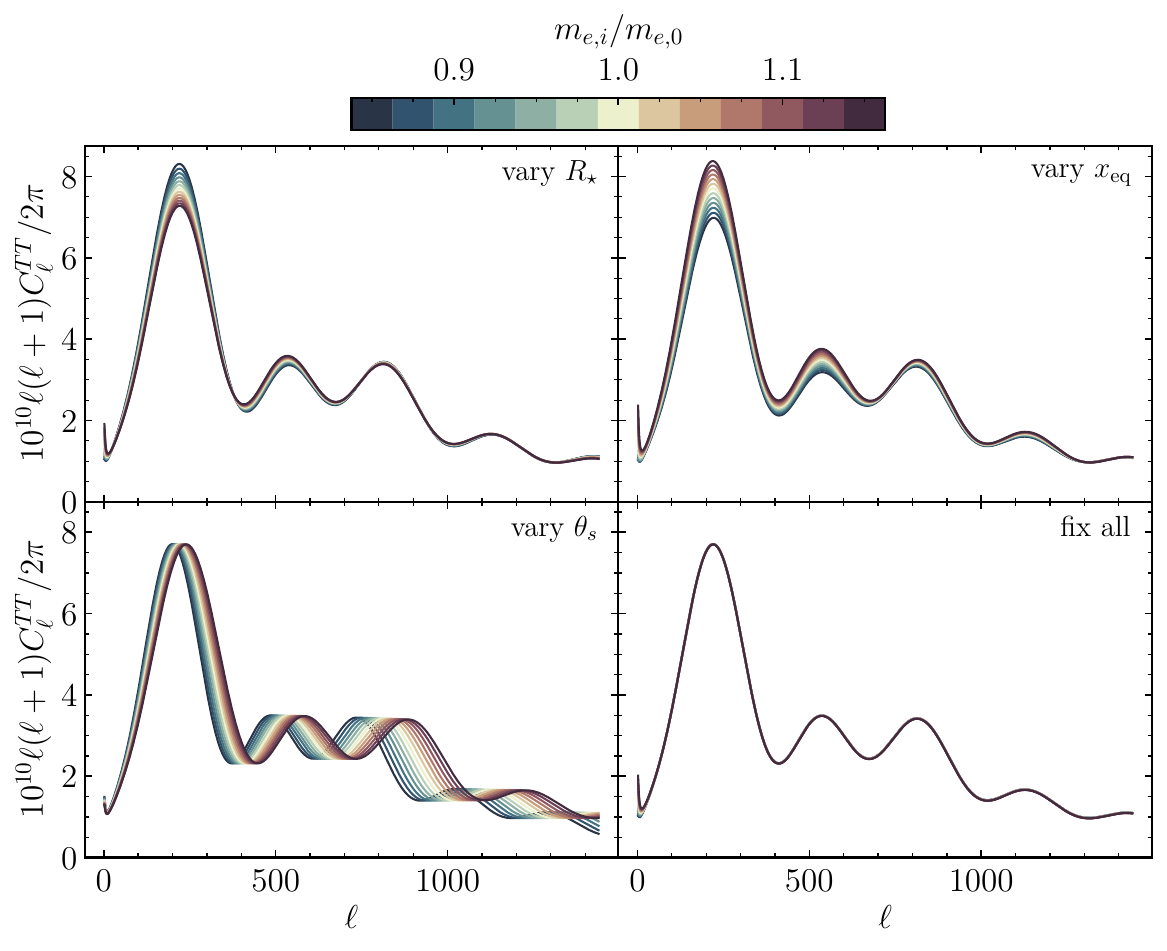}
    \caption{
        CMB temperature power spectrum in cosmologies varying the electron mass $m_{e, i}$ by color
        and holding various combinations of $R_\star$ [\cref{eqn:baryon-to-photon-ratio-ito-x}],
        $x_\mathrm{eq}$ [\cref{eqn:def-x-eq}], and $\theta_s$ [\cref{eqn:theta-s-degeneracy}] fixed
        to their values in the best-fit \LCDM{} cosmology~\cite{Planck:2018vyg}.
        The top-left and top-right panels fix all but
        $R_\star$ and $x_\mathrm{eq}$, respectively, illustrating the impact of changing ratios of energy densities at recombination.
        The bottom-left panel varies $\theta_s$, shifting the spectrum to larger multipoles when recombination occurs earlier.
        The bottom-right panel fixes them all, leaving the spectrum largely unchanged.
        The angular size of the sound horizon is set by choosing $h$ via numerical optimization; the
        bottom-left panel simply fixes $h$.
        Each panel also fixes the helium mass fraction $Y_\mathrm{He}$, the combination
        $A_s a_\star^{1 - n_s}$, and the optical depth to reionization $\tau_\mathrm{reion}$.
    }
    \label{fig:cmb-TT-degeneracies-vary-m_e}
\end{figure}

\Cref{fig:cmb-TT-degeneracies-vary-m_e} illustrates the phenomenological impact of varying
$R_\star$, $x_\mathrm{eq}$, and $\theta_s$ on the CMB temperature power spectrum $C_\ell^{TT}$ and
in particular demonstrates the degeneracy achieved under varying $m_{e, i}$.
The role played by baryons in acoustic oscillations is clear in the panel varying $R_\star$: the
relative heights of the first peaks and troughs are reduced at larger $m_{e, i}$, for which
$R_\star \propto a_\star \propto m_{e, i}^{-1}$, and therefore the amplitude of acoustic oscillations, is
smaller.
Next, if recombination happens closer to matter-radiation equality (as it does if a larger $m_{e, i}$
triggers recombination earlier, all else equal), radiation has a larger effect on gravity; around
recombination, the metric potentials therefore decay relatively faster at horizon entry, increasing
the driving of acoustic oscillations.
In addition, the decay of the potentials due to radiation enhances the early ISW effect for earlier
recombination, since they are constant in a Universe with only cold dark matter.
Finally, \cref{fig:cmb-TT-degeneracies-vary-m_e} shows that holding both $R_\star$ and
$x_\mathrm{eq}$ fixed but not $\theta_s$ (i.e., holding $h$ constant instead) effectively shifts the
entire spectrum to larger multipoles when recombination happens earlier.

\begin{figure}[th!]
    \centering
    \includegraphics[width=\textwidth]{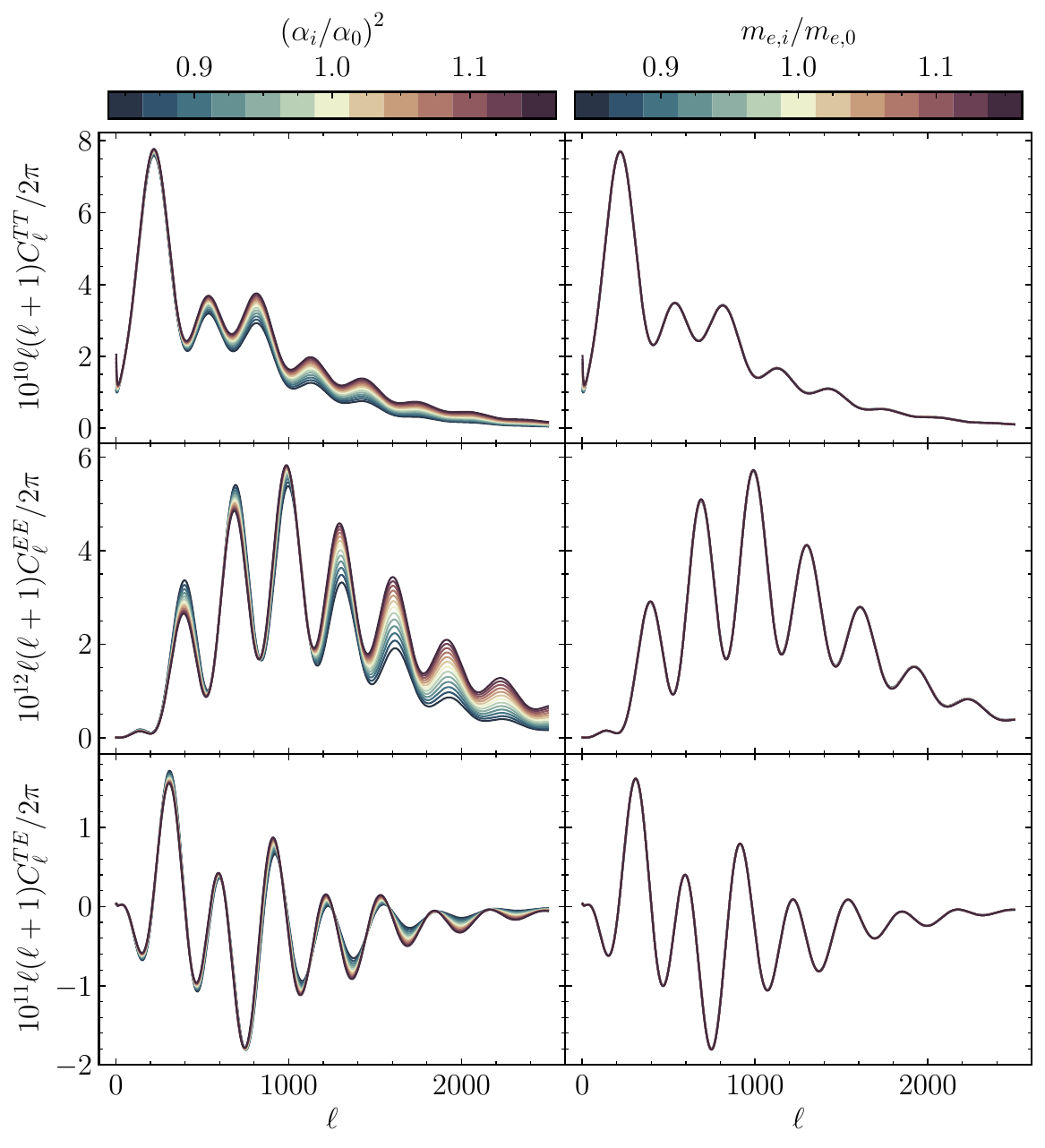}
    \caption{
        CMB temperature (top panels) and $E$-mode (middle panels) auto power spectra
        and their correlation (bottom panels) in cosmologies varying the
        fine-structure constant $\alpha_i$ (left panels) or the electron mass $m_{e, i}$ (right
        panels) by color.
        The impact on small-scale damping and polarization is much more significant under variations of
        $\alpha_i$ than of $m_{e, i}$, as encoded in \cref{eqn:constrained-parameter-combos}.
        Each curve holds all of $R_\star$, $x_\mathrm{eq}$, $\theta_s$, $Y_\mathrm{He}$,
        $A_s a_\star^{1 - n_s}$, and $\tau_\mathrm{reion}$ fixed.
    }
    \label{fig:cmb-all-spectra-degeneracies-vary-both}
\end{figure}

The impact of the remaining parameters on CMB polarization and on small-scale anisotropies is
illustrated well when varying $\alpha_i$, shown in
\cref{fig:cmb-all-spectra-degeneracies-vary-both}.
The impact of small-scale damping on both temperature and polarization anisotropies is dramatic: the
Thomson scattering rate increases more rapidly than the Hubble rate with the fine-structure
constant, pushing the effects of diffusion damping to smaller scales (and therefore larger
multipoles).
In addition, recalling \cref{fig:ionization-visibility-varying-constants}, the dimensionless
visibility function also narrows with larger $\alpha_i$, reducing the degree of Landau damping.
A narrower visibility function also reduces the amplitude of polarization, which is most evident at
$\ell \lesssim 800$ in the middle-left and lower-left panels of
\cref{fig:cmb-all-spectra-degeneracies-vary-both}; at larger multipoles, the reduction in damping
overcomes this suppression.
The $m_{e, i}$ independence of diffusion damping, Landau damping, and the generation of polarization
at fixed $R_\star$, $x_\mathrm{eq}$, and $\theta_s$ is evident in the right panels of
\cref{fig:cmb-all-spectra-degeneracies-vary-both}, where the changes to the CMB spectra are
imperceptible even for $15\%$ changes to the early-time electron mass.

The preceding discussion facilitates the interpretation of degeneracies in a variety of proposed
extensions to \LCDM{}.
The degeneracies in scenarios featuring early recombination via varying constants, discussed in
Ref.~\cite{Sekiguchi:2020teg}, fix $R_\star$ and $x_\mathrm{eq}$ by taking constant $a_\star
\omega_b$ and $a_\star \omega_m$, as discussed in \cref{sec:photon-baryon-plasma}.
Per \cref{eqn:theta-s-degeneracy}, holding $\theta_s$ constant then requires
$h \propto a_{\star}^{-3.18}$, as found in Ref.~\cite{Sekiguchi:2020teg}.
As demonstrated in Ref.~\cite{Sekiguchi:2020teg} and \cref{sec:damping}, no additional parameter
freedom is required to preserve the diffusion damping scale only when varying the electron mass, for
which $a_\star \propto m_{e, i}^{-1}$.
Furthermore, the ionization history is effectively invariant (as a function of $a / a_\star$) only
under varying $m_{e, i}$, in which case the diffusion and Landau damping scales imprint at the same
angular scale on the sky (i.e., $r_{s, \star} k_{D, \star}$ and $r_{s, \star} k_L$ invariant) and
polarization is generated with the same amplitude, as evident in
\cref{fig:cmb-all-spectra-degeneracies-vary-both}.
We therefore expect that early-time variations in the fine-structure constant are much more severely
constrained by primary CMB data than the electron mass.

Another class of early-recombination models postulates substantial small-scale inhomogeneities in
the baryon distribution, which enhance the recombination rate~\cite{Jedamzik:2020krr,
Rashkovetskyi:2021rwg, Thiele:2021okz, Lee:2021bmn}.
These changes, however, do not preserve the angular extent of the diffusion damping
feature~\cite{Rashkovetskyi:2021rwg} as is possible with a varying electron mass; presumably the
shape of the visibility function is also affected, with consequences described in
\cref{sec:damping,sec:polarization}.
Similarly, though \cref{eqn:theta-s-degeneracy} shows that $\theta_s$ is preserved when increasing
$N_\mathrm{eff}$ (i.e., $\omega_r / \omega_\gamma$) and $h$ simultaneously, the diffusion damping
rate [\cref{eqn:damping-combo-ito-constants}] is
not~\cite{Wyman:2013lza,Dvorkin:2014lea,Bernal:2016gxb}.
Changes to the damping tail may be partially absorbed by the primordial tilt $n_s$ but not without
involving the overall amplitude as well as the baryon abundance; varying the fraction of the
radiation density that is collisionless has additional, well-studied physical effects on the
CMB~\cite{Bashinsky:2003tk, Hou:2011ec, Baumann:2015rya, Pan:2016zla, Ge:2022qws}.
\Cref{eqn:constrained-parameter-combos-K-star} shows that the effect of a change to the total
radiation density on the damping rate may be compensated not just by adjusting the helium yield
$Y_\mathrm{He}$, a commonly studied degeneracy, but also by altering the early-time values of the
fine-structure constant or the hydrogen mass.

The scale factor of recombination would also be different if we instead suppose $T_0$ differs from
its measured value.
Reference~\cite{Ivanov:2020mfr} phrases the CMB's dependence on $\omega_b$,
$\omega_c$, and $T_0$ as being via the combinations $\omega_b / T_0^3$ and $\omega_c / T_0^3$.
Since $T_0^3 \propto \omega_\gamma / T_0 = \omega_\gamma / a_\star T_\star$, holding
$\omega_b / T_0^3$, $\omega_c / T_0^3$, and $T_\star$ fixed also fixes $R_\star$ and $x_\mathrm{eq}$.
Per \cref{eqn:damping-combo-ito-constants}, the diffusion damping parameter $K_\star$ is invariant
along this direction because both of $\sqrt{\omega_\gamma} / a_\star^2 \propto T_\star^2$ and
$\omega_\gamma / \omega_r$ are unchanged.
Since changing $T_0$ does not affect physics during recombination (but rather the amount of
expansion that has elapsed since then), the visibility function is also unchanged [as a function of
$\ln (a / a_\star)$].
\Cref{eqn:theta-s-degeneracy} shows that $\theta_s$ is fixed if $h \propto T_0^{-1.18}$, matching
the degeneracy direction identified in Ref.~\cite{Ivanov:2020mfr}.
The effects of $T_0$ and $m_{e, i}$ are only distinguished by the latter's effect on the total
radiation density at recombination, explaining why $h$ scales differently with $T_0$ and $m_{e, i}$
at fixed $\theta_s$.

The degeneracies discussed in this section are grounded in a scaling invariance of cosmological
observables.
Because the dynamics of perturbations in the early Universe is linear, the Boltzmann and Einstein
equations governing their evolution depend only on the ratio of scales, i.e., are invariant under a
common rescaling of the length scales appearing in the problem.
Moreover, the initial conditions preferred by current data are, aside from being nearly scale
invariant, also scale free: any rescaling of length scales may be absorbed into the over
normalization of the primordial power spectrum.
Because most cosmological observables are measured as features on the sky, i.e., as dimensionless
ratios of length scales, they do not measure the value of any individual scale.
The parameters of \cref{eqn:constrained-parameter-combos} enumerate the general requirements for a
common rescaling of all length scales specifically relevant to early-Universe dynamics, which are
met by varying the scale factor of recombination (via $m_{e, i}$ or $T_0$ per the above discussion).

A distinct class of models, in which the scale factor of recombination is unchanged, was explored in
detail by Refs.~\cite{Cyr-Racine:2021oal, Ge:2022qws, Greene:2023cro, Greene:2024qis} by taking
$\omega_i \propto \lambda^2$ for all species $i$, $\sigma_T n_e \propto \lambda$, and
$A_s \propto \lambda^{1 - n_s}$.
A simple realization thereof is to vary both $T_0$ and $m_{e, i}$ proportionately, in which scaling the
density of all species $\propto T_0^4$ preserves all of the parameter combinations in
\cref{eqn:constrained-parameter-combos} without requiring changes to any other parameters.
Given that direct measurements of $T_0$ preclude this possibility, Refs.~\cite{Cyr-Racine:2021oal,
Ge:2022qws, Greene:2023cro, Greene:2024qis} invoke a ``mirror'' dark sector with the same particle
content and parameters as the SM to allow for arbitrarily increasing the abundances $\omega_i$,
meaning the abundance of the visible sector does not scale with $\lambda$.
The proportional scaling of the total matter and radiation densities preserves $x_\mathrm{eq}$ as
well as $\theta_s$ (since $h^2 \propto \sum_i \omega_i$), while the nonscaling of the visible photon
and baryon densities preserves $R_\star$.
Constancy of the damping parameter $K_\star$, however, requires an additional ingredient.
[Note that, in \cref{eqn:K-star}, $\omega_\gamma$ and $Y_\mathrm{He}$ are those for the visible sector,
while $\omega_r$ is the radiation density of the entire Universe.]
References~\cite{Cyr-Racine:2021oal, Ge:2022qws} invoke a differing helium yield $Y_\mathrm{He}$ to
achieve the requisite scaling of the Thomson scattering rate, while Refs.~\cite{Greene:2023cro,
Greene:2024qis} instead takes $m_{e, i} \propto \alpha_i^{-2} \propto \lambda^{-1/3}$ to
simultaneously fix $K_\star$ and $a_\star$.

The symmetry of the Einstein-Boltzmann equations is broken by nonequilibrium effects during
recombination that alter the ionization history, but only mildly so.
(The symmetry is also broken by nonzero neutrino masses, but since they are only bounded and not
measured, one is free to rescale them as well.)
However, each of the aforementioned possible realizations is constrained by direct measurements of
either $T_0$ or $Y_\mathrm{He}$, including that varying
$m_{e, i} \propto \alpha_i^{-2} \propto \lambda^{-1/3}$ because the Big Bang nucleosynthesis prediction for the helium yield is
not invariant under this transformation~\cite{particle-paper}.
Moreover, the arguments of this section demonstrate that the full rescaling symmetry is more
restrictive than necessary to keep \textit{early-time} dynamics invariant.
Namely, distinct degeneracy directions are afforded by freeing the scale factor of recombination
$a_\star / a_0$, whether by varying $T_0$ or $T_\star$.
But such scenarios do not achieve the full rescaling symmetry of the Einstein-Boltzmann system.
They are distinguished by effects arising at late times, which we turn to now.

\subsection{Late-Universe signatures}\label{sec:late-universe-signatures}

The dynamics of the late Universe are driven by the dark sector and so do not depend explicitly on the fundamental constants.
However, given the constraining power of the early Universe observables discussed in the previous subsection, \LCDM{} parameters---specifically the relative contributions of matter and dark energy to the total energy budget---change along the degeneracy direction as fundamental constants are varied.
The discussion thus far has entirely neglected dark energy beyond its effect on the distance to last scattering and the sound horizon, \cref{eqn:theta-s-degeneracy}.
We now focus on the late-time effects of dark energy.

In models that are well described by flat \LCDM{} cosmology at late times, the shape of the
late-time expansion history is only characterized by the scale factor of matter--dark-energy
equality,
\begin{align}
    \amL
    &\equiv \sqrt[3]{\omega_m / \omega_\Lambda}.
\end{align}
Along the degeneracy direction that preserves the primary CMB spectra,
$\omega_m \propto \omega_r /a_\star$
but $h \approx \sqrt{\omega_m + \omega_\Lambda} \propto \sqrt{\omega_r} / a_\star^{3.18}$.
The dark energy density thus increases much more steeply than the matter density along the
degeneracy, drastically altering when matter--dark-energy equality occurs and the physical effects
that depend upon it.
Matter perturbations grow more slowly in a Universe dominated by dark energy than one dominated by
matter; \cref{sec:isw-lensing} discusses the resulting impact on secondary signatures in the CMB and
\cref{sec:matter-clustering} that on matter clustering.
Datasets that measure distances at low redshift directly trace out the shape of the expansion
history and provide independent constraints; these are discussed in \cref{sec:bao,sec:supernova}.

\subsubsection{Late-time integrated Sachs-Wolfe effect and CMB lensing}\label{sec:isw-lensing}

While dark energy is entirely negligible near recombination when most of the CMB anisotropies
are seeded, at late times its presence alters the dynamics of metric perturbations, sourcing a
nonzero integrated Sachs-Wolfe effect and altering the degree to which the primary CMB anisotropies
are gravitationally lensed.
In a Universe dominated by cold dark matter, the metric potentials (in Newtonian gauge, for
instance) are constant.
As CMB photons traverse potential wells, the amount by which they blueshift upon entering is
precisely compensated by the corresponding redshift as they exit.
If large-scale metric potentials instead decay with time, which they do to an extent controlled by
the relative abundance of dark energy~\cite{Sachs:1967er,Weller:2003hw}, this cancellation is
incomplete, enhancing CMB power.
The earlier the dark energy era begins (i.e., the smaller $\amL$ is), the larger this effect.
Because matter--dark-energy equality occurs rather near the present, this late ISW effect
affects only the largest observable scales---at multipoles $\ell \lesssim 30$ where cosmic variance strongly limits the CMB's constraining power.

\begin{figure}
    \centering
    \includegraphics[width=\textwidth]{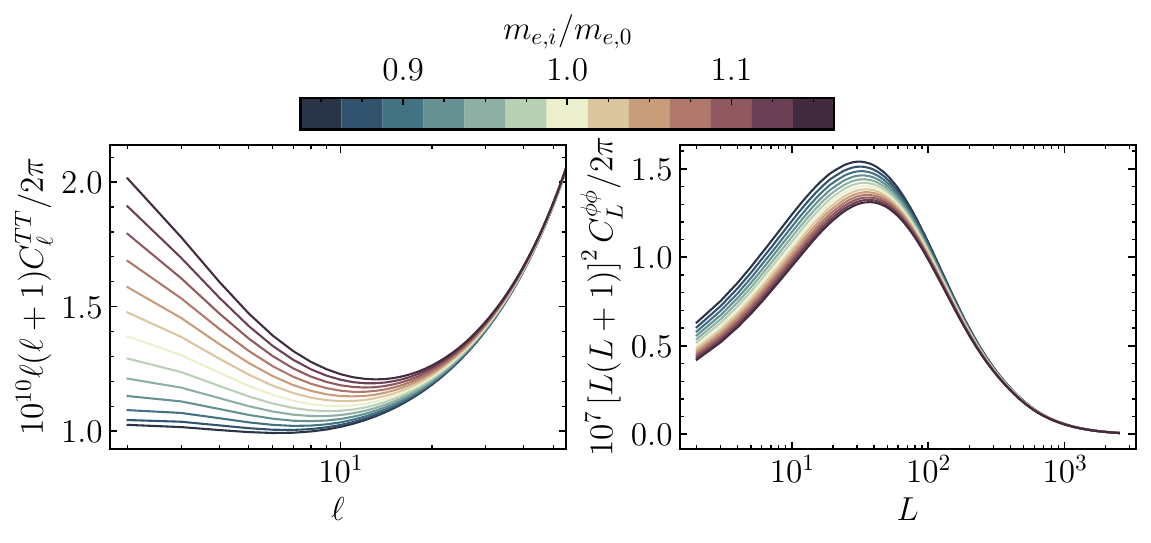}
    \caption{
        Large-scale CMB temperature (left) and lensing (right) power spectra in cosmologies varying
        the electron mass $m_{e, i}$ by color, holding all of $R_\star$, $x_\mathrm{eq}$, $\theta_s$,
        $Y_\mathrm{He}$, $A_s a_\star^{1 - n_s}$, and $\tau_\mathrm{reion}$ fixed.
        The changes in the power spectra arise due to the correlation of $\amL$ with
        $a_\star \propto 1/m_{e, i}$ rather than directly from processes that depend on the constants'
        values.
        Therefore, these results are broadly unchanged for commensurate variations in $\alpha_i^2$ as
        for $m_{e, i}$ and are effectively identical under variations in $T_0$ that yield the same values
        of $\amL$ when holding $\theta_s$ and $x_\mathrm{eq}$ fixed.
    }
    \label{fig:cmb-isw-lensing-vary-me}
\end{figure}

The CMB is also lensed by matter fluctuations at late times~\cite{Lewis:2006fu}, which grow more
after horizon crossing in Universes where matter--dark-energy equality occurred later.
Whereas an increased absolute change in metric perturbations over time results in an enhanced ISW
signal on large scales, the lensing deflection spectrum scales in proportion to the power in matter
fluctuations.
Both effects are depicted in \cref{fig:cmb-isw-lensing-vary-me}.
The disproportionate increase in the dark energy density required to decrease the distance to last
scattering in proportion to the sound horizon [\cref{eqn:theta-s-degeneracy}] makes the dark energy
era begin earlier, leading to the suppression of lensing power and enhancement of the late ISW
effect evident in \cref{fig:cmb-isw-lensing-vary-me}.

\subsubsection{Baryon acoustic oscillations}\label{sec:bao}

Late-time changes to the Universe's energy budget may also be constrained directly via distance measurements.
First, baryon acoustic oscillations (BAO) imprint an oscillatory pattern in the matter correlation
function that may be treated as a standard ruler.
The BAO scale specifically corresponds to a local maximum of the matter density correlation
function, but it can be accurately modeled by simply evaluating the comoving sound horizon evaluated
at the baryon drag epoch, $r_\mathrm{d} \equiv r_s(a_\mathrm{drag})$~\cite{Thepsuriya:2014zda,
Schoneberg:2019wmt}.
Baryons decouple slightly after recombination, $a_\mathrm{drag} > a_\star$, because the photon
number density is so much greater than that of the baryons, so $r_\mathrm{d}$ is slightly larger than $r_{s, \star}$.

The angular BAO scales are extracted from various galaxy surveys and reported as
\begin{subequations}\label{eqn:theta-bao}
\begin{align}
    \theta_\mathrm{BAO}(a)
    &\equiv \frac{r_\mathrm{d}}{D_V(a)}
    \label{eqn:theta-bao-DV}
    \\
    \theta_{\mathrm{BAO}, \parallel}(a)
    &\equiv \frac{r_\mathrm{d}}{c H(a)^{-1}}
    \label{eqn:theta-bao-parallel}
    \\
    \theta_{\mathrm{BAO}, \perp}(a)
    &\equiv \frac{r_\mathrm{d}}{D_M(a)}.
    \label{eqn:theta-bao-perp}
\end{align}
\end{subequations}
(We retain factors of $c$ in this subsection to clearly specify units.)
In \cref{eqn:theta-bao-DV},
\begin{align}
    D_V(a)
    &= \sqrt[3]{
            \frac{D_M(a)^2}{H(a) / c}
            \left( 1 / a - 1 \right)
        }
\end{align}
is an effective, comoving-volume--averaged distance to a galaxy sample and the transverse distance
$D_M$ in \cref{eqn:theta-bao-perp} is defined in \cref{eqn:transverse-distance}.
While these measurements rely on template modeling in a fiducial cosmology,
Ref.~\cite{Bernal:2020vbb} showed that any bias in application to modified cosmological models is
negligible for existing datasets.

Substituting \cref{eqn:comoving-distance-matter-Lambda} for the comoving distance in
\cref{eqn:theta-bao-perp} shows that BAO distances depend on $H_0$ only through an overall
scaling in the combination $r_\mathrm{d} H_0$:
\begin{align}
    \theta_{\mathrm{BAO}, \perp}(a)
    &= \frac{r_\mathrm{d} h}{c / H_{100}}
        \frac{
            \left[ F_M(1; \amL) - F_M(a; \amL) \right]^{-1}
        }{
            2 \sqrt{1 + \amL^{-3}}
        }.
    \label{eqn:theta-bao-perp-flat-lcdm}
\end{align}
The dependence on the scale factor $a$ is through a (nonlinear) function depending on cosmological
parameters only via $\amL$.
A sufficient number of BAO measurements, spanning a sufficient range of scale factors about
$\amL$, could therefore constrain $\amL$ independently of $r_\mathrm{d}$ and $H_0$
individually.
(The same conclusions apply for $\theta_{\mathrm{BAO}}$ and $\theta_{\mathrm{BAO}, \parallel}$.)
Neglecting the small difference between $r_\mathrm{d}$ and $r_{s, \star}\propto a_\star/\sqrt{\omega_r}$
[\cref{eqn:matter-radiation-rs-power-law}], the overall scaling of
\cref{eqn:theta-bao-perp-flat-lcdm} then requires that $h \propto \sqrt{\omega_r} / a_\star$, which
differs substantially from $h \propto \sqrt{\omega_r} / a_\star^{3.18}$ as required by the CMB
[\cref{eqn:theta-s-degeneracy}].
For models that only vary the scale factor of recombination (whether via $T_0$ or early-time
variations in fundamental constants), BAO data therefore offer constraining power in the
$h$--$a_\star$ plane complementary to that of the CMB.

\subsubsection{Low-redshift distances from supernovae}\label{sec:supernova}

Type Ia supernovae (SNe Ia) occur at a characteristic mass that yields a standard luminosity suitable
for inferring distances.
The apparent magnitude of each SN Ia (labeled $i$) is parametrized
as~\cite{Pan-STARRS1:2017jku,Brout:2022vxf, Scolnic:2021amr}
\begin{align}
    m_i
    &= 5 \log_{10} \left[
            \frac{a_i}{a_{i, \mathrm{hel}}}
            \frac{D_L(a_i)}{10~\mathrm{pc}}
        \right]
        + M_B
    ,
    \label{eqn:apparent-magnitude}
\end{align}
where $M_B$ is the fiducial magnitude of an SN Ia (a parameter that itself must be sampled
over) and $D_L$ the luminosity distance, $D_L(a) = D_M(a) / a$.
The ratio of the CMB-frame ($a_i$) and heliocentric ($a_{i, \mathrm{hel}}$) redshifts inside the
logarithm corrects for peculiar velocities~\cite{Davis:2010jq,Davis:2019wet,Carr:2021lcj};
SNe distance datasets tabulate both these redshifts and either the apparent magnitude $m_i$
or the distance modulus $m_i - M_B$.
Absorbing the peculiar velocity correction into the apparent magnitude via the definition
$\hat{m}_i \equiv m_i - 5 \log_{10} (a_i / a_{i, \mathrm{hel}})$, we may rearrange
\cref{eqn:apparent-magnitude} into the apparent brightness, which may be interpreted as a
``luminosity angle'' in analogy to the BAO angles \cref{eqn:theta-bao}:
\begin{align}
    \theta_L(a_i)
    \equiv \frac{10~\mathrm{pc} \cdot 10^{- M_B / 5}}{D_L(a_i)}
    &= 10^{- \hat{m}_i / 5}.
    \label{eqn:luminosity-angle}
\end{align}
The numerator of the middle expression acts as a standardized or calibrated luminosity distance.

Akin to BAO distances, without external constraints on the fiducial magnitude $M_B$ (from, e.g.,
Cepheids~\cite{Riess:2016jrr,Riess:2019cxk,Riess:2021jrx} or the tip of the red giant
branch~\cite{Freedman:2019jwv,Freedman:2020dne,Freedman:2021ahq,Freedman:2023jcz}), SN Ia distances
only probe the shape of the expansion history, not its absolute scale (i.e., $H_0$).
The ``luminosity angle'' [\cref{eqn:luminosity-angle}] factorizes in the same manner as
\cref{eqn:theta-bao-perp-flat-lcdm}:
\begin{align}
    \theta_L(a_i)
    &= \frac{10^{\log_{10} h - M_B / 5}}{c~\mathrm{s} / \mathrm{m}}
        \frac{
            a_i \left[ F_M(1; \amL) - F_M(a_i; \amL) \right]^{-1}
        }{
            2 \sqrt{1 + \amL^{-3}}
        }.
\end{align}
The Hubble constant is only constrained via the combination $\log_{10} h - M_B / 5$, and as for BAO
a sufficient number of SN Ia distance measurements may constrain the relative amount of dark energy
and matter independent of their overall calibration (and of $H_0$).

\subsubsection{Matter clustering}\label{sec:matter-clustering}

The clustering of matter is sensitive to the late-Universe expansion history, in addition to
nonstandard dynamics of dark matter and dark energy.
The amplitude of structure is often quantified by the present-day, root-mean-squared overdensity in
spheres of radius $R$, $\sigma(R)$, defined by
\begin{align}
    \sigma(R)^2
    &= \int \ud \ln k \, W(k, R)^2 \frac{k^3}{2 \pi^2} P_\mathrm{lin}(k),
    \label{eqn:sigma-R-def}
\end{align}
where $P_\mathrm{lin}$ is the linear matter power spectrum and
\begin{align}
    W(k, R)
    &= \frac{3}{(k R)^3} \left[ \sin (k R) - k R \cos (k R) \right]
\end{align}
is the Fourier transform of a top-hat function with radius $R$.
A standard metric is $\sigma_8 \equiv \sigma(8 \, h^{-1}~\mathrm{Mpc})$, which is most sensitive
to scales that entered the horizon in the radiation era; constraints, however, are often quoted as
combinations of $\sigma_8$ and $\Omega_m$ that are best constrained by particular probes; for
example, $S_8 \equiv \sqrt{\Omega_m / 0.3} \, \sigma_8$ for cosmic
shear~\cite{Jain:1996st, DES:2015gax, Hildebrandt:2016iqg}.

The clustering of matter is sensitive to the expansion history via two effects.
First, the onset of the dark energy era suppresses structure growth; the earlier matter--dark-energy
equality ($\amL$) occurs, the stronger the effect.
Second, even at fixed $\amL$ increasing the total matter density $\omega_m$ pushes the
comoving horizon at matter-radiation equality (which determines the turnover in the matter power
spectrum from growing with $\sim k$ to falling off as $\sim k^{-3}$) to smaller scales
$k_\mathrm{eq} \propto \omega_m$ [per \cref{eqn:comoving-hubble-mr}], regardless of $a_\star$.
When fixing $\amL$ by taking $\omega_\Lambda \propto \omega_m$, $h$ scales with
$\sqrt{\omega_m}$ and $k_\mathrm{eq} \cdot 8 h^{-1}~\mathrm{Mpc} \propto \sqrt{\omega_m}$.
In other words, the earlier the matter era begins, the less suppressed the scales that contribute to
\cref{eqn:sigma-R-def} are by their subhorizon evolution during the radiation era, enhancing
$\sigma_8$.
As such, absent modifications to the dynamics of dark matter or the early-Universe expansion
history, $h$ and $\sigma_8$ are generally positively correlated.

%% file: cosmo-scalars.tex
\section{Cosmology with coupled, hyperlight scalars}\label{sec:cosmology-hyperlight-scalars}

In the previous section, we discuss expanded phenomenology of cosmologies with varying fundamental
constants without considering a concrete model.
A ``bottom-up'' approach promotes a fundamental constant to a dynamical scalar described by an
action principle; the scalar's value dictates that of the constants and it carries
stress energy~\cite{Terazawa:1981ga,Bekenstein:1982eu,Damour:2010rm,Damour:2010rp}.
Such scalars also arise in theories with extra dimensions as so-called dilaton or moduli
fields~\cite{Brans:1961sx,Scherk:1974ca,Green:2012pqa,Taylor:1988nw,Cho:1987xy,Damour:1994ya,Dimopoulos:1996kp,Kaplan:2000hh,Gasperini:2001pc,Damour:2002mi}.
To realize the scenarios discussed in \cref{sec:varying-constants} we consider one such simple
model, that of a massive scalar whose amplitude modulates the strength of electromagnetism $\alpha$
and the mass of the electron $m_e$.
We take the field to be so light as to have a cosmologically frozen amplitude until some time after
recombination, leading to the fundamental constants taking on time-independent values prior to
recombination that differ from those measured today.

In \cref{sec:scalar-impact-on-cosmological-background}, we review the gravitational impact of the
scalar on the background (i.e., spatially homogeneous) cosmology.
In particular, we identify a new degeneracy in cosmological parameters when one allows for an
increase in the total energy density of matter after recombination relative to that which was
present at earlier times.
While such a ``late dark matter'' scenario is necessarily implemented by a hyperlight scalar field
that modifies the value of fundamental constants at earlier times, it could also be realized via
other mechanisms; massive neutrinos are an additional example.
In \cref{sec:scalar-impact-on-cosmological-perturbations} we expose how the necessary impact of the
scalar's spatial perturbations on the growth of structure ultimately affects cosmological
observables.
Finally, in \cref{sec:hyperlight-scalar-parameter-space}, we connect the phenomenological parameters
of the modified cosmology (energy densities and fundamental constants) with microphysical model
parameters (the mass and amplitude of the field). In combination with Ref.~\cite{particle-paper}, we find that varying-constant cosmologies can be implemented with microphysical parameters that are as yet unconstrained by other probes.

Studying the cosmological impact of a scalar field requires solving the covariant Klein-Gordon
equation in the perturbed, expanding spacetime [\cref{eqn:scalar-ele}]:
\begin{align}
    \nabla^\mu \nabla_\mu \phi
    &= \frac{\ud V(\phi)}{\ud \phi}.
    \label{eqn:kg-main}
\end{align}
Numerical solutions for massive scalar fields [i.e., with potential $V(\phi) = m_\phi^2 \phi^2$]
provide a unique challenge by introducing a fixed timescale, i.e., the scalar's oscillation period
$1/m_\phi$.
As we detail in full in \cref{app:scalar-implementation}, we extend \textsf{CLASS} by implementing a
scheme recently devised in Ref.~\cite{Passaglia:2022bcr} that solves the Klein-Gordon equation for
some number of oscillations before switching to an effective fluid approximation scheme.
The prescription in Ref.~\cite{Passaglia:2022bcr} for matching onto fluid variables and determining
the fluid's effective equation of state and sound speed provides systematically improved precision
compared to prior approaches.
As described in \cref{app:scalar-implementation}, we extend Ref.~\cite{Passaglia:2022bcr}'s results,
which were applied only to scalars that begin oscillating in radiation domination, to a generalized
background expansion rate~\cite{class-uls}.

In addition, fundamental constant variations due to a scalar field inevitably acquire spatial
dependence, at least to some degree.
The resulting fifth-force effects on the dynamics of electron density perturbations may also be
ignored because electron density perturbations themselves, again neglecting their suppressed
contribution to the Einstein equations, only enter the Boltzmann equation for photons at second
order in inhomogeneities~\cite{Senatore:2008vi, Khatri:2008kb}.
That is, anisotropies generated by Thomson scattering are already first order because scattering in
a homogeneous plasma generates no anisotropy; inhomogeneities in the Thomson scattering rate
(whether due to inhomogeneities in the electron number density or in the Thomson cross section
itself) therefore enter at second order.

\subsection{Impact on the cosmological background}\label{sec:scalar-impact-on-cosmological-background}

To solve for the classical evolution of the scalar perturbatively, we expand the scalar into a
homogeneous component and a small perturbation as
$\phi(t, \three{x}) = \bar{\phi}(t) + \delta \phi(t, \three{x})$; the homogeneous component of the
Klein-Gordon equation [\cref{eqn:kg-main}] is then
\begin{align}
    \ddot{\bar{\phi}}
    + 3 H \dot{\bar{\phi}}
    + m_\phi^2 \bar{\phi}
    &= 0.
    \label{eqn:homogeneous-klein-gordon}
\end{align}
The scalar contributes to the Friedmann equations an energy density
$\bar{\rho}_\phi = \dot{\bar{\phi}}^2 / 2 + m_\phi^2 \bar{\phi}^2 / 2$ and a pressure
$\bar{P}_\phi = \dot{\bar{\phi}}^2 / 2 - m_\phi^2 \bar{\phi}^2 / 2$.

If the field is sufficiently light compared to the Hubble rate at recombination, $H_\star \approx 3
\times 10^{-29}~\eV$, the scalar's effect on the fundamental constants is effectively time
independent over the Universe's entire history up to (and including) recombination.
Note that when a scalar's couplings to matter are large enough, its equation of motion can be
significantly modified from \cref{eqn:homogeneous-klein-gordon}.
Here we implicitly assume that these contributions are negligible; we consider the possibility and
its impact on scalar field phenomenology in Ref.~\cite{particle-paper}.
Because the late-time probes discussed in Ref.~\cite{particle-paper} provide relatively strong
constraints on fundamental-constant variations at late times and the present day, the scalar (if it
shifts the early-time constants from their present-day values) cannot be too light---it must be
rolling at some point so that its oscillation amplitude redshifts by a sufficient amount by the
present day.
Since in standard cosmology recombination occurs after matter-radiation equality and
matter--dark-energy equality was relatively recent, the relevant parameter space for the scenarios
we discuss requires the scalar to begin oscillating in the matter-dominated era.

More concretely, at early times ($H \gg m_\phi$) the scalar is approximately frozen at its initial value
$\bar{\phi}_i$, i.e., the appropriate initial condition for \cref{eqn:homogeneous-klein-gordon} is
$\bar{\phi} \to \bar{\phi}_i$ and $\dot{\bar{\phi}} \to 0$ for $m_\phi t \to 0$.
In a matter-dominated Universe with $H = 2 / 3 t$, the solution to
\cref{eqn:homogeneous-klein-gordon} with these initial conditions is
$\bar{\phi}(t) = \bar{\phi}_i \sin(m_\phi t) / m_\phi t$.
This solution motivates defining the time when oscillations begin by
$H(t_\mathrm{osc}) = 2 / 3 t_\mathrm{osc} = 2 m_\phi / 3$.
We further define $a_\mathrm{osc}$ as the scale factor and $z_\mathrm{osc}$ as the redshift at that
time.
Well before $t_\mathrm{osc}$, the scalar has negligible kinetic energy compared to that of its
potential and therefore gravitates (at the background level) like dark energy.
However, $\bar{\rho}_\phi / \bar{\rho}_m \propto (a / a_\mathrm{osc})^3$ in this period, so its
contribution to the Einstein equations is entirely negligible until just before $t_\mathrm{osc}$.
Noting that $a(t) \propto t^{2/3}$ in the matter era, after $t_\mathrm{osc}$ the scalar's
oscillation-averaged energy density and pressure are, to leading order in $1 / m_\phi t$,
\begin{align}
    \bar{\rho}_\phi
    &\approx \frac{m_\phi^2 \bar{\phi}_i^2}{2 (a / a_\mathrm{osc})^3}
    \label{eqn:scalar-rho-matter-era}
    \\
    \bar{P}_\phi
    &\approx 0,
\end{align}
just like nonrelativistic matter.
(\Cref{app:effective-fluid} discusses the leading corrections to these results.)
We define $\omega_\phi = \bar{\rho}_\phi(t_0) / 3 H_{100}^2 \Mpl^2$ as for the other density parameters.

\begin{figure}[t!]
\begin{centering}
    \includegraphics[width=0.85\textwidth]{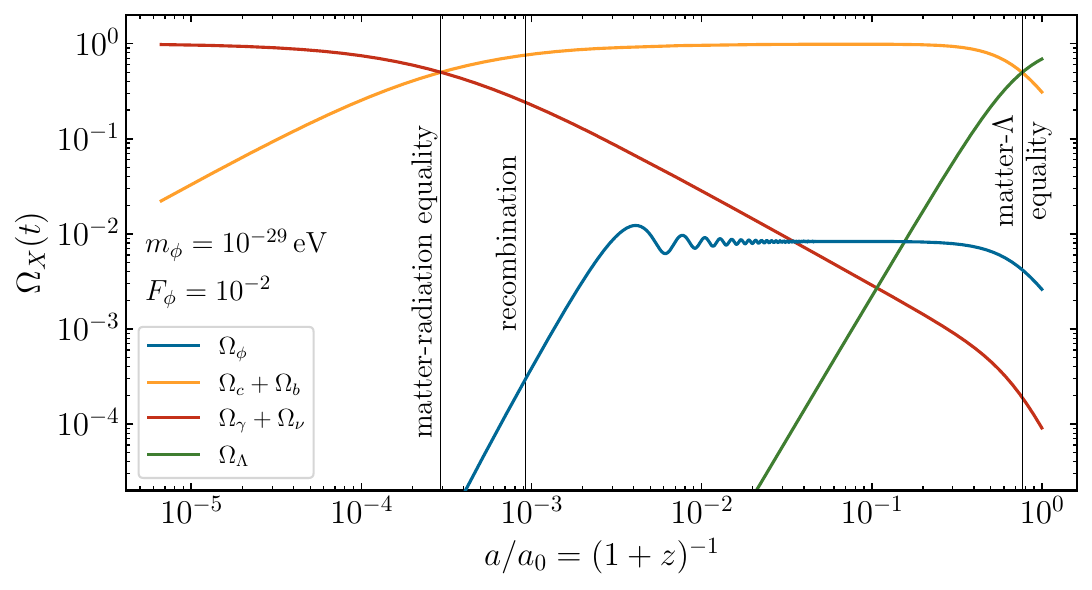}
    \caption{
        Relative abundances as a function of time, $\Omega_X(t) = \bar{\rho}(t) / 3 H^2 \Mpl^2$, of
        matter (orange), radiation (red), dark energy (green), and a hyperlight scalar field (blue).
        Here the scalar has a mass $m_\phi = 10^{-29}~\eV$ and an abundance
        $\fphi \equiv \bar{\rho}_{\phi, 0} / \bar{\rho}_{c, 0} = 10^{-2}$ relative to that of CDM.
    }
    \label{fig:abundance-evolution}
\end{centering}
\end{figure}

The most important gravitational effect of the scalar on the background cosmology is therefore its
delayed contribution to the matter density at late times, illustrated in \cref{fig:abundance-evolution}.
So long as the scalar begins rolling well before matter--dark-energy equality, any cosmological
distance may effectively be computed in a pure \LCDM{} cosmology that counts the scalar field's
energy density toward the total CDM density.
Certainly any distance measurement at $a \ll a_\mathrm{osc}$ is well described in this
approximation.
The integral that defines the comoving distance, \cref{eqn:comoving-distance}, has most of its
support where $1 / a H$ is largest; in the matter- and dark-energy--dominated limits,
$1 / a H \propto a^{1/2}$ and $a^{-1}$, respectively, placing the peak of the integrand at the epoch
of equality.
More concretely, the small-$a$ limit of \cref{eqn:matter-lambda-distance-function},
$F_M(a, \amL) \propto a^{7/2}$, quantifies the (un)importance of early time deviations from
a pure \LCDM{} cosmology on the full comoving distance in
\cref{eqn:comoving-distance-matter-Lambda}.
As such, the comoving distance to recombination may also be approximated by simply augmenting the
CDM density to include the scalar's, ignoring its differing dynamics before
$a_\mathrm{osc} \ll \amL$.
(The scale factor of equality $\amL$ is also implicitly defined with reference to this
combined, late-time matter density.)

Defining the late-to-early matter density ratio $A_m$ and inserting
$\omega_{m, \mathrm{late}} = A_m \omega_{m, \mathrm{early}}$ for the matter density in the
comoving distance to recombination [\cref{eqn:matter-radiation-chi-power-law}], the parameter
dependence of $\theta_s$ in \cref{eqn:theta-s-degeneracy} extends to
\begin{align}
    \theta_s
    &\propto \left(
            \frac{a_\star^{3.18} h A_m^{2.09}}{\sqrt{\omega_r}}
            x_\mathrm{eq}^{-0.810} R_\star^{-0.499}
        \right)^{1/5.173}
    .
    \label{eqn:theta-s-degeneracy-scalar}
\end{align}
The additional freedom afforded by $A_m$ enables preserving both
$a_\star^{3.18} h A_m^{2.09} / \sqrt{\omega_r}$ and
\begin{align}
    \amL
    &= \left( \frac{A_m \omega_{m, \mathrm{early}}}{\omega_\Lambda} \right)^{1/3}
    = \left( \frac{1}{h^2 / A_m \omega_{m, \mathrm{early}} - 1} \right)^{1/3}
    \label{eqn:a-mL-with-scalar}
\end{align}
at any value of $a_\star$, i.e., $h$ and $A_m$ provide two parameters to satisfy both conditions.
(We take $\omega_r$ to be fixed here, though our arguments generalize straightforwardly to scenarios
that vary $T_0$ rather than $T_\star$.)
The latter condition is more algebraically tractable when phrased as requiring
$\Omega_m = A_m \omega_{m, \mathrm{early}} / h^2 \propto A_m / a_\star x_\mathrm{eq} h^2$ to be
fixed.
The resulting degeneracy direction has $A_m \propto a_\star^{-1.034}$ and
$h \propto a_\star^{-1.017}$.

This degeneracy direction \emph{also} nearly exactly preserves $h r_\mathrm{d}$ as
constrained by BAO distance measurements when $\amL$ is also fixed [per
\cref{eqn:theta-bao-perp-flat-lcdm}].
Measurements of the BAO scale at various redshifts---including that by the CMB at
recombination---trace out the expansion history using a common ruler; distances probing either side
of matter--dark-energy equality may only jointly agree at an arbitrary size of the sound horizon if
$\amL$ remains the same.
In other words, the absolute scale (via $h$) and shape (via $\amL$) of the late-Universe
expansion history are both free to change to satisfy that any individual BAO measurement (or
$\theta_s$), but fitting multiple simultaneously requires the shape ($\amL$) to be
unchanged.
Furthermore, as argued in \cref{sec:supernova} SN Ia distance measurements (sans calibration)
themselves are only sensitive to $\amL$ (within flat the \LCDM{}-like models considered
here).

Prior literature studying electron mass variation as a solution to the Hubble
tension~\cite{Sekiguchi:2020teg,Schoneberg:2021qvd,Khalife:2023qbu} considered allowing for spatial
curvature to address the inconsistency between the distance to recombination and late-Universe
distances.
The analysis here points to additional matter at late times as a more natural solution, even on an
\textit{ad hoc} basis, which in addition avoids confrontation with the expectation in inflationary
cosmologies that spatial curvature is negligible.
Whether models with scalar fields fulfilling these cosmologically motivated requirements are
theoretically and phenomenologically tenable first requires mapping the parametrization of their
cosmological effects ($A_m$ and the early-time $m_{e, i}$) to their fundamental parameters.

The background-level effect of the scalar field closely resembles that due to massive neutrinos.
Namely, a single neutrino species with mass $m_\nu$, temperature
$T_\nu = \sqrt[3]{4/11} \sqrt[4]{N_\mathrm{eff} / 3} T_\gamma$, and average momentum
$p_\nu \approx 3.15 T_\nu$ (as for a Fermi-Dirac distribution) becomes exactly semirelativistic at a
redshift
\begin{align}
    z_\mathrm{NR} + 1
    = \frac{m_\nu}{3.15 T_\nu(a_0)}
    = 113 \,
        \frac{m_\nu}{0.06~\eV}
        \left( \frac{N_\mathrm{eff}}{3.046} \right)^{-1/4}
    \label{eqn:massive-neutrino-z-NR}
\end{align}
and has a present-day density of\footnote{
    For simplicity of illustration, these results take the idealized limit of instantaneous neutrino
    decoupling~\cite{Mangano:2005cc, Akita:2020szl, Froustey:2020mcq, Bennett:2020zkv,
    Cielo:2023bqp}.
}
\begin{align}
    \omega_{\nu, \mathrm{massive}}
    = \frac{3 \zeta(3)}{2 \pi^2} \frac{m_\nu T_\nu^3}{3 H_{100}^2 \Mpl^2}
    &= \frac{m_\nu}{93~\eV} \left( \frac{N_\mathrm{eff}}{3.046} \right)^{3/4}.
    \label{eqn:massive-neutrino-density}
\end{align}
Neutrino oscillation experiments require the sum of the neutrino masses to be greater than
$0.06~\eV$~\cite{deSalas:2020pgw, Esteban:2020cvm, Capozzi:2017ipn}, while \Planck{} data place an
upper limit of $0.24~\eV$ (within \LCDM{})~\cite{Planck:2018vyg}; the heaviest neutrino species thus
becomes nonrelativistic no earlier than $z \sim 450$ but no later than $z \sim 40$.
Massive neutrinos therefore contribute to the matter density only well after recombination (but
still well before matter--dark-energy equality) with an abundance proportional to their mass,
achieving the same effect encoded by $A_m$ in
\cref{eqn:theta-s-degeneracy-scalar,eqn:a-mL-with-scalar}.
The neutrino mass therefore plays the same role as $\omega_\phi$ for the scalar field; if the sum of
the neutrino masses is taken as a free parameter (rather than being fixed to a fiducial, minimal
value as in most analyses) in addition to the early-time electron mass, we would expect the two to
be positively correlated when considering CMB data in conjunction with low-redshift distance
datasets.
This correlation was indeed observed in Ref.~\cite{Khalife:2023qbu}; its physical origin is
explained by the arguments of this section.

However, the neutrinos' masses determine both their abundance and the redshift at which they become
matterlike (whereas the scalar's abundance is set by both its mass and its initial misalignment),
which means massive neutrinos can only contribute so much extra matter density without becoming
nonrelativistic before recombination; concretely,
\cref{eqn:massive-neutrino-z-NR,eqn:massive-neutrino-density} combine into
\begin{align}
    \omega_{\nu, \mathrm{massive}}
    &= 6.45 \times 10^{-4} \frac{z_\mathrm{NR} + 1}{113} \frac{N_\mathrm{eff}}{3.046}.
    \label{eqn:massive-neutrino-density-today}
\end{align}
Thus, if all three neutrino flavors becomes semirelativistic soon after recombination, their
present-day density is of order $10\%$ of the CDM density.
A massive scalar has the potential to make a greater contribution to the late-time matter density
while still doing so only well after recombination.

In otherwise standard cosmological scenarios, the background-level effect of hyperlight scalars and
massive neutrinos also reduces the Hubble constant inferred from CMB data: the increase in density
in the matter era makes the Universe smaller without changing the sound horizon, requiring a
compensatory reduction in the dark energy density to maintain the distance to last scattering.
\Cref{eqn:theta-s-degeneracy-scalar} shows that $h \propto A_m^{-2.09}$ at fixed $a_\star$
(and $x_\mathrm{eq}$ and $R_\star$).
Per \cref{eqn:massive-neutrino-density-today}, a massive neutrino species with fiducial mass
$m_\nu = 0.06~\eV$ increases the total matter density by $\approx 0.5\%$ relative to that from
baryons and CDM ($\omega_c + \omega_b \approx 0.143$), explaining why the CMB prefers values of $h$
about a percent lower in cosmologies with versus without massive neutrinos.
An anticorrelation between $\omega_\phi$ and $h$ is likewise evident in the results of
Ref.~\cite{Hlozek:2014lca} for ultralight axions with mass below $\sim 10^{-29}~\eV$.

If the scalar's potential is steeper than quadratic about its minimum, then the scalar's energy
density redshifts faster than that of matter once it begins oscillating.
In this case, its gravitational effects would be important only in the brief interval near the
beginning of its oscillations.
While such effects are substantial in, e.g., early dark energy scenarios where a scalar starts
oscillating around matter-radiation equality~\cite{Poulin:2018cxd, Poulin:2018dzj, Smith:2019ihp,
Hill:2020osr, Poulin:2023lkg}, the impact of a massless scalar that does so after recombination
should be much less important.

\subsection{Impact on cosmological perturbations}\label{sec:scalar-impact-on-cosmological-perturbations}

A hyperlight scalar's effect on the expansion history may be well understood as an
enhancement of the late-time CDM density, as described in the previous subsection; the dynamics of its perturbations, however, are qualitatively
distinct from CDM.
A scalar's mass $m_\phi$ associates a Jeans scale to its perturbations, which is parametrically $k_J
\sim a \sqrt{H m_\phi}$ at any given time; fluctuations on larger scales cluster like CDM, but on
smaller scales they oscillate~\cite{Hu:2000ke,Amendola:2005ad,Marsh:2015xka,Passaglia:2022bcr}.
Perturbations in CDM then grow more slowly on scales where not all of the matter content clusters,
again in close analogy to massive neutrinos~\cite{Bond:1980ha,Lesgourgues:2006nd} (which effectively
behave as warm dark matter and have a free-streaming scale associated to their temperature).
On scales where a fraction $f$ of matter does not cluster, the growth rate of CDM density
fluctuations $\delta_c \propto a^{p}$ is reduced from $p = 1$ to
$p \approx 1 - 3 f / 5$~\cite{Amendola:2005ad}.
\Cref{app:scalar-dynamics} provides more detailed exposition on the dynamics of scalar fields, their
perturbations, and their contribution to the Einstein equations.

In the mass range of interest, $m_\phi \lesssim 10^{-28}~\eV$, the Jeans scale is
cosmologically macroscopic: writing $H \sim m_\phi / (a / a_\mathrm{osc})^{3/2}$ for scalars that
begin oscillating in the matter era, the Jeans scale evolves as
$k_J \sim a_\mathrm{osc} m_\phi \sqrt[4]{a / a_\mathrm{osc}}$. \
Before oscillations begin, the scalar's energy density is small relative to the total density
(evident in \cref{fig:abundance-evolution}), so it has a negligible impact on the dynamics of CDM
and baryons.
Because of the weak scaling with $a$, a hyperlight scalar suppresses the growth of structure roughly
on scales smaller than the horizon when oscillations begin,
$k_\mathrm{osc} \equiv a_\mathrm{osc} m_\phi \sim (120 ~\Mpc)^{-1}\sqrt[3]{m_\phi/10^{-28}~\eV}$, at
all times after it begins oscillating.
The modifications to the growth of structure affect the CMB via the ISW effect and gravitational
lensing; the latter is constrained both by direct reconstruction of the CMB lensing potential and by
smearing of the acoustic peaks.

\begin{figure}
    \centering
    \includegraphics[width=\textwidth]{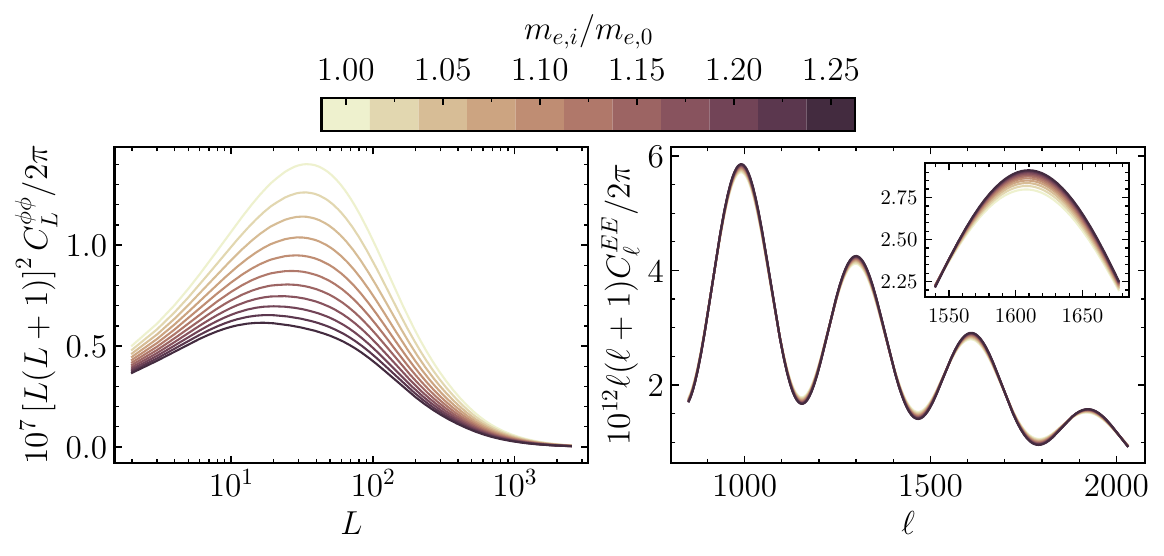}
    \caption{
        CMB lensing (left) and $E$-mode (right) power spectra in cosmologies varying the early-time
        electron mass $m_{e, i}$ with a scalar field, by color.
        Each curve holds all of $R_\star$, $x_\mathrm{eq}$, $\theta_s$, $Y_\mathrm{He}$,
        $A_s a_\star^{1 - n_s}$, and $\tau_\mathrm{reion}$ fixed. The scalar
        field has mass $m_\phi = 10^{-30}~\eV$ and a density chosen to (approximately) fix the  scale factor of matter--dark-energy equality via the degeneracy derived from
        \cref{eqn:a-mL-with-scalar}.
        The plot inset in the right panel displays the $E$-mode power spectrum enlarged to highlight
        the decreased smearing of one of the acoustic peaks.
    }
    \label{fig:cmb-lensing-ee-peaks-vary-me-with-scalar}
\end{figure}

\Cref{fig:cmb-lensing-ee-peaks-vary-me-with-scalar} depicts the impact of a hyperlight scalar field
on CMB lensing.
The results follow the degeneracy identified in \cref{sec:scalar-impact-on-cosmological-background}
with varying $m_{e, i}$, setting the scalar field energy density to fix $\amL$ as discussed
below \cref{eqn:a-mL-with-scalar}.
The specific parameter dependence is made concrete in \cref{sec:hyperlight-scalar-parameter-space};
the scalar's present energy density relative to that of CDM is approximately
$\omega_\phi / \omega_c \approx 1.2 (m_{e, i} / m_{e, 0} - 1)$ [see
\cref{eqn:Xi-a-star-degeneracy} below].
The lensing spectrum is greatly suppressed by the presence of the scalar on all scales, but
especially the smaller scales which, at late times, lens the primary CMB anisotropies, smearing the
acoustic peaks.
The latter effect is subtle but most evident in the $E$-mode power spectra displayed in
\cref{fig:cmb-lensing-ee-peaks-vary-me-with-scalar}, where the cases with the largest early-time
electron mass $m_{e, i}$ (and therefore the largest contribution from the scalar field) reach the
point where the peaks are essentially unaltered by structure at late times.
These findings are similar to those in Ref.~\cite{Rogers:2023ezo}, which studied the purely
gravitational effects of ultralight axions.\footnote{
    Note that the gravitational effects of pseudoscalars/axions and scalars are identical,
    up to the choice of the potential.
    Constraints on axions like those in
    Refs.~\cite{Hlozek:2014lca, Hlozek:2016lzm, Hlozek:2017zzf, Lague:2021frh,Rogers:2023ezo}
    therefore apply equally well to scalars without additional couplings.
}
Observe also that the suppression of the lensing potential is far more drastic (at fixed $m_{e, i} /
m_{e, 0}$) along the degeneracy direction with a hyperlight scalar than for that without one
(\cref{fig:cmb-isw-lensing-vary-me}).

\begin{figure}
    \centering
    \includegraphics[width=\textwidth]{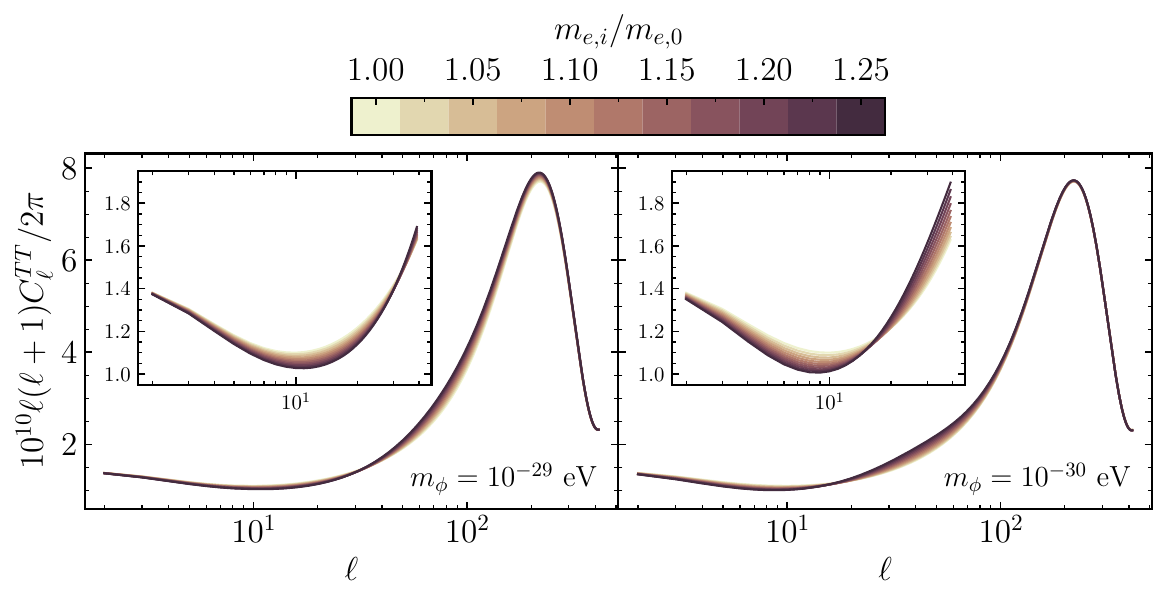}
    \caption{
        Large-scale CMB temperature power spectrum in cosmologies varying
        the electron mass $m_{e, i}$ with a scalar field, by color.
        Each curve holds all of $R_\star$, $x_\mathrm{eq}$, $\theta_s$, $Y_\mathrm{He}$,
        $A_s a_\star^{1 - n_s}$, and $\tau_\mathrm{reion}$ fixed and additionally includes a scalar
        field with mass $m_\phi = 10^{-29}~\eV$ (left) or $10^{-30}~\eV$ (right) with a density
        chosen to (approximately) fix the scale factor of matter--dark-energy equality via the
        degeneracy derived from \cref{eqn:a-mL-with-scalar}.
        The inset panels enlarge $\ell < 40$.
    }
    \label{fig:cmb-isw-vary-me-with-scalar}
\end{figure}

As discussed in \cref{sec:isw-lensing}, in Universes dominated by cold matter the ISW contribution
to CMB anisotropies vanishes, a feature contingent on density fluctuations growing exactly as
$\delta \rho / \bar{\rho} \propto a$.
The ISW effect sourced by a hyperlight scalar is depicted in \cref{fig:cmb-isw-vary-me-with-scalar}
for two different masses.
With $m_\phi = 10^{-29}~\eV$, the scalar begins oscillating at $z_\mathrm{osc} \approx 390$,
which is near the earliest it can while remaining effectively frozen during recombination (i.e.,
such that any shift in fundamental constants is constant during recombination).
However, this is still early enough to noticeably enhance the temperature power spectrum at the
first acoustic peak (at $\ell \sim 200$).
With a lighter scalar, e.g., with $m_\phi = 10^{-30}~\eV$ (and $z_\mathrm{osc} \approx 86$),
the ISW effect is irrelevant at $\ell > 100$ but is slightly more pronounced at smaller multipoles.
Indeed, in both cases the ISW effect has nontrivial scale dependence, even suppressing power on
large scales rather than enhancing it (as in smaller scales).
The same features are again evident in cosmologies with ultralight axions~\cite{Hlozek:2014lca}.
The scalar's full behavior---its homogeneous oscillations and dark-energy--like behavior beforehand,
and the growth of fluctuations on large scales---has effects beyond the suppression of structure
below the Jeans scale discussed here.
The dynamics of the scalar's perturbations therefore provide a possible means to constrain its mass
through gravitational effects alone; see Refs.~\cite{Hlozek:2014lca, Hlozek:2016lzm, Hlozek:2017zzf,
Lague:2021frh, Rogers:2023ezo} for more detailed exploration.

Scalars in the mass range we consider also suppress structure on scales that
contribute to $\sigma_8$, defined in \cref{sec:matter-clustering}.
At present, the CMB-derived value of $\sigma_8$ (within \LCDM{}) is mildly discrepant with that
inferred from large-scale structure probes~\cite{Abdalla:2022yfr}.
While the discrepancy is not nearly so severe as that of $H_0$ between \Planck{} and the
SH0ES-calibrated distance ladder, it is worth noting that hyperlight scalars could in principle
reconcile it, since those with masses $m_\phi \lesssim 10^{-24}~\eV$ do not cluster on
the scales to which $\sigma_8$ is sensitive.
Reference~\cite{Rogers:2023ezo} explored this possibility in detail, including constraints from galaxy
clustering.
However, since $h$ is negatively correlated with $\omega_\phi$ at fixed $a_\star$ (as explained in
\cref{sec:scalar-impact-on-cosmological-background}),
$h$ and $\sigma_8$ are positively correlated when increasing $\omega_\phi$.
The exacerbation of the discrepancy between $\sigma_8$ inferred from large-scale structure and CMB
datasets in cosmologies that raise the CMB's preference for $H_0$ is cited as a common feature of
\LCDM{} extensions.
On the other hand, a hyperlight scalar that also modifies the fundamental constants and triggers
early recombination has the potential to simultaneously reduce the CMB's preference of $\sigma_8$
and raise that of $H_0$.
Similar conclusions were drawn for the ``axi-Higgs'' model in Ref.~\cite{Luu:2021yhl}, in which an
ultralight axion shifts the electron mass by shifting the Higgs vacuum expectation value (VEV) at early times.
Ignoring the effect of the Higgs VEV on the proton mass (which is not a valid approximation, as
argued in Ref.~\cite{particle-paper}), the phenomenologies of the ``axi-Higgs'' scenario and of a
hyperlight scalar coupled directly to the electron are effectively identical.

Finally, ultra- and hyperlight components of dark matter can
suppress structure on nonlinear scales in a manner that may not be well captured by standard methods
to estimate nonlinear corrections to the matter power spectrum~\cite{Vogt:2022bwy, Lague:2023wes}.
Scalars in the mass range we consider, however, do not cluster on observationally relevant scales or
redshifts~\cite{Marsh:2016vgj, Hlozek:2016lzm, Bauer:2020zsj}.
Moreover, nonlinear structure growth substantially impacts the CMB lensing potential only on
scales smaller than those \Planck{} measures~\cite{Lewis:2006fu}.
Since we consider no other CMB lensing observations, we therefore follow Refs.~\cite{Hlozek:2014lca,
Hlozek:2016lzm, Hlozek:2017zzf, Lague:2021frh, Rogers:2023ezo} in neglecting any nonlinear effects
in the matter power spectrum.

\subsection{Parameter space}\label{sec:hyperlight-scalar-parameter-space}

The mass of a scalar that begins oscillating in the matter-dominated era is related to the matter
density as
\begin{align}
    m_\phi^2
    = \frac{9 H(t_\mathrm{osc})^2}{4}
    \approx \frac{3}{4}
        \frac{\bar{\rho}_c(t_0) + \bar{\rho}_b(t_0)}{a_\mathrm{osc}^3 \Mpl^2},
\end{align}
neglecting the scalar's own contribution to the total energy density at that time.
Using \cref{eqn:scalar-rho-matter-era}, the scalar's energy density relative to that of CDM (at
$t > t_\mathrm{osc}$) is therefore
\begin{align}
    \fphi
    &\equiv \frac{\bar{\rho}_\phi}{\bar{\rho}_c}
    = \frac{m_\phi^2 \bar{\phi}_i^2 / 2 (a / a_\mathrm{osc}^3)}{\bar{\rho}_c(t_0) / a^3}
    = \frac{3 \left( 1 + \omega_b / \omega_c \right) \bar{\varphi}_i^2}{4},
    \label{eqn:Xi-ito-varphi}
\end{align}
where $\bar{\varphi}_i$ is the initial homogeneous component of the rescaled field
\begin{align}
    \varphi
    &\equiv \frac{\phi}{\sqrt{2} \Mpl}
    \label{eq:normalization}
    .
\end{align}
Observe that this expression is independent of the mass $m_\phi$ and holds for masses between (3/2
times) the Hubble rate at matter-radiation equality,
$3 H_\mathrm{eq}/2 \approx 3.4 \times 10^{-28}~\eV$
and that at matter--dark-energy equality,
$3 H_{m-\Lambda}/2 \approx 2.6 \times 10^{-33}~\eV$.
The matter-era solution is then
\begin{align}
   \bar{\phi}(t) &\equiv \sqrt{2} \Mpl \varphiamp(t) \sin m_\phi t,
    \label{eq:field_amplitude}
\end{align}
where the time-dependent, squared oscillation envelope for at $t > t_\mathrm{osc}$ is
\begin{align}
    \varphiamp(t)^2
    &= \frac{4}{3} \frac{\fphi}{1 + \omega_b / \omega_c}
        \left( \frac{a(t)}{a_\mathrm{osc}} \right)^{-3}.
    \label{eq:field_redshift}
\end{align}

In Ref.~\cite{particle-paper}, the shift in a fundamental parameter $\lambda$ (i.e., $\alpha$ or
$m_e$) is parametrized in terms of the scalar field by
\begin{align}
    \lambda(\varphi)
    &\approx \lambda(0) \left[ 1 + g_\lambda(\varphi) \right],
\end{align}
where $\lambda(0)$ is the vacuum value of the parameter, i.e., when the scalar is at the minimum of
its potential.
(For consistency with prior literature, we replace the subscript $\alpha$'s with $e$'s for the
photon couplings, $g_e \equiv g_\alpha$.)
A shift in the inverse scale factor of recombination, with fundamental-constant dependence given by
\cref{eqn:a-star-proportionality}, may therefore be written in terms of the scalar's couplings to
the SM as
\begin{align}
    \frac{\Delta a_\star^{-1}}{a_{\star}^{-1}}
    \approx
        \frac{m_{e, i} - m_{e, 0}}{m_{e, 0}}
        + 2 \frac{\alpha_i - \alpha_0}{\alpha_0}
    &= g_{m_e}(\bar{\varphi}_i)
        + 2 g_{e}(\bar{\varphi}_i).
    \label{eqn:variation-a-star-inverse-ito-couplings}
\end{align}
One may translate the $\alpha_i$ dependence of any other quantity in
\cref{eqn:constrained-parameter-combos} (e.g., those parametrizing the effects on small-scale
damping and CMB polarization in \cref{sec:damping,sec:polarization}) analogously.
For a scalar field that is frozen during recombination, these relationships, combined with
\cref{eqn:Xi-ito-varphi}, fully specify the correspondence between the phenomenological parameters
($\fphi$ and the early-time values of $m_e$ and $\alpha$) to the fundamental parameters of the model
(those parametrizing $g_{m_e}$ and $g_e$ in addition to the initial condition
$\bar{\varphi}_i$).

To phrase the degeneracy direction identified in \cref{sec:scalar-impact-on-cosmological-background} in terms of
$\fphi$, write $\omega_{m, \mathrm{late}} = (1 + \fphi) \omega_c + \omega_b$, such that
$A_m = 1 + \fphi / (1 + \omega_b / \omega_c)$.
When $A_m \propto a_\star^{-1.034}$, which ensures that matter--dark-energy equality occurs at the
same scale factor regardless of $a_\star$ (along the identified degeneracy),
\begin{align}
    \frac{\fphi}{1 + \omega_b / \omega_c}
    \approx 1.034 \frac{\Delta a_\star^{-1}}{a_{\star}^{-1}}
    \label{eqn:Xi-a-star-degeneracy}
\end{align}
for small shifts $\Delta a_{\star}^{-1}$.
The scalar must therefore make a relative contribution to the matter density at late times that is
commensurate with the shift in recombination.
This condition is approximately independent of the scalar's mass, so long as it begins oscillating
after recombination (and therefore after the matter era begins) but sufficiently long before
matter--dark-energy equality.
More precisely, lighter scalars begin contributing to the matter abundance later and affect the
distance to last scattering over a shorter interval.
Scenarios with lighter scalars therefore require a slightly larger increase in response to an
increase in $a_\star^{-1}$ than given by the coefficient $1.034$ in \cref{eqn:Xi-a-star-degeneracy}
(but still order unity), which is derived in \cref{sec:scalar-impact-on-cosmological-background}
under the assumption that the scalar contributes to the matter density at all times after
recombination.

Per \cref{eqn:Xi-ito-varphi}, to realize the condition of \cref{eqn:Xi-a-star-degeneracy} the scalar
thus must have a near-Planckian initial condition.
Combining with \cref{eqn:variation-a-star-inverse-ito-couplings},
\begin{align}
    \bar{\varphi}_i^2
    = \frac{\bar{\phi}_i^2}{2 \Mpl^2}
    &\approx \frac{4}{3} \cdot 1.034
        \left[
            g_{m_e}(\bar{\varphi}_i)
            + 2 g_{\alpha}(\bar{\varphi}_i)
        \right].
    \label{eqn:degeneracy-ito-fund}
\end{align}
Reference~\cite{particle-paper} motivates models in which the leading contributions to $g_\lambda$ are
quadratic: $g_\lambda(\varphi) \approx d_\lambda^{(2)} \varphi^2 / 2$.
Because the scalar's energy contribution is \emph{also} quadratic in the field's value, the
degeneracy direction \cref{eqn:degeneracy-ito-fund} lies at a fixed value of the dimensionless
coupling coefficient.
In this case, the early-time variation in fundamental constants
[$\lambda_i \equiv \lambda(\bar{\varphi}_i)$] may be written as
\begin{align}
    \frac{\lambda_i}{\lambda(0)}
    - 1
    &= g_\lambda(\bar{\varphi}_i)
    = d_\lambda^{(2)} \bar{\varphi}_i^2 / 2
    = \frac{2 d_\lambda^{(2)} \fphi}{3 \left( 1 + \omega_b / \omega_c \right)}.
    \label{eqn:dlambda-i-ito-Xi}
\end{align}
For instance, when $g_e(\bar{\varphi}_i)$ is negligible, we expect the quadratic coupling
$d_{m_e}^{(2)} \approx 1.45$ to quantify the degeneracy direction in cosmological data. Interestingly, we show in Ref.~\cite{particle-paper} that such $\mathcal{O}(1)$ quadratic scalar couplings to electrons and photons are allowed by all other experimental probes.

The cosmological data we consider depend only on the early-time values of $m_e$ and $\alpha$, i.e.,
on $g_{m_e}(\bar{\varphi}_i)$ and $g_{e}(\bar{\varphi}_i)$ regardless of their functional form
(again, when the field remains frozen until after recombination).
The scalar's cosmological abundance is conveniently parametrized like other species via $\fphi$,
since $\omega_\phi \equiv \fphi \omega_c$.
The simplest parametrization of the model for parameter inference using cosmological data is
therefore simply that of the early-time constants, $m_{e, i}$ and $\alpha_i$, and $\fphi$.
However, physically motivated priors may only be (directly) phrased in terms of
the fundamental parameters.
In particular, while $\fphi$ depends only on $\bar{\varphi}_i^2$, variations in $m_{e, i}$ and $\alpha_i$ are
given by the product of (some power of) the field amplitude $\bar{\varphi}_i$ and dimensionless
coupling coefficients.
Any fixed shift in a fundamental constant may be accommodated by arbitrarily small $\bar{\varphi}_i$
with a large enough coupling.
Within this class of models and taking quadratic couplings as an example, the limit of a
gravitationally negligible scalar, $\bar{\varphi}_i^2 \ll 1$ and $d_{\lambda}^{(2)} \gg 1$---the
limit of strong coupling, in other words---corresponds to that of prior work that considered
variations in fundamental constants only at a phenomenological level.

Aside from neglecting the potential cosmological importance of the late dark matter contribution of a hyperlight scalar motivated in
\cref{sec:scalar-impact-on-cosmological-background}, the strong-coupling limit runs afoul of the
considerations of microphysical models in Ref.~\cite{particle-paper}. The effective scalar potential sourced by matter
becomes substantial and can affect the scalar's dynamics at early times, even before recombination.
Shifts in fundamental constants that are time independent through the end of recombination, as
studied by prior work on a phenomenological basis, thus require that the responsible scalar
contributes to gravity to some extent.
We exclude regimes in which the constants evolve before recombination from our analysis here but
discuss the possibility in more detail in Ref.~\cite{particle-paper}.

The opposite, weakly coupled limit is specified by $\bar{\varphi}_i \gtrsim 1$.
One might exclude this regime on the grounds that the scalar then traverses a super-Planckian field
distance~\cite{Ooguri:2006in}, but such a regime is surely entirely excluded observationally, given
that a scalar with $m_\phi \gtrsim H_0$ would have come to dominate the energy density of the
Universe before the start of coherent oscillations and cause a period of inflation when
$\bar{\varphi}_i \gtrsim 1$.
One might then wonder whether some or all of the Universe's present-day dark energy can in fact
made up of the scalar (e.g. $m_\phi \lesssim H_0$).
By its very nature, a species that behaves like dark energy or a slowly varying cosmological
``constant'' (e.g., a quintessence field) must have a nearly flat bare potential (which is true of
the $m_\phi^2 \phi^2$ potential when $m_\phi \ll H_0$).
Such a requirement would preclude any significant variation of the field between the present day and
earlier times.
However, the effective potential sourced by the cosmological bath of SM matter and radiation (or
even dark matter~\cite{Olive:2001vz,Olive:2007aj}) can easily dominate the bare potential at early
times.
We therefore cannot outright exclude the possibility of significant variations in fundamental
constants at early times from a field that is today dark-energy--like without considering concrete
models of its interactions and their constraints from other probes.
We restrict our analysis to scalars as subcomponents of the present-day dark matter for simplicity.

%% file: cosmo-results.tex
\section{Cosmological parameter inference}\label{sec:parameter-inference}

We now test the models and hypothesized parameter degeneracies discussed in
\cref{sec:varying-constants,sec:cosmology-hyperlight-scalars} using cosmological data.
Insofar as the degeneracy directions identified in
\cref{sec:varying-constants,sec:cosmology-hyperlight-scalars} (which are calculated about the
best-fit \LCDM{} cosmology) hold throughout parameter space, we expect \LCDM{} extended with a
hyperlight scalar coupled to the electron to satisfy CMB, BAO, and SN Ia data at any $m_e$ with $h
\propto m_e$ and $\fphi \equiv \omega_\phi / \omega_c \approx m_{e, i} / m_{e, 0} - 1$.
Moreover, beyond simply reproducing the best-fit \LCDM{} prediction, the additional freedom afforded
by the electron-coupled scalar can in principle enable precise changes in predictions that better
fit the data but are inaccessible within the \LCDM{} parameter space.
The scalar's matter contribution could also enhance the consistency of these datasets when $\alpha$
is varied, but the impact on small-scale damping and CMB polarization presumably remains as the
strongest limitation on the scenario.
Quantitatively testing these questions requires an analysis of those data using full numerical
solutions to the Einstein-Boltzmann equations (and the Klein-Gordon equation).

We first establish baseline expectations for phenomenological varying-constants scenarios (i.e.,
without a scalar field) in \cref{sec:results-varying-constants}, analyzing in detail the individual
and joint constraining power of CMB and low-redshift distance measurements.
We study the degeneracies of scenarios with electron mass variations and the secondary physical
effects in the CMB that mildly break them.
We also show how \Planck{} data constrain the fine-structure constant via its effects on
polarization and small-scale damping, which will be probed even more precisely by future
high-resolution CMB experiments.
We next show how BAO and SNe data combine with CMB data to break degeneracies with \LCDM{}
parameters and tighten constraints on varying $m_e$ and $\alpha$ scenarios.
\Cref{sec:results-scalars} then studies the impact of consistently including a coupled scalar field
as  a microphysical realization of time-varying fundamental constants.
We first derive constraints from the CMB alone on both coupled and uncoupled scalar fields.
We then  test the extent to which the novel degeneracy hypothesized to arise in scenarios with late
dark matter components (\cref{sec:cosmology-hyperlight-scalars}) is observed in the combination of
early- and late-time datasets.
For illustration, we compare results with hyperlight scalar fields to those that instead implement
late dark matter by varying the neutrino mass.

Finally, \cref{sec:concordance} discusses the implications of our results for cosmological tensions.
In particular, we highlight for the first time the disparity of results in varying-$m_e$ scenarios
(with or without a scalar field) among different BAO and SNe datasets; we identify the origin of
this disagreement in their discrepant preferences for when the dark energy era began (i.e., $\amL$).
We quantify the resulting impact on concordance of early and late datasets in the context of varying
constant and hyperlight scalar models for both the expansion history (via $h$ and $\amL$) and the
amplitude of structure ($\sigma_8$).
We argue more generally that concordance between late- and early-time datasets must be assessed in
the full parameter space of \LCDM{} cosmologies---not just the Hubble constant, parametrizing the
overall scale of the Universe, but also the shape of the late-time expansion history as encoded by
$\amL$ (or, equivalently, the matter fraction $\Omega_m$).

Before proceeding, we describe the methods and datasets we employ in our analysis.
In parameter inference, we sample over the standard set of \LCDM{} parameters with broad,
uninformative priors: the present baryon and CDM densities,
$\omega_b \sim \mathcal{U}(0.005, 0.035)$ and $\omega_c \sim \mathcal{U}(0.01, 0.25)$; the Hubble
rate $h \sim \mathcal{U}(0.25, 1.1)$; the tilt and amplitude of the scalar power spectrum,
$n_s \sim \mathcal{U}(1.61, 3.91)$ and $A_s$ [the latter via the combination as
$\ln(10^{10} A_s) \sim \mathcal{U}(1.61, 3.91)$]; and the redshift of reionization
$z_\mathrm{reion} \sim \mathcal{U}(4, 12)$.
Here we use $\mathcal{U}(a, b)$ to denote a uniform prior between $a$ and $b$.
In some circumstances it proves more computationally efficient to sample over the angular scale of
the sound horizon $\theta_s \sim \mathcal{U}(0.9, 1.1)$ in place of $h$; we comment on the impact of
this choice where it is used.
We exclude the possibility of a negative dark energy density (which is otherwise accessible in
allowed parameter space in some situations).
We sample over $m_{e, i} / m_{e, 0}$ and $\alpha_i / \alpha_0 \sim \mathcal{U}(0.7, 1.3)$ whenever
considering variations of each parameter.
We fix a single massive neutrino with the standard, minimal mass $m_\nu = 0.06~\eV$;
representing the other two by a relativistic fluid with an effective number of degrees of freedom
of $2.0308$ yields a total early-time $N_\mathrm{eff}$ [\cref{eqn:omega-nu-omega-gamma}] of $3.044$.
\Cref{app:sampling} details our methods for sampling via Markov chain Monte Carlo (MCMC).

In the following, we consider a variety of datasets of the types described in \cref{sec:varying-constants} and discuss their
implementations in more detail in \cref{app:likelihoods}.
We use the 2018 \Planck{} likelihoods via the \texttt{Plik\_lite} variants that are marginalized over
the parameters of the foreground models~\cite{Planck:2018vyg,Planck:2019nip}.
These likelihoods require an additional parameter $y_\mathrm{cal}$ characterizing the overall map
calibration, sampled with the standard Gaussian prior with mean $1$ and standard deviation
$0.0025$~\cite{Planck:2019nip}.
We use BAO measurements from a common combination of surveys: the Six-degree Field Galaxy Survey
(6dFGS)~\cite{Beutler:2011hx} (with an effective sample redshift $z = 0.106$), the SDSS Main Galaxy
Sample DR7~\cite{Ross:2014qpa} ($z = 0.15$), the Baryon Oscillation Spectroscopic Survey
(BOSS) DR12 galaxies ($z = 0.38$ and $0.51$), and the Extended BOSS (eBOSS) DR16 luminous red galaxies (LRG)~\cite{eBOSS:2020lta, eBOSS:2020hur}
($z = 0.70$).
Note that the constraining power of this combination is dominated by eBOSS measurements, at least
for the models we consider; we refer to this commonly used combination as the SDSS+ BAO dataset.
We separately use recent BAO measurements from DESI DR1~\cite{DESI:2024mwx, DESI:2024lzq,
DESI:2024uvr}, using data from all tracers; we note that these results are yet unpublished at the
time of writing.
Finally, we use supernova distance measurements from the Pantheon~\cite{Pan-STARRS1:2017jku},
Pantheon+~\cite{Brout:2022vxf, Scolnic:2021amr}, DES 5YR~\cite{DES:2024tys}, and
Union3~\cite{Rubin:2023ovl} results; we sample over the overall normalization via the degenerate
combination $\log_{10} h - M_B / 5 \sim \mathcal{U}(3.7, 3.73)$ that is well constrained
independently of external measurements of the fiducial magnitude $M_B$ (or of $h$, for that matter).
That is, we do not include calibration from, e.g.,
Cepheids~\cite{Riess:2016jrr,Riess:2019cxk,Riess:2021jrx} or the tip of the red giant
branch~\cite{Freedman:2019jwv,Freedman:2020dne,Freedman:2021ahq,Freedman:2023jcz}.

\subsection{Varying constants}\label{sec:results-varying-constants}

To set the stage, we begin by analyzing phenomenological varying-constant models in detail, first
with CMB data alone (\cref{sec:results-varying-constants-planck}) and then in combination with
low-redshift distance data (\cref{sec:results-varying-constants-low-z}).

\subsubsection{\Planck{} data alone}\label{sec:results-varying-constants-planck}

\begin{figure}[ht!]
    \centering
    \includegraphics[width=\textwidth]{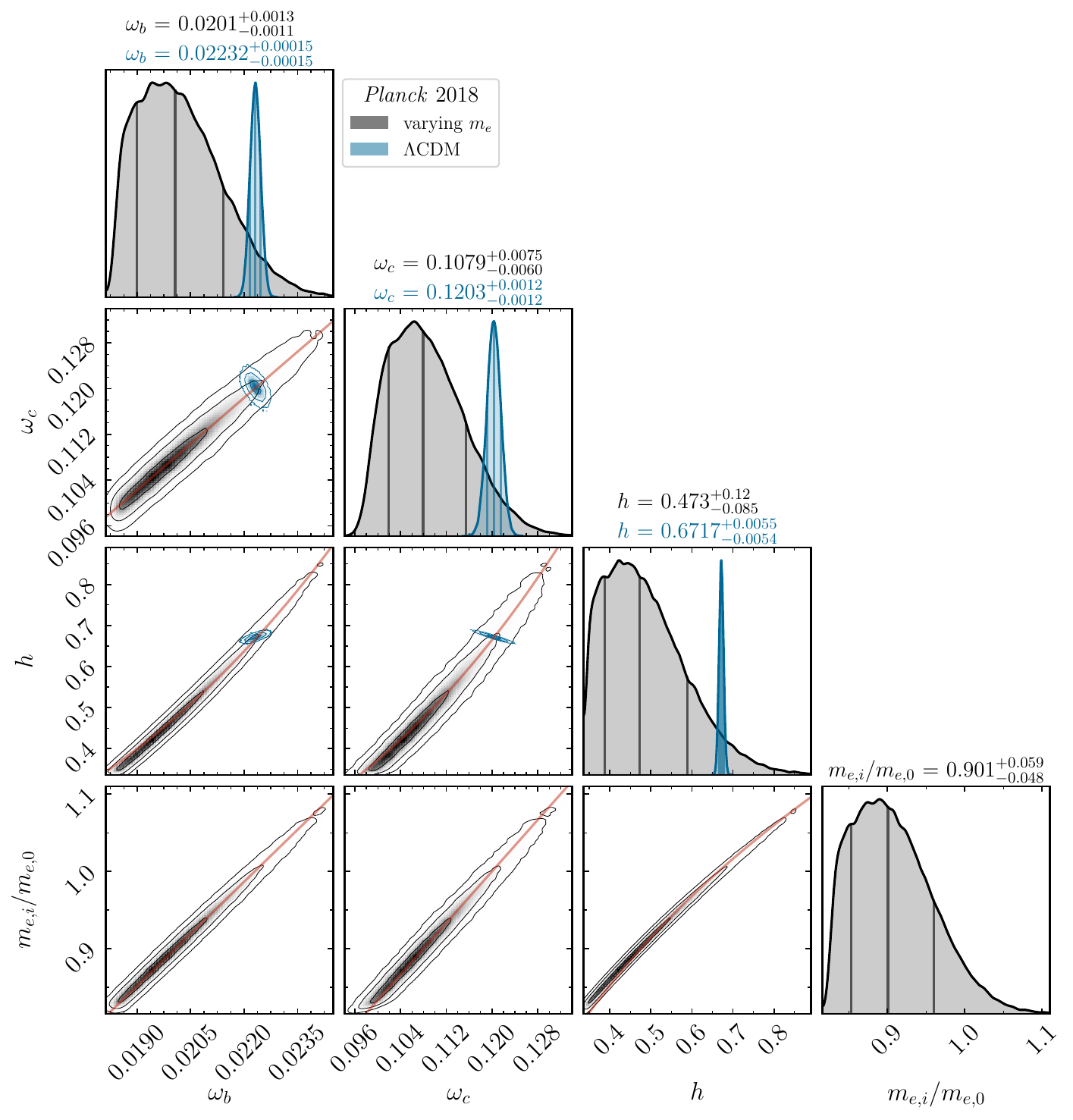}
    \caption{
        Posterior distribution over a subset of the parameters of the varying electron mass model
        without a scalar field (gray) which exhibits extended degeneracies compared to standard \LCDM{} (blue), using the likelihood for the
        full set of \Planck{} 2018 data including lensing (see \cref{app:likelihoods}).
        The panels along the diagonal depict kernel density estimates of marginalized,
        one-dimensional posteriors in each parameter, with vertical lines marking the median and
        $16$th and $84$th percentiles.
        These one-dimensional posteriors are normalized relative to their peak value to facilitate
        comparison.
        The median and corresponding $\pm 1 \sigma$ uncertainties for each parameter are reported
        above the diagonal panels.
        The lower panels display the marginalized, two-dimensional joint posterior density for pairs
        of parameters as a histogram and the 1, 2, and $3 \sigma$ contours thereof
        (i.e., the $39.3\%$, $86.5\%$, and $98.9\%$ mass levels).
        Contours are slightly smoothed to mitigate noise due to the binning of such narrow
        distributions, but not to an extent that artificially broadens their extent.
        Red lines in each panel depict the expected degeneracy direction derived in
        \cref{sec:angular-power-spectra} which is largely born out in the data:
        $\omega_b$ and $\omega_c \propto m_{e, i}$ and $h \propto m_{e, i}^{3.18}$.
    }
    \label{fig:vary-me-corner-wb-wc-h-me}
\end{figure}

We first seek to test whether the parameter degeneracies argued for in \cref{sec:varying-constants}
bear out in a complete analysis of CMB data.
\Cref{fig:vary-me-corner-wb-wc-h-me} displays the posterior distribution for \LCDM{} with a varying
early-time electron mass, evaluated with \Planck{} 2018 CMB and lensing likelihoods.
The results indeed display a strong degeneracy, which the red curves show correspond closely to the
direction $\omega_b \propto m_{e, i}$, $\omega_c \propto m_{e, i}$, and $h \propto m_{e, i}^{3.18}$.
The posterior is cut off at low values of these parameters almost exclusively by requiring
$h^2 \geq \omega_b + \omega_c$ (i.e., requiring a nonnegative dark energy density).
The posteriors are therefore asymmetric, falling off at values of
$m_{e, i} / m_{e, 0} \gtrsim 0.92$.
Namely, arbitrarily large values of the electron mass are \emph{not} as favored as those with
$m_{e, i} / m_{e, 0} < 1$.
As a consequence, $97.5\%$ of the posterior lies where $h < 0.72$---not the resounding solution to
the Hubble tension the results of, e.g., \cref{fig:cmb-all-spectra-degeneracies-vary-both} might
promise.
Moreover, the median of the marginalized posterior over the Hubble constant is $h = 0.47$, and that
of the early-time electron mass is $m_{e, i} / m_{e, 0} = 0.90$.

\begin{figure}[t!]
    \centering
    \includegraphics[width=\textwidth]{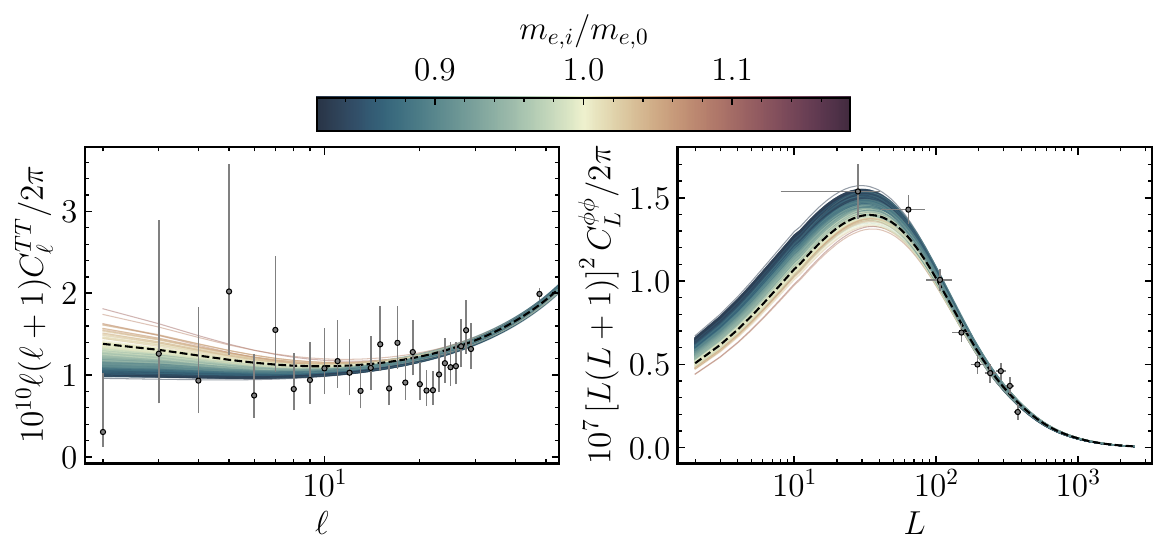}
    \caption{
        Large-scale CMB temperature (left) and lensing (right) power spectra for $1000$
        cosmologies sampled from the posterior in \cref{fig:vary-me-corner-wb-wc-h-me}, i.e., that
        for the varying electron mass model (without a scalar field) subject to \Planck{} 2018 data.
        Each curve is colored according to its value of $m_{e, i} / m_e$ as indicated by the color
        bar.
        The binned \Planck{} 2018 spectra and standard errors are superimposed in gray, and the result
        for the \LCDM{} best-fit cosmology is depicted in dashed black.
        These results may be compared with those in \cref{fig:cmb-isw-lensing-vary-me} (though the
        vertical scales are not identical).
    }
    \label{fig:isw-lensing-spaghetti-vary-me-cmb-only}
\end{figure}

We identify several effects that disfavor larger $m_{e, i}$ (and therefore larger $h$).
As argued in \cref{sec:varying-constants}, the degeneracy in the varying electron mass model is
effectively perfect so far as early-time effects are concerned, but the CMB is sensitive to the late
Universe via the ISW effect and lensing (\cref{sec:isw-lensing}).
In \cref{fig:isw-lensing-spaghetti-vary-me-cmb-only} we display the low-$\ell$ temperature power
spectrum and the lensing power spectrum evaluated over a representative sample of the posterior.
The early-time electron mass correlates strongly to the ISW enhancement of the temperature power
spectrum at low multipoles because $h$ is strongly constrained to increase with $m_{e, i}$ in order
to correctly locate the acoustic peaks.
As explained in \cref{sec:isw-lensing}, $h$ must decrease so rapidly with decreasing $m_{e, i}$ that
matter--dark-energy equality is pushed to much later times (even to the future,
$\amL > 1$).
As illustrated as well by \cref{fig:cmb-isw-lensing-vary-me}, matter overdensities therefore decay
less, enhancing the lensing spectrum relative to \LCDM{} and reducing the ISW enhancement of
large-scale temperature power.
The data in the left panel of \cref{fig:isw-lensing-spaghetti-vary-me-cmb-only} suggest that the ISW
effect by itself prefers late recombination to dark energy, given that most of their central values
skew low, even compared to curves at the lower boundary of the posterior (which corresponds to
negligible dark energy density).
Indeed, a lack of power on large angular scales is a persistent and unexplained feature of CMB
observations~\cite{WMAP:2003elm, Efstathiou:2009di, Copi:2010na, Copi:2013jna, Copi:2016hhq,
Planck:2019evm}.

In addition, lensing data also appear to favor the enhanced peak evident in
\cref{fig:isw-lensing-spaghetti-vary-me-cmb-only}, but it is challenging to compare the relative
constraining power of the first few data points that skew high to those at higher multipoles that
skew low (and have smaller error bars).
More likely, the enhanced lensing power is favored by the ``lensing anomaly'' of \Planck{}
data~\cite{Calabrese:2008rt, Planck:2016tof, Planck:2018vyg, Motloch:2019gux} in which the acoustic
peaks exhibit greater lensing-induced smoothing than predicted in \LCDM{}.
Such a conclusion was drawn in Ref.~\cite{Ivanov:2020mfr}, which studied CMB power spectrum
constraints on its present-day temperature $T_0$; as argued in \cref{sec:degeneracies}, varying $T_0$
and $m_{e, i}$ should exhibit nearly identical phenomenology.

Finally, the shape of a chosen prior is not necessarily negligible over posteriors so broad as those
in \cref{fig:vary-me-corner-wb-wc-h-me}.
For the results in \cref{fig:vary-me-corner-wb-wc-h-me} (and any results using CMB data alone), we
sample over $\theta_s$ instead of $h$ (using numerical optimization to solve for $h$ in this case),
as is conventional in MCMC analyses of CMB data.
In scenarios with varying electron mass, uniformly sampling over $h$ and $m_{e, i}$ covers a broad
range of values for $\theta_s$, most of which lie far from the narrow range of $\theta_s$ values
allowed by \Planck{}.
The angular size of the sound horizon $\theta_s$, suitably calculated, provides a measure of a
broadly model-independent feature in the data that is well constrained (the angular locations of the
acoustic peaks) and whose posterior allows for more efficient MCMC sampling (in terms of effective
sample size per MCMC sample).
But $\theta_s$ and $h$ are nonlinearly related, meaning a uniform prior for one results in a
nonuniform prior for the other.
\Cref{eqn:theta-s-degeneracy} (which is evaluated near the \LCDM{} best-fit parameters) allows us to
estimate that the effective prior on $h$ from sampling uniformly in $\theta_s$ is
$\partial \theta_s / \partial h \sim h^{-0.81}$, a trend that is roughly reproduced by direct
samples from our prior.
This scaling does not explain all (nor even most) of the trend displayed in
\cref{fig:vary-me-corner-wb-wc-h-me}, however.
Such a penalty on larger $h$ is entirely negligible for \LCDM{} with contemporary datasets, where
$h$ is constrained at the $1\%$ level or better, but it is of at least marginal consequence for
attempts to address the Hubble tension with models that achieve broader posteriors on $h$ (like that
in \cref{fig:vary-me-corner-wb-wc-h-me}).\footnote{
    As both $h$ and $\theta_s$ are derived parameters, a uniform prior in one is not necessarily any
    more motivated than in the other; one could equally well choose the dark energy density
    $\omega_\Lambda$.
    A more theoretically grounded prior (and choice of parametrization) would require a concrete
    model of dark energy.
}

\begin{figure}[t!]
    \centering
    \includegraphics[width=\textwidth]{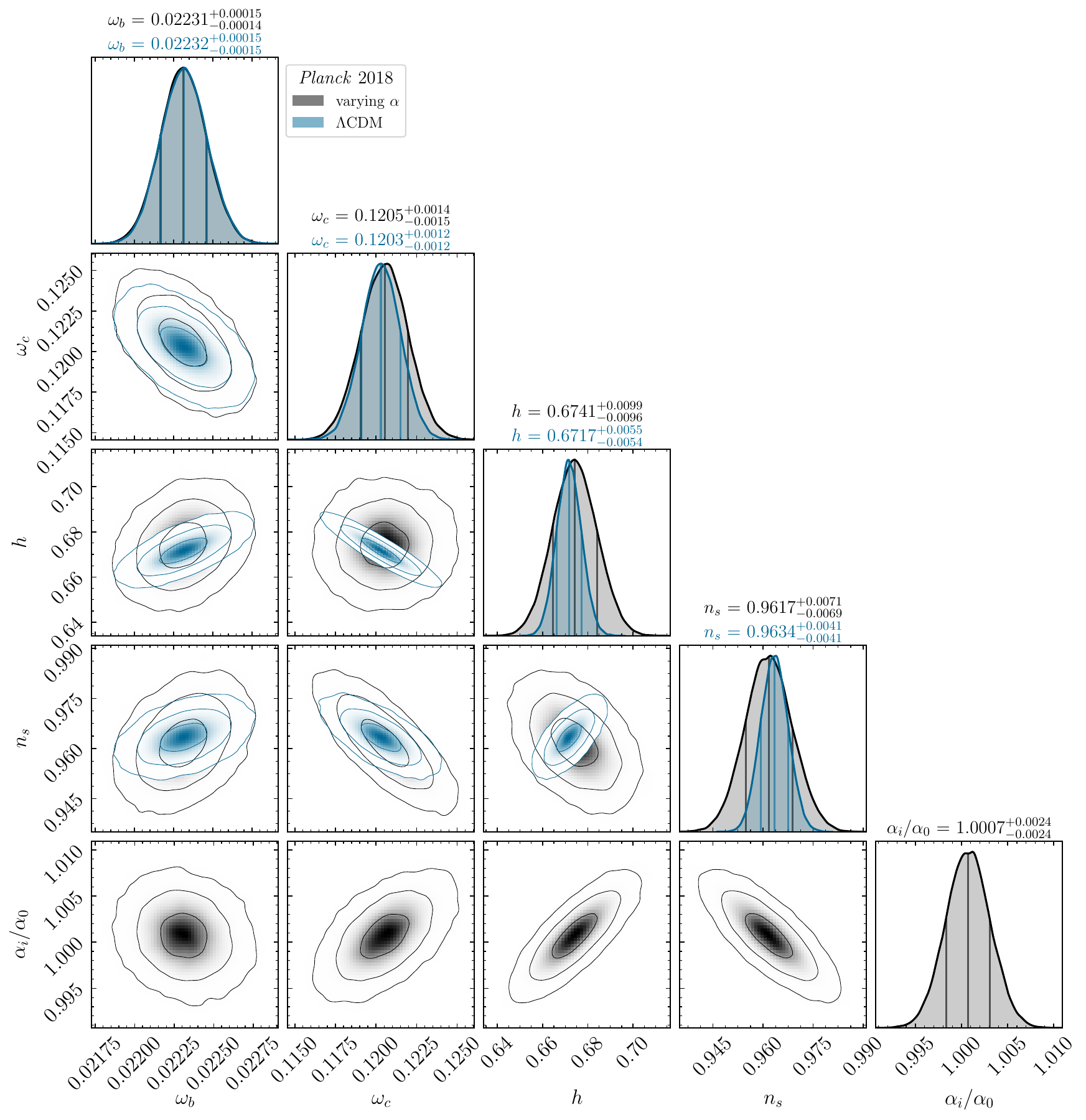}
    \caption{
        Posterior distribution over a subset of the parameters of the varying fine-structure
        constant model (without a scalar field), using the likelihood for the full set of \Planck{}
        2018 data including lensing (see \cref{app:likelihoods}).
        Results are depicted as described in the caption of \cref{fig:vary-me-corner-wb-wc-h-me},
        additionally plotting the scalar spectral tilt $n_s$ due to its correlation with
        $\alpha_i / \alpha_0$.
    }
    \label{fig:vary-alpha-corner-wb-wc-h-ns-alpha}
\end{figure}

Turning to variations in the fine-structure constant, \cref{fig:vary-alpha-corner-wb-wc-h-ns-alpha}
presents parameter constraints from \Planck{} 2018 data alone in analogy to
\cref{fig:vary-me-corner-wb-wc-h-me}.
Due to the strong dependence of small-scale damping (\cref{sec:damping}) and of the amplitude of the
polarization spectrum (\cref{sec:polarization}) on $\alpha_i$, illustrated in
\cref{fig:cmb-all-spectra-degeneracies-vary-both}, no extended degeneracy of \LCDM{} parameters with
$\alpha_i$ appears.
Because $h$ is even more sensitive to $\alpha_i$ than to $m_{e, i}$ at fixed $\theta_s$
[\cref{eqn:theta-s-degeneracy}], the posterior distribution over $h$ broadens slightly compared to
\LCDM{}, with a standard deviation roughly twice as large.
In contrast to $m_{e, i}$, the fine-structure constant is also correlated with the tilt of the
scalar power spectrum $n_s$ [defined in \cref{eqn:def-scalar-power-spectrum}].
Though diffusion and Landau damping effectively have exponential dependence on scale, \Planck{}'s
observations cover only a limited range of scales where they are relevant, enabling a partial
degeneracy with the power-law tilt of the initial power spectrum.
Because the pivot scale $k_\mathrm{p}$ that parametrizes the primordial power spectrum
[\cref{eqn:def-scalar-power-spectrum}] is larger than the inverse sound horizon at recombination, a
bluer spectrum (larger $n_s$) reduces the amplitude of the first peak and requires a larger
amplitude $A_s$ to compensate.
The degeneracy between $A_s$ and $\alpha_i$ is evident in our results, but more weakly than that
between $n_s$ and $\alpha_i$, and also does not affect the marginalized posterior over $A_s$.
Finally, we note that neither the lensing nor low-$\ell$ anomalies discussed around
\cref{fig:isw-lensing-spaghetti-vary-me-cmb-only} for varying $m_{e, i}$ are relevant in the
parameter space available when varying $\alpha_i$.

\begin{figure}
    \centering
    \includegraphics[width=\textwidth]{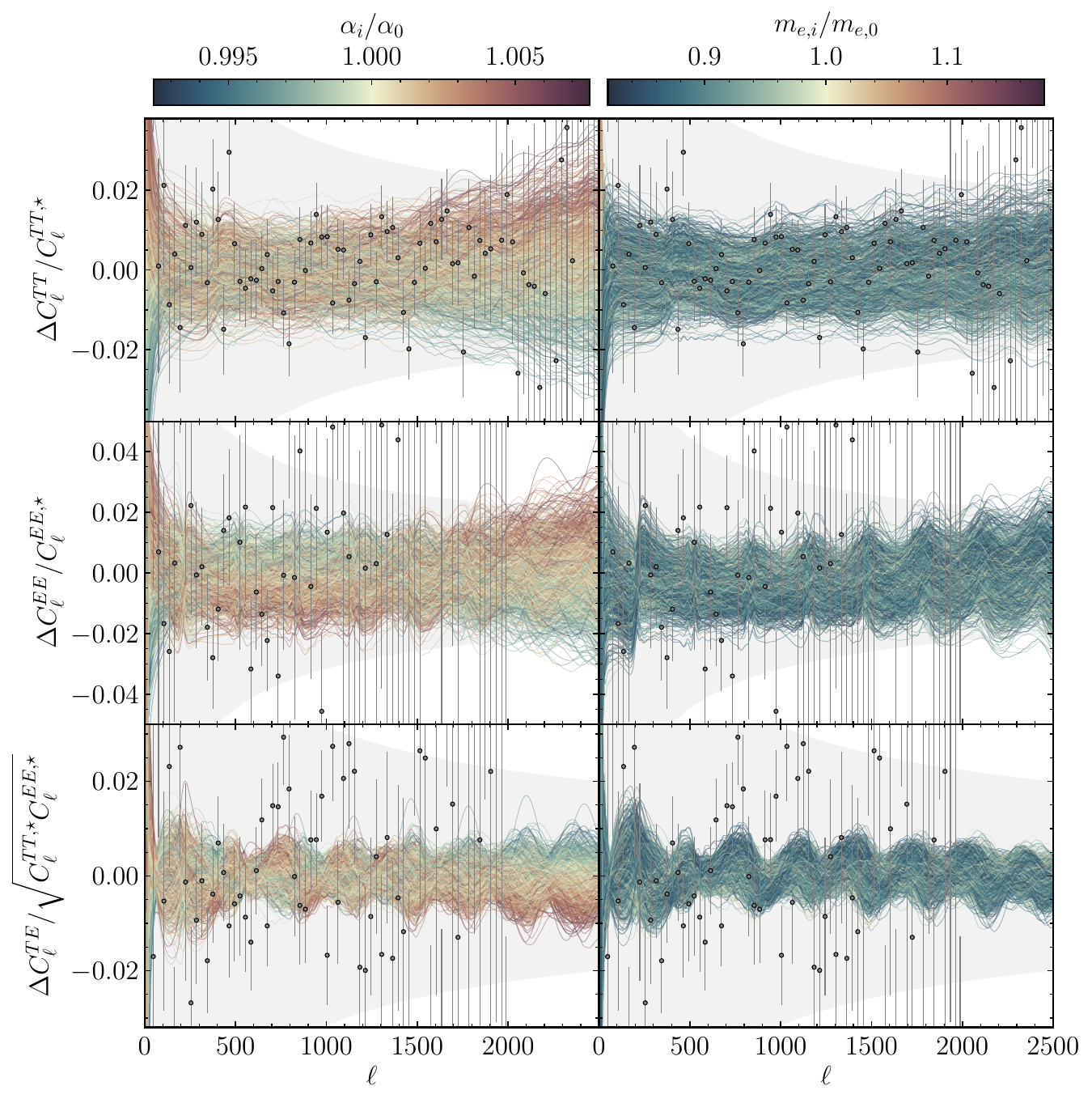}
    \caption{
        Residuals of the CMB power spectrum relative to the best-fit \LCDM{} result,
        $\Delta C_\ell^{XY} = C_\ell^{XY} - C_\ell^{XY, \star}$, evaluated for $1000$ cosmologies
        sampled from the posterior of the varying-$\alpha$ (left panels) and -$m_e$ (right panels)
        models with the \Planck{} 2018 likelihoods alone.
        In order to illustrate the correlation between parameters and features in the residuals,
        each curve is colored by its value of the fundamental constant that is varied as indicated
        by the respective color bar.
        The top and middle panels depict relative difference of the temperature and $E$-mode auto
        power spectra, respectively, while the bottom panels depict the residual of the
        $TE$ cross correlation normalized by $\sqrt{C_\ell^{TT, \star} C_\ell^{EE, \star}}$.
        The binned \Planck{} 2018 spectra and standard errors are superimposed in gray, and the shaded
        gray regions show the extent of cosmic variance for each multipole $\ell$ [i.e., relative
        errors of $\sqrt{2 / (2 \ell + 1)}$].
        (\Planck{}'s error bars can be smaller because they are binned over a range of multipoles.)
        Note that the color scales are symmetric about one even when the posterior is not (as is the
        case for $m_{e, i} / m_{e, 0}$ in the right panels).
    }
    \label{fig:cmb-all-spectra-spaghetti-vary-both}
\end{figure}

\Cref{fig:cmb-all-spectra-spaghetti-vary-both} displays residuals of the CMB power spectrum relative
to the \LCDM{} best fit, evaluated for a representative sample of the posterior for the
varying-$m_e$ and varying-$\alpha$ scenarios using only the \Planck{} 2018 likelihoods.
Beyond the differences exhibited at low $\ell$ in \cref{fig:isw-lensing-spaghetti-vary-me-cmb-only},
the residuals in the varying-$m_e$ scenario exhibit no clear trend with $m_{e, i}$, nor any easily
identifiable systematic effect.
Aside from noticeable oscillations in the $TE$ cross correlation residuals, the lack of particularly
distinguishable features is consistent with the expectation from \cref{sec:varying-constants} that
no important physical effects correlate uniquely with the electron mass (within the full extended
\LCDM{} parameter space).

The residuals for the varying-$\alpha$ scenario display several features in line with
the expectations from the effect of the fine structure constant on the damping tail and polarization discussed in \cref{sec:varying-constants} and
\cref{fig:cmb-all-spectra-degeneracies-vary-both}.
The correlation of damping with $\alpha_i$ is particularly evident at high multipoles, with larger
(red) and smaller (blue) $\alpha_i / \alpha_0$ displaying increased and decreased small-scale power,
respectively.
In addition, the enhancement of moderate-scale ($\ell \sim 500 - 1200$) polarization power for
\emph{smaller} $\alpha_i / \alpha_0$ (in contrast to the trend at larger $\ell$) due to the
broadened visibility function is also evident in \cref{fig:cmb-all-spectra-spaghetti-vary-both}, as
anticipated in \cref{fig:cmb-all-spectra-degeneracies-vary-both}.
More recent CMB polarization observations, like the Atacama Cosmology Telescope~\cite{ACT:2020gnv}
and the South Pole Telescope~\cite{SPT-3G:2021eoc,SPT-3G:2022hvq} would likely marginally improve
upon \Planck{}'s ability to constrain $\alpha_i$ via the effect with cosmic-variance--limited
measurements at $\ell \gtrsim 1000$.
However, the ultimate constraining power of features on moderate scales is limited by cosmic
variance; current and future high-resolution CMB experiments clearly have great potential to improve
upon the fine-structure constant constraints of \Planck{}.
In particular, \cref{fig:cmb-all-spectra-degeneracies-vary-both} suggests that high-resolution
observations can break the degeneracy of $\alpha_i$ and $n_s$ evident in
\cref{fig:vary-alpha-corner-wb-wc-h-ns-alpha}.
While the residuals in the varying-$m_e$ scenario also exceed cosmic variance on scales beyond
\Planck{}'s resolution, they do not appear to correlate with $m_{e, i}$ to any noticeable degree; this
suggests that high-resolution CMB observations may not substantially improve bounds on electron mass
variation.

\subsubsection{Impact of low-redshift distance measurements}
\label{sec:results-varying-constants-low-z}

We now consider the impact of constraints on the late-Universe expansion history on the findings of
the previous section.
In particular, the CMB itself provides very weak constraints on the early-time electron mass that
are mostly driven by \Planck{}'s mild (and possibly anomalous) preference for late recombination in a
Universe with a much lower dark energy abundance than in \LCDM{}.
However, \cref{sec:bao} shows that the BAO scale at late times exhibits a distinct degeneracy
between $h$ and $a_\star$ (at fixed $\amL$).
To provide effective constraints using BAO data alone, we impose an additional prior on $R_\star$
and $x_\mathrm{eq}$ derived from the posteriors in \cref{fig:vary-me-corner-wb-wc-h-me}.
In particular, a bivariate normal distribution with means $(0.620, 0.321)$,
autocovariances $(1.7, 0.66) \times 10^{-5}$, and cross-covariance $4.1 \times 10^{-6}$ effectively
reproduces posteriors for $R_\star$ and $x_\mathrm{eq}$ using \Planck{} 2018 data for \emph{both}
\LCDM{} and varying-$m_e$ cosmologies (as \cref{sec:varying-constants} argued should be the case).
This prior effectively includes the information from the shape of the CMB as measured by \Planck{}
that drives constraints on $\omega_b$ and $\omega_c$ marginalized over the angle subtended by the
sound horizon, $\theta_s$.
Without this information, BAO data alone would be unable to pick out a unique degeneracy between $h$
and $a_\star$.

\begin{figure}[t!]
    \centering
    \includegraphics[width=0.495\textwidth]{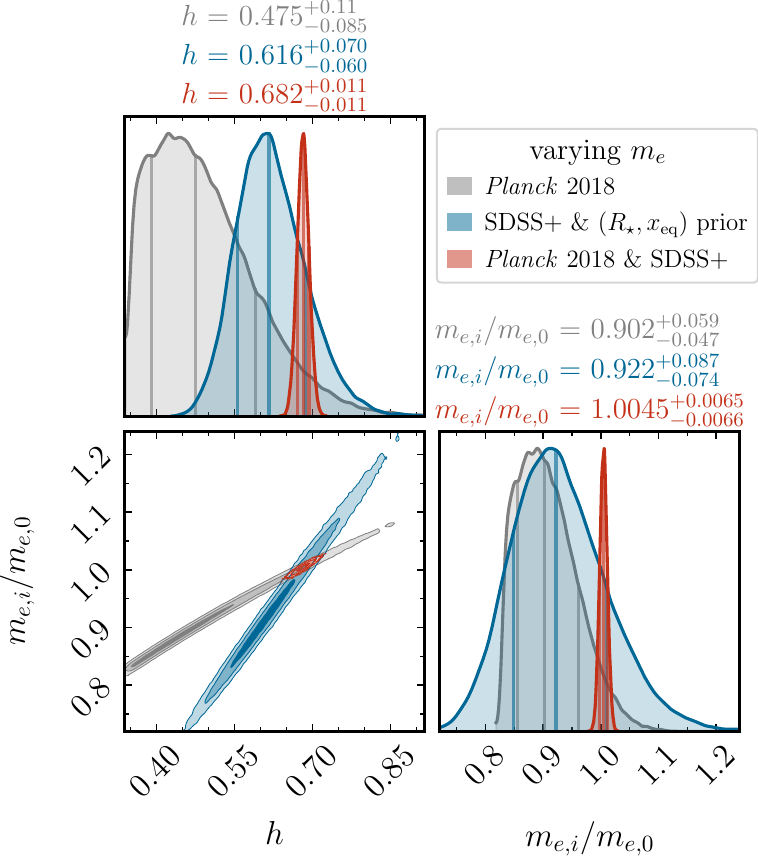}
    \includegraphics[width=0.495\textwidth]{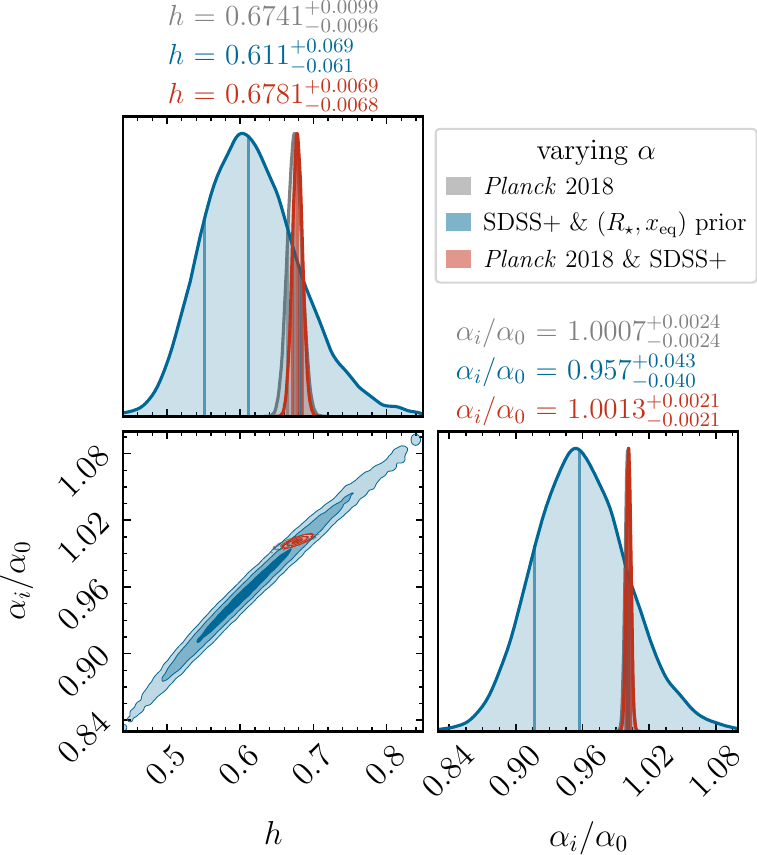}
    \caption{
        Posterior distribution over $h$ and $m_{e, i} / m_{e, 0}$ (left) or $\alpha_i / \alpha_0$
        (right) in cosmologies with varying constants without a scalar field.
        The same quantities are depicted in
        \cref{fig:vary-me-corner-wb-wc-h-me,fig:vary-alpha-corner-wb-wc-h-ns-alpha}, comparing
        results using the \Planck{} 2018 likelihoods (gray); the BAO dataset from eBOSS, SDSS, and
        6dFGS described in the text (blue); and both simultaneously (red).
        The diagonal panels display kernel density estimates of the one-dimensional, marginalized
        posterior relative to their peak value, and the lower panels depict 1, 2, and $3 \sigma$
        mass levels of the two-dimensional joint posterior, again slightly smoothed to facilitate
        comparison.
    }
    \label{fig:planck-vs-bao}
\end{figure}

\Cref{fig:planck-vs-bao} compares the posteriors for $h$ and $m_{e, i}$ or $\alpha_i$ when using
\Planck{} 2018 data and the SDSS+ BAO dataset (including the above-described prior) separately
and jointly.
Without any additional mechanism, varying constant scenarios may only simultaneously fit both
\Planck{} and BAO data where their individual posteriors have overlapping support.
Their combined constraints lie modestly far from the peak of either posterior (and in fact is
localized at larger values of $h$ and $m_{e, i}$ than either individual posterior).
The $2 \sigma$ mass levels for both cases in fact hardly overlap, further highlighting that \Planck{}
data prefer quite a different late-time Universe via the shape of the CMB than preferred by
by BAO measurements or by measurements of the distance to last scattering in \LCDM{}.
Nevertheless, the posterior is localized in the $h$-$m_{e, i}$ plane about
$m_{e, i} \approx m_{e, 0}$ and $h$ near its \LCDM{} region.
Under variations in the fine-structure constant, a broad degeneracy with $h$ is allowed by BAO data,
since it and the prior on $R_\star$ and $x_\mathrm{eq}$ are only sensitive to $\alpha_i$ via its
effect on $a_\star$.
Comparing these results with those including \Planck{} data highlights how precisely CMB observations
can measure the early-time fine structure constant.
The datasets combined constrain the early-time electron mass to deviate at only the percent level
and the fine-structure constant at the $0.2 \%$ level from their values measured today.

\begin{figure}[t!]
    \centering
    \includegraphics[width=0.495\textwidth]{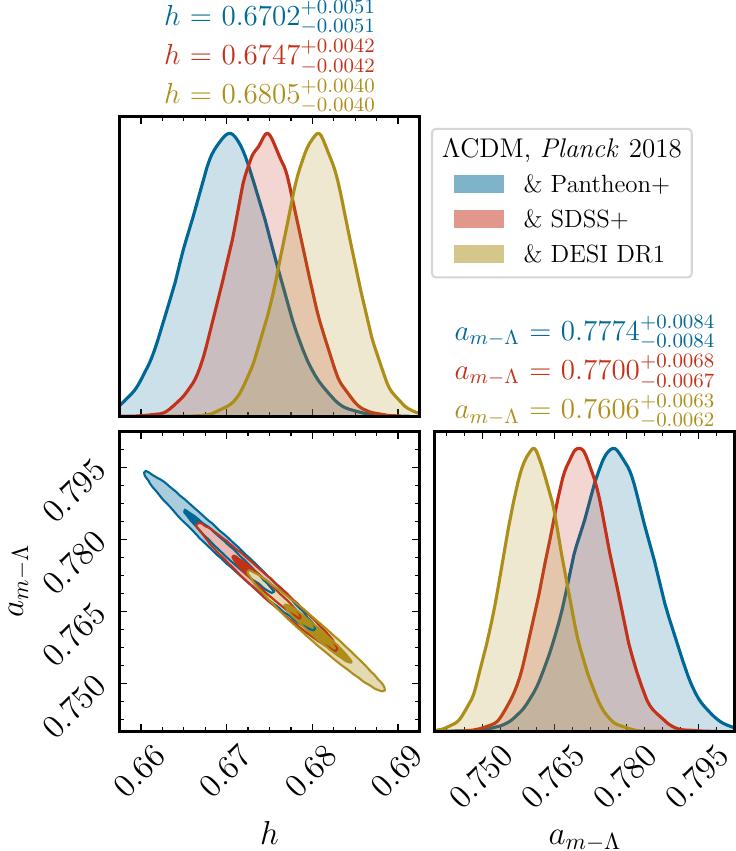}
    \includegraphics[width=0.495\textwidth]{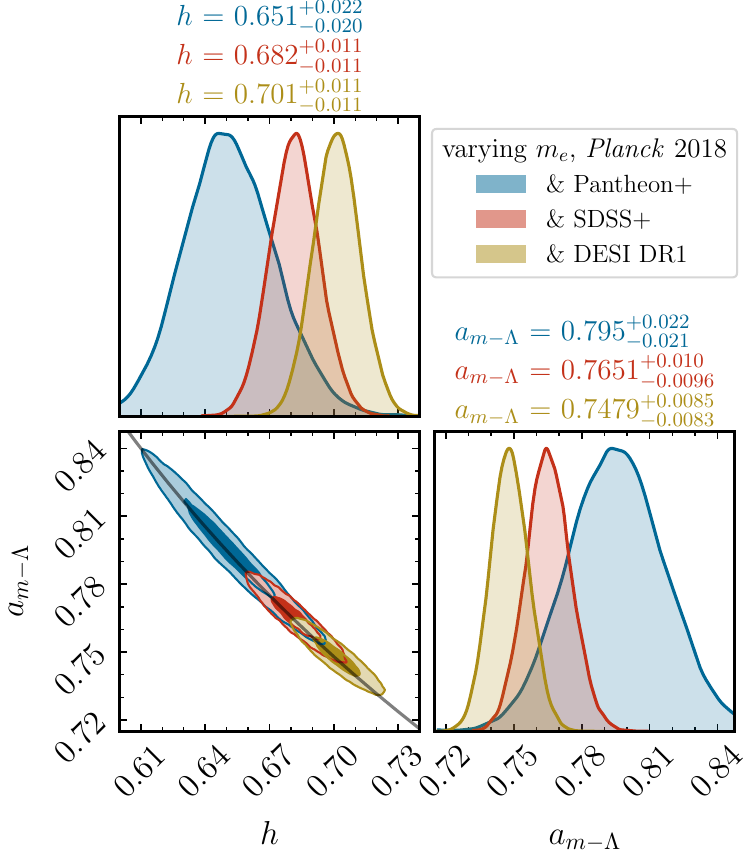}
    \caption{
        Posterior distribution over $h$ and $\amL$ (a derived parameter) in \LCDM{} (left) and
        varying-$m_e$ scenarios (right), comparing results that use the \Planck{} 2018 likelihoods
        combined with one of the Pantheon+ (blue), SDSS+ BAO (red), and DESI DR1 BAO (gold)
        datasets.
        The diagonal panels display kernel density estimates of the one-dimensional, marginalized
        posterior relative to their peak value, and the lower panels depict 1 and $2 \sigma$ mass
        levels of the two-dimensional joint posterior.
        Note that the left panels span a much broader range in $h$ and $\amL$ than do the
        right ones.
        The gray line in the lower left panel depicts the expected degeneracy inferred from
        \cref{eqn:theta-s-degeneracy}, i.e., with
        $\omega_m \propto m_{e, i}$ and $h \propto m_{e, i}^{3.18}$.
    }
    \label{fig:h-amL-low-z-compare}
\end{figure}

For neither varying $\alpha_i$ nor $m_{e, i}$ do the posteriors in \cref{fig:planck-vs-bao} using
only BAO data lie precisely along $h \propto 1/a_\star$, as would be expected from \cref{sec:bao} if
$\amL$ were well constrained.
The BAO measurements we employ in reality only cover a relatively narrow range of redshifts and are
insufficient in number and precision to fully measure $\amL$ independently of the overall
normalization in \cref{eqn:theta-bao-perp-flat-lcdm}, i.e., they remain slightly correlated.
While eBOSS provides additional BAO measurements from tracers beyond LRGs like quasars, emission
line galaxies, and Lyman-$\alpha$ absorption~\cite{eBOSS:2020yzd}, cosmological constraints from
these are known to be mildly inconsistent~\cite{BOSS:2014hhw, BOSS:2014hwf, Addison:2017fdm,
Cuceu:2019for}.
However, DESI DR1's results from various spanning redshifts $\sim 0.3$ to $2.33$ may be more
justifiably combined~\cite{DESI:2024mwx}.
We compare to results with \Planck{} combined instead with DESI data in
\cref{fig:h-amL-low-z-compare}; the posterior shifts over the $h-\amL$ degeneracy to larger values
of $h$ and smaller $\amL$.
This shift is driven not just by DESI's preference for an early onset of dark-energy domination
(evident in its \LCDM{} results as well~\cite{DESI:2024mwx}) but also for a slightly larger
uncalibrated amplitude $h r_\mathrm{d}$.
However, we note that the posterior of $h r_\mathrm{d}$ shifts to lower values by several
percent when excluding DESI DR1's LRG results in redshift bins $z = 0.51$ and $0.71$; these data
appear to be mild outliers with a fair amount of leverage when interpreted within \LCDM{}
cosmology~\cite{DESI:2024mwx}, so we caution against interpreting this result too seriously.

Type Ia supernova datasets can also measure the scale factor of matter--dark-energy equality
$\amL$, and they do so independently of the sound horizon and with distance measurements
out to slightly higher redshifts than LRG-derived BAO measurements.
Since the degeneracy direction that preserves $\theta_s$ does not hold $\amL$ fixed (i.e.,
has $\amL$ strongly correlated with $m_{e, i}$), these datasets should yield different
preferences for $m_{e, i}$ (and therefore $h$) if they prefer different $\amL$.
\Cref{fig:h-amL-low-z-compare} shows that this is the case, to an extent: the medians of the
marginalized posteriors for $h$ vary by $3-8\%$ percent when \Planck{} data are combined with either those of
the SDSS+ BAO dataset, DESI's DR1 BAO measurements, or the Pantheon+ dataset.
Because $h$ varies rather sensitively with $m_{e, i}$ along the posteriors, the corresponding
variation in inferred values for $m_{e, i}$ is less than $2.5\%$.
Further, the posteriors in the $h$-$\amL$ plane show a strong correlation as expected, affirming that
constraints from the SN datasets (which include no external calibration on the fiducial brightness
magnitude $M_B$) derive from their measurement of the shape of the expansion history.

The \LCDM{} posteriors in \cref{fig:h-amL-low-z-compare} are more mutually consistent across dataset
combinations in spite of being much narrower than those that vary the electron mass, since in
\LCDM{} $h$ is constrained to simultaneously satisfy \Planck{}'s precise measurement of $\theta_s$.
The varying-$m_e$ model, not so restricted, allows the posteriors on $\amL$ to be mostly driven by
late-time datasets.
Indeed, the constraints on $\amL$ in the left panels of \cref{fig:h-amL-low-z-compare} nearly match
those each individually infers for the flat \LCDM{}
model~\cite{Brout:2022vxf,eBOSS:2020yzd,DESI:2024mwx}.
Similar conclusions apply to the fine-structure constant, since it also affects the scale factor of
recombination, but to a lesser degree given that the CMB constrains $\alpha_i$ independently of its
effect on $a_\star$.

\subsection{Coupled, hyperlight scalars}\label{sec:results-scalars}

We now study the effect of consistently accounting for a hyperlight scalar hypothesized to couple to
the electron or photon, realizing the fundamental-constant variations in
\cref{sec:results-varying-constants}.
We again begin by studying constraints with CMB data alone before studying combinations with BAO and
SNe data.
We numerically implement the scalar field's dynamics as described in
\cref{sec:cosmology-hyperlight-scalars} (with further details in \cref{app:scalar-implementation}).
In addition to the parameters specified in \cref{sec:parameter-inference}, we sample over the
scalar's present energy density relative to that of CDM, i.e.,
$\fphi = \omega_\phi / \omega_c \sim \mathcal{U}(0.0, 0.3)$.
Directly sampling over the scalar's mass $m_\phi$ is challenging because observables are only weakly
sensitive to its value (within the mass range we consider); again following
Refs.~\cite{Hlozek:2014lca, Hlozek:2016lzm, Hlozek:2017zzf, Lague:2021frh, Rogers:2023ezo}, we
therefore fix $m_\phi$ to a single value in parameter inference.
We take a fiducial mass $m_\phi = 10^{-30}~\eV$, commenting on other choices in
\cref{sec:results-scalars-low-z}.
Because we only consider scalars that begin oscillating in the matter era, $\fphi$ uniquely determines
the scalar's initial misalignment $\bar{\phi}_i$ via \cref{eqn:Xi-ito-varphi}.
We take purely adiabatic initial conditions, assuming any isocurvature perturbations in $\phi$
generated during inflation are negligible.

\subsubsection{\Planck{} data alone}
\label{sec:results-scalars-planck}

The arguments of \cref{sec:cosmology-hyperlight-scalars} that cosmological data should favor a
hyperlight scalar with energy contribution proportional to the increase in the electron mass hinge
upon the assumption that \LCDM{} provides an effectively optimal fit to features determined by
late-time dynamics.
In particular, \cref{sec:cosmology-hyperlight-scalars} shows that when \emph{increasing} $h$ to
compensate for a smaller sound horizon at larger $m_e$, the scalar's contribution to the matter
density can restore the relative amount of dark energy and matter (i.e., the scale factor at which
they are equal, $\amL$).
Though \Planck{} data by itself does not favor the parameter space in which this effect is relevant, a
hyperlight scalar still opens up additional freedom in the late-time expansion history that could
allow for larger $m_{e, i}$ without altering $\theta_s$.
Constraining the model with \Planck{} 2018 likelihoods alone therefore tests the extent to which
they can tolerate hyperlight subcomponents of dark matter (or whether any features in the data are
actually better explained by hyperlight scalar fields).

\begin{figure}[t!]
    \centering
    \includegraphics[width=0.495\textwidth]{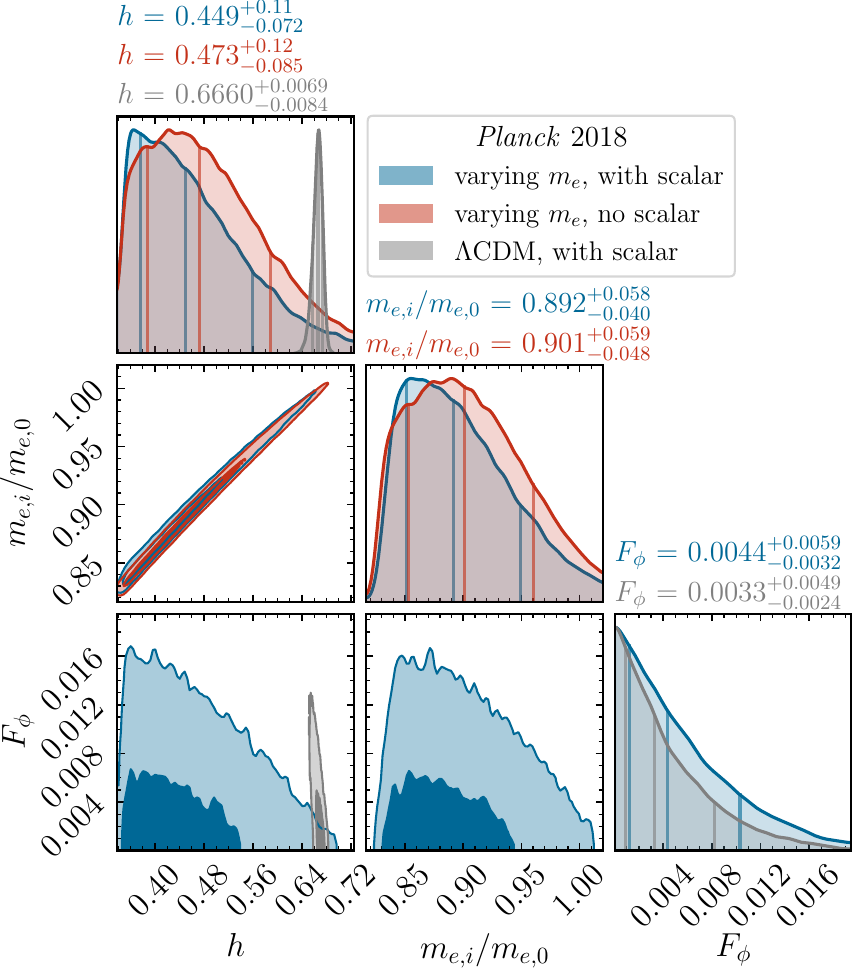}
    \includegraphics[width=0.495\textwidth]{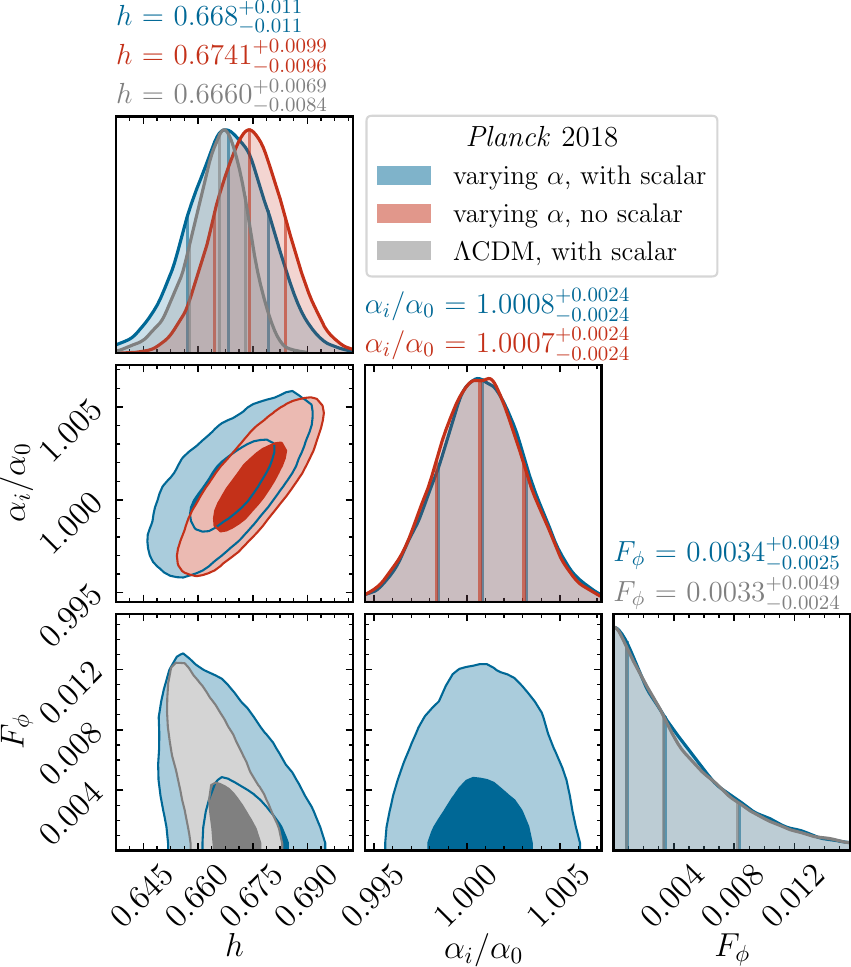}
    \caption{
        Posterior distribution over $h$, $m_{e, i} / m_{e, 0}$ (left) or $\alpha_i / \alpha_0$
        (right), and the fraction $\fphi$ of the CDM density contributed by a
        scalar field with mass $10^{-30}~\eV$.
        Depicted are results for varying constant scenarios without (blue) and with (red) scalar
        fields as well as for \LCDM{} with a scalar field (gray), all using \Planck{} 2018 likelihoods
        alone.
        The diagonal panels display kernel density estimates of the one-dimensional, marginalized
        posterior relative to their peak value, and the lower panels depict 1 and $2 \sigma$ mass
        levels of the two-dimensional joint posterior, again slightly smoothed to facilitate
        comparison.
    }
    \label{fig:Xi-posteriors-me-al-lcdm}
\end{figure}

\Cref{fig:Xi-posteriors-me-al-lcdm} shows that \Planck{} 2018 data only allows a hyperlight scalar
to marginally affect the results without a scalar (\cref{sec:results-varying-constants-planck}).
The posteriors over $h$ and $m_{e, i}$ peak at slightly smaller values, but the degeneracy between
$h$ and $m_{e, i}$ is largely unmodified by including a scalar field.
The posterior over $\fphi$ in this case is also marginally broader than that without any SM couplings,
but only at lower $h$.
At these smallest values of $h$, the late ISW effect due to the onset of the dark energy era is
negligible, opening up room for a hyperlight scalar to induce more of a ISW contribution to the CMB
on large scales.
The $95$th percentile of the marginalized posteriors over $\fphi$ thus increase from $0.0127$ when
no constants vary to $0.0152$ when $m_{e, i}$ varies.
(We note that our results without varying constants are consistent with those from
Ref.~\cite{Rogers:2023ezo}, which obtains bounds of order $10^{-2}$ for masses
$10^{-30} \lesssim m_\phi / \eV \lesssim 10^{-28}$.)
Such strong constraints demonstrate how sensitive \Planck{} is to the dynamics of dark matter
perturbations.

For scalars with photon couplings (which are already strongly constrained by high-multipole \Planck{}
measurements) we might expect the constraints on $\fphi$ to correspond closely to constraints on
hyperlight scalar fields without varying constants.
\Cref{fig:Xi-posteriors-me-al-lcdm} corroborates this expectation: the marginalized posterior over
$\fphi$ is unchanged relative to \LCDM{}.
The results in \cref{fig:Xi-posteriors-me-al-lcdm} do, on the other hand, exhibit a slight
anticorrelation between $\fphi$ and $h$ in all cases, even for \LCDM{} with a hyperlight scalar field.
This effect may in part be again understood via the arguments of
\cref{sec:scalar-impact-on-cosmological-background} regarding the scalar's effect on the distance to
last scattering: at constant $a_\star$, \cref{eqn:theta-s-degeneracy-scalar} shows that fixing the
size of the sound horizon $\theta_s$ requires $h \propto A_m^{-2.09} \sim (1 + \fphi)^{-2.09}$.
For the varying-$m_e$ scenario, the impact of this effect is negligible compared to the degeneracy
between $h$ and $m_{e, i}$.
As such, the CMB residuals depicted in \cref{fig:cmb-all-spectra-spaghetti-vary-both} are
qualitatively unchanged by the inclusion of a scalar, and similar conclusions may be drawn.

\subsubsection{Impact of low-redshift distance measurements}
\label{sec:results-scalars-low-z}

While \Planck{}'s preference is for $m_{e, i} < m_{e, 0}$, preempting the regime in which the scalar's
could realize a novel degeneracy direction, the features in the data that drive it hold only
marginal constraining power.
We might therefore only expect this degeneracy to bear out when including direct measurements of the
late-time expansion history that would yield stronger constraints on $\amL$.
In particular, per \cref{sec:scalar-impact-on-cosmological-background}, fixing $\theta_s$ requires a
disproportionate increase in $\omega_\Lambda$ relative to that in $\omega_c$ and $\omega_b$ (imposed
by the structure of the acoustic peaks); late dark matter provides a means to restore the relative
amount of dark energy and matter at late times.
Including a hyperlight scalar field in the cosmological background thus opens up a degeneracy that
can satisfy both CMB and BAO measurements of the angular size of the sound horizon at any value of
$m_{e, i}$.
On the other hand, \cref{fig:Xi-posteriors-me-al-lcdm} show that the CMB disfavors a substantial
contribution from a hyperlight scalar, even without varying constants; the arguments of
\cref{sec:scalar-impact-on-cosmological-perturbations} suggest this is due to its effects on the
dynamics of perturbations.
We now test whether including low-redshift distance data drives the former effect to outweigh the
CMB's constraints on the latter and realizes the degeneracy anticipated in
\cref{sec:scalar-impact-on-cosmological-background}.
Given that BAO data has a substantial impact on $m_{e, i}$ constraints (\cref{fig:planck-vs-bao}),
we also assess whether other low-redshift distance datasets independently exhibit similar
preferences.

\begin{figure}[t!]
    \centering
    \includegraphics[width=\textwidth]{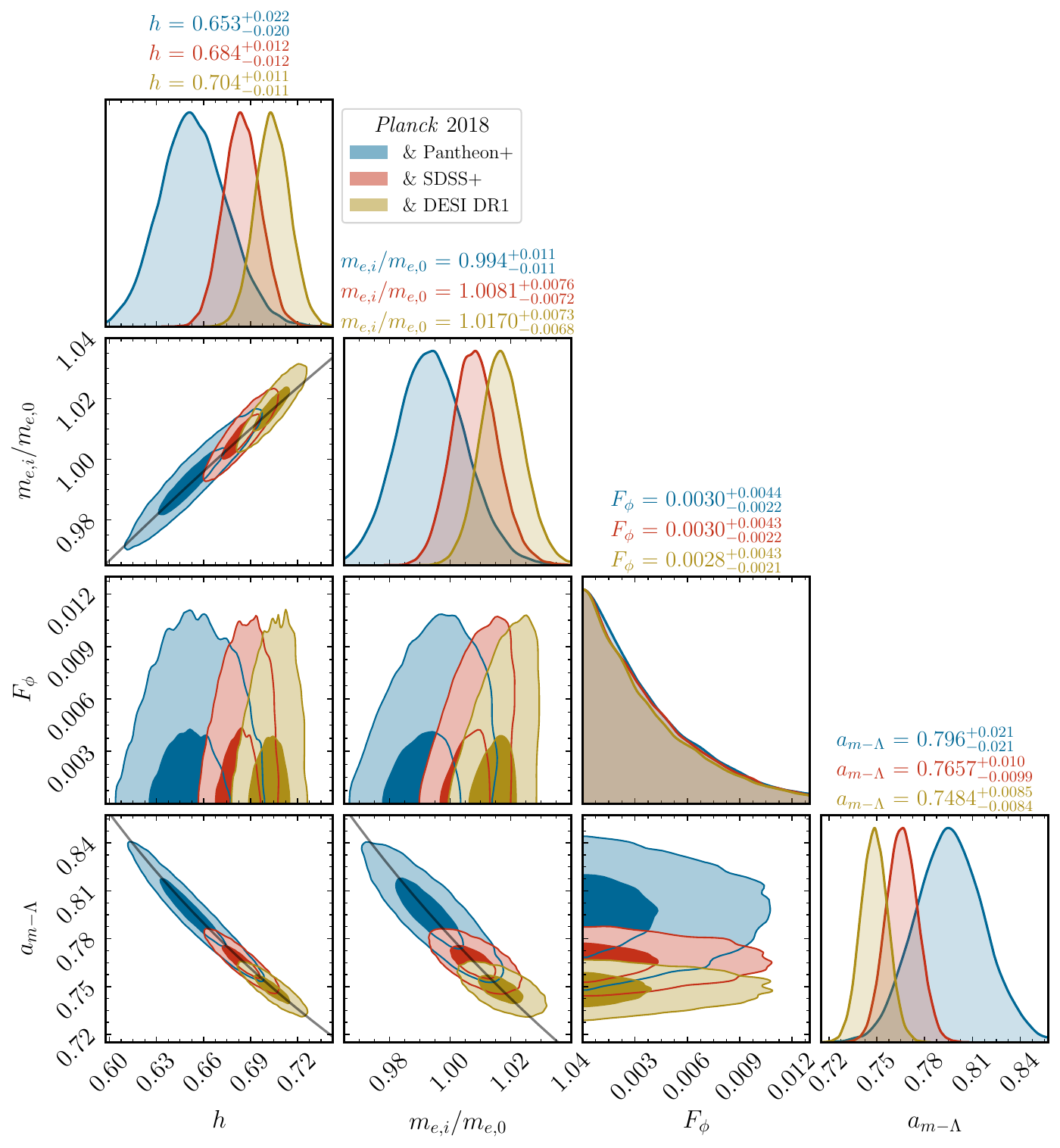}
    \caption{
        Posterior distribution over a subset of the parameters of the electron-coupled scalar model
        (with $m_\phi = 10^{-30}~\eV$) and the scale factor of matter--dark-energy equality $\amL$.
        Posteriors use the likelihood for the full set of \Planck{} 2018 data separately combined
        with the Pantheon+ (blue), SDSS+ BAO (red), and DESI DR1 BAO (gold) datasets.
        The 1 and $2 \sigma$ contours of the marginalized two-dimensional posteriors are depicted as
        in \cref{fig:Xi-posteriors-me-al-lcdm}.
        Gray lines depict the degeneracy inferred from \cref{eqn:theta-s-degeneracy} for scenarios
        \emph{without} a scalar field, i.e., with $\omega_m \propto m_{e, i}$ and $h \propto m_{e,
        i}^{3.18}$.
    }
    \label{fig:me-scalar-bao-sn-corner-h-me-Xi-amL}
\end{figure}

\Cref{fig:me-scalar-bao-sn-corner-h-me-Xi-amL} indicates that low-redshift distance data do not
outweigh \Planck{}'s constraining power on nonclustering subcomponents of dark matter: the degeneracy
of \cref{sec:scalar-impact-on-cosmological-background} where $h \propto m_{e, i}$ is not realized.
The posteriors in the $h$-$m_{e, i}$ plane still lie along the same parameter direction as for
phenomenological varying constants (\cref{fig:planck-vs-bao,fig:h-amL-low-z-compare}).
The joint posteriors over $m_{e, i}$ and $\fphi$ \emph{do}, however, display some degree of
correlation, indicating that the relationship expected in
\cref{sec:scalar-impact-on-cosmological-background} would hold more clearly if the CMB did not
disfavor the effect of hyperlight scalars on spatial perturbations.
Indeed, the marginalized posteriors over $\fphi$ are identical among the four cases, and the
constraints are (marginally) sharpened compared to that for \Planck{} alone
(\cref{fig:Xi-posteriors-me-al-lcdm}).

Finally, we note that our electron-coupling results are nearly insensitive to the scalar's mass
$m_\phi$ in the range $10^{-31}$ to $10^{-29}~\eV$.
In particular, the marginalized posteriors over the standard \LCDM{} parameters and $m_{e, i}$
do not change to a discernible degree for any choice of dataset combination.
Constraints on $\fphi$ are slightly more sensitive to $m_\phi$, but in a manner entirely consistent
with constraints on ultralight axions of Ref.~\cite{Rogers:2023ezo}.
For $m_\phi / \mathrm{eV} = 10^{-29}$, $10^{-30}$, and $10^{-31}$, we obtain $95\%$ upper limits of
$10^2\,\fphi = 1.01$, $1.12$, and $1.79$, respectively, from \Planck{} 2018 combined with the
SDSS+ BAO dataset.
These results again indicate that the constraints on $\fphi$ are largely driven by effects that are
not significantly correlated with any other cosmological parameter.\footnote{
    Reference~\cite{Rogers:2023ezo} shows that constraints from \Planck{} 2018 data improve upon the
    2015 data release due to an improved measurement of $\tau_\mathrm{reion}$, which breaks a
    degeneracy between $A_s e^{- 2 \tau_\mathrm{reion}}$ and $\fphi$ that arises from a hyperlight
    scalar's suppression of large-scale power.
}
Given that the scalar's gravitational effect is predominantly on large scales, we have no reason to
suspect different findings for scalars coupled instead to the photon.

These results illustrate the power of the CMB to discriminate between models of small subcomponents
of dark matter, even independent of their effect on the background cosmology.
Though hyperlight scalars are not a successful realization of late dark matter, other possibilities
with different dynamics could well be.
In fact, the Standard Model provides one such candidate: massive neutrinos, which we consider now.

\subsubsection{Comparison to massive neutrinos}\label{sec:comparison-to-massive-neutrinos}

\begin{figure}[th!]
    \centering
    \includegraphics[width=\textwidth]{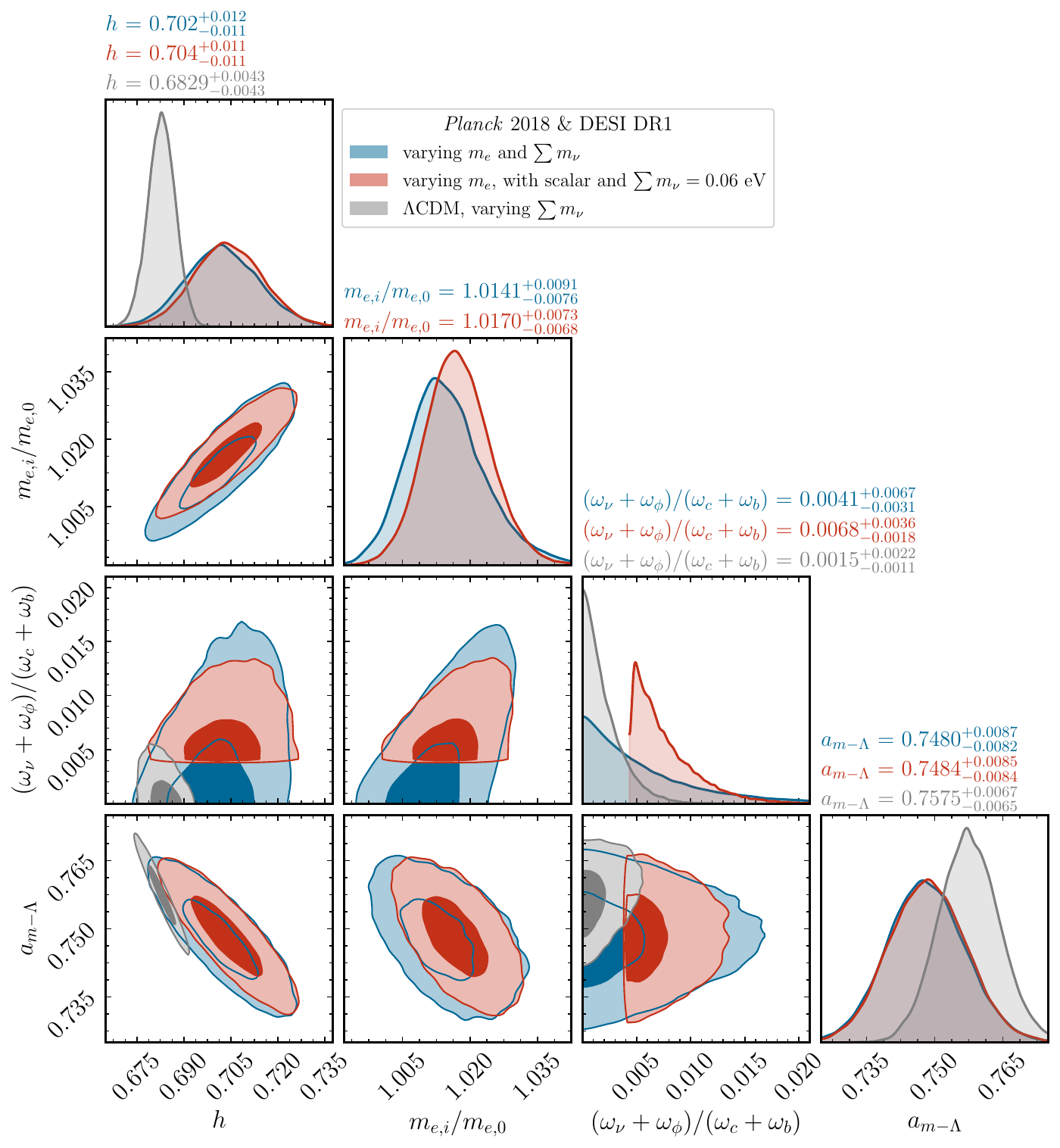}
    \caption{
        Comparison of posterior distributions in scenarios varying the electron mass and the
        abundance of a late dark matter component from either massive neutrinos (blue) or a new,
        hyperlight scalar with mass $10^{-30}~\eV$ (red).
        Results for \LCDM{} with a varying neutrino mass are depicted in gray for comparison.
        Each scenario takes a single massive neutrino state whose mass either is a free parameter
        or, in the case that includes a scalar, is fixed to $0.06~\eV$.
        The posterior is depicted over the Hubble constant $h$, the early-time electron mass
        $m_{e, i} / m_{e, 0}$, the net abundance of late dark matter relative to baryons and CDM
        $(\omega_\nu + \omega_\phi) / (\omega_c + \omega_b)$, and the scale factor of
        matter--dark-energy equality $\amL$.
        All results use the likelihood for the full set of \Planck{} 2018 data combined with DESI
        DR1 BAO data.
        The 1 and $2 \sigma$ contours of the marginalized two-dimensional posteriors are depicted as
        in \cref{fig:Xi-posteriors-me-al-lcdm} and the diagonal panels depict the one-dimensional
        marginal posteriors, each normalized to unity.
    }
    \label{fig:me-scalar-vs-mnu-desi-corner}
\end{figure}

Having established that hyperlight scalars modify structure growth too severely to make a sizeable
contribution to the late-time matter abundance, we consider another realization of late dark matter
that is contained within the SM: massive neutrinos.
One could simultaneously vary the neutrino masses and include a hyperlight scalar as a microphysical
realization of varying constants, but the results are not materially different.
As discussed in \cref{sec:scalar-impact-on-cosmological-background}, within \LCDM{} massive
neutrinos increase the matter density after recombination, requiring a reduced dark energy density
(and so a smaller $h$) to offset the decrease in distance to last scattering (just as for a
hyperlight scalar).
A BAO measurement of $h r_\mathrm{d}$ (see \cref{sec:bao}) breaks this degeneracy, yielding the strongest
current constraints on neutrino masses from cosmological data~\cite{Planck:2018vyg, DESI:2024mwx}.
DESI's first data release, already reaching the level precision of prior surveys, yields stronger
constraints on $\sum m_\nu$ that are driven by its preference for $h r_\mathrm{d}$ several percent (or $\sim
1 \sigma$) larger relative to SDSS's~\cite{DESI:2024mwx}.\footnote{
Part of this shift is due to DESI LRG datapoints, which are in up to a $3 \sigma$ tension with SDSS
in some redshift bins~\cite{DESI:2024mwx}; future data releases may clarify whether this discrepancy
is statistical.}

Since late dark matter allows for early recombination without altering the relative amount of dark
energy and matter at late times, $h$ and $\sum m_\nu$ are instead positively correlated.
\Cref{fig:me-scalar-vs-mnu-desi-corner} compares results in varying-$m_e$ scenarios featuring late
dark matter in the form of a massive neutrino or a hyperlight scalar.
We sample over the neutrino mass with a uniform prior between $0$ and $1~\eV$.
The tail of the posterior density over $(\omega_\nu + \omega_\phi) / (\omega_c + \omega_b)$, i.e.,
the increase in the matter density at late times, demonstrates the \Planck{} tolerates a marginally
larger abundance of late dark matter in the form of neutrinos than in a hyperlight scalar.
The posteriors for the hyperlight scalar scenario are centered at slightly larger $h$
and $m_{e, i}$ only because this case takes a fiducial neutrino mass $m_\nu = 0.06~\eV$, which
effectively imposes a prior $(\omega_\nu + \omega_\phi) / (\omega_c + \omega_b) > 5 \times 10^{-3}$.
Because (uncalibrated) BAO distances depend only on the background cosmology at late times, the
marginal posteriors over $\amL$ are effectively identical.
Ultimately, hyperlight scalars and massive neutrinos are comparably constrained as late dark matter
components because their dynamics are qualitatively similar (as discussed in
\cref{sec:cosmology-hyperlight-scalars}).

The $95$th percentile of the posterior over $\sum m_\nu$ increases from $0.073~\eV$ in \LCDM{} to
$0.239~\eV$ when varying the early-time electron mass, similar to that found in
Ref.~\cite{Khalife:2023qbu}.
The latter result precisely matches the constraint within \LCDM{} from \Planck{} 2018 data
alone (i.e., without BAO data)~\cite{Planck:2018vyg}, which again establishes that the primary
constraints on early recombination with late dark matter derive from the new component's effects on
structure and not those on the expansion history.
This result also highlights the sensitivity of cosmological inference of the neutrino masses to the
underlying cosmological model: the \LCDM{} result nearly rules out the inverted mass hierarchy, in
early recombination scenarios cosmological data provide no discriminatory power.

\subsection{Implications for cosmological concordance}\label{sec:concordance}

The findings of the previous sections for varying-$m_e$ cosmologies---the impact of consistently
including a scalar field as a microphysical realization of the scenario and the lack of strong
concordance among datasets---have important implications for the interpretation of contemporary
tensions between cosmological datasets.
Prior literature pointed to the degeneracy of the varying-$m_e$ model as a preferred means to infer
a larger Hubble constant from CMB data~\cite{Hart:2019dxi, Sekiguchi:2020teg, Schoneberg:2021qvd,
Khalife:2023qbu} that could be (relatively more) consistent with that measured via the distance
ladder~\cite{Riess:2016jrr,Riess:2019cxk,Riess:2021jrx}.
The results of \cref{fig:vary-me-corner-wb-wc-h-me}, however, suggest that the CMB by itself prefers
negligible dark energy in models where its abundance is not predominantly fixed by the
distance to last scattering, because such models have the additional freedom to better fit the
deficit in temperature power on large scales and the excess lensing of the acoustic peaks (relative
to \LCDM{}) evident in \Planck{} 2018 data.

On its own, \Planck{} 2018 data therefore prefer late recombination: a lighter early-time electron
mass and an even lower value of $h$ than in \LCDM{}, regardless of whether one accounts for the
gravitational effect of a scalar field coupled to the electron.
This finding, though evident in prior work~\cite{Hart:2019dxi, Schoneberg:2021qvd} (and to a
slightly lesser extent in results using previous \Planck{} data releases~\cite{Planck:2014ylh,
Hart:2017ndk}), has not been emphasized in discussions of the model as a solution to the Hubble
tension, nor has its physical origin been explored.
Though the broad posteriors superficially appear to reduce the tension, in no region of parameter
space does the model \emph{simultaneously} improve the fit to
both CMB data and the calibrated distance-ladder datasets that prefer larger $h$.\footnote{
    Reference~\cite{Cortes:2023dij} succinctly makes the general point that a proper Bayesian
    interpretation of a model's compatibility with multiple datasets requires marginalizing
    likelihoods over parameter space---i.e., metrics beyond the improvement to the best fit point (a
    point in parameter space with vanishing posterior mass) or the number of standard deviations
    between inferred parameter values (which provides little information on whether a model better
    explains different datasets in the same part of parameter space).
}
Setting aside questions of whether the lensing and low-$\ell$ anomalies have a statistical or
systematic origin, we conclude that a varying electron mass alone does not meaningfully restore
concordance of these datasets.

\Planck{}'s preference for late recombination derives from features that can be better explained by
delaying the transition to dark-energy domination, but SNe and BAO distances directly measure the
shape of the expansion history (i.e., $\amL$) in this epoch.
\Cref{fig:me-scalar-bao-sn-corner-h-me-Xi-amL} shows that even these direct probes (each combined
with \Planck{} and without external calibration for SNe) are far from concordant within the
varying-$m_e$ model.
Because $h$ and $m_{e, i}$ correlate so strongly with $\amL$, their marginal posteriors only agree
to the extent of the datasets' individual preferences for $\amL$.\footnote{
    Constraints on the flat \LCDM{} model are usually quoted in terms of $\Omega_m$.
    We center our discussion on $\amL = \sqrt[3]{\Omega_m / (1 - \Omega_m)}$ both because it more
    directly connects to the shape of the late-time expansion history and because a late dark matter
    component (like a hyperlight scalar) changes the late-time matter abundance but not the
    early-time one (\cref{sec:scalar-impact-on-cosmological-background}).
}
\begin{figure}[t!]
    \centering
    \includegraphics[width=\textwidth]{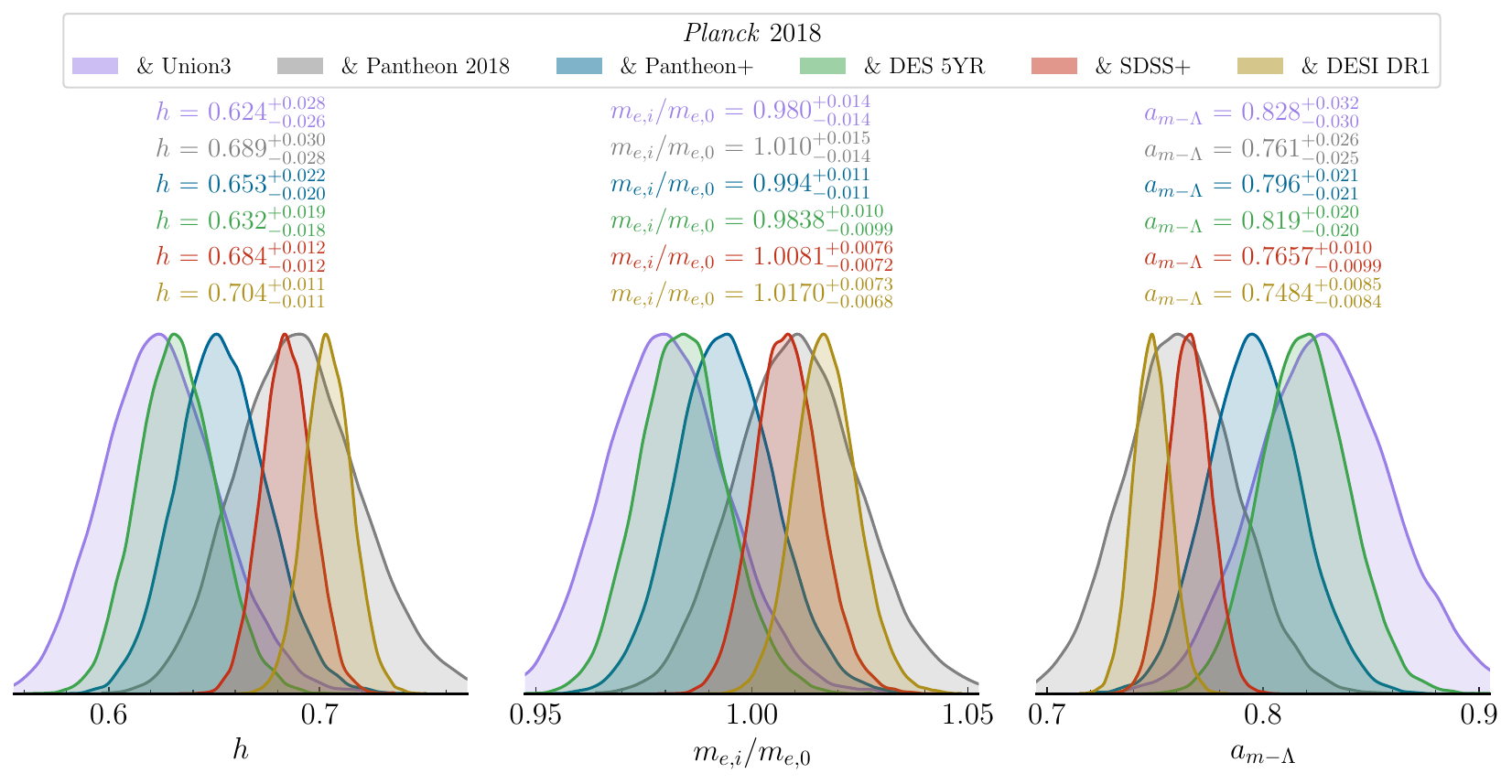}
    \caption{
        Posterior distributions over $h$, $m_{e, i} / m_{e, 0}$, and $\amL$ for the electron-coupled
        scalar model (with $m_\phi = 10^{-30}~\eV$), using the likelihood for the full set of
        \Planck{} 2018 data separately combined with the Pantheon 2018 (gray), Pantheon+ (blue),
        DES 5YR (green), SDSS+ BAO (red), and DESI DR1 BAO (gold) datasets.
        Titles report the corresponding median and $\pm 1 \sigma$ quantiles of each distribution.
        Note that results for the varying-$m_e$ model without a scalar field are nearly identical,
        with posteriors over $h$ and $m_{e, i}$ shifted marginally to lower values.
    }
    \label{fig:me-scalar-all-low-z-compare-h-me-amL}
\end{figure}
Results inferred from current SNe and BAO datasets are particularly discrepant, as evident in
\cref{fig:me-scalar-all-low-z-compare-h-me-amL}.
These divergent preferences are especially pronounced for the very recent Union3 and DES 5YR SNe
datasets, which both prefer a transition to dark-energy domination even later than does the
Pantheon+ result, yielding correspondingly smaller $h$ and $m_{e, i}$.
Curiously, the posteriors from the 2018 Pantheon dataset in
\cref{fig:me-scalar-all-low-z-compare-h-me-amL}, though much broader than the others, agree much
better with \Planck{} combined with either BAO dataset.

These discrepancies in late-time datasets yield highly divergent interpretations of concordance.
At one end, the na\"ive Gaussian tension between \Planck{} combined with DES
($h \approx 0.632 \pm 0.018$) remains above the $5 \sigma$ level with Pantheon+ and SH0ES
($h \approx 0.735 \pm 0.01$)~\cite{Brout:2022vxf}, despite the broader posteriors over $h$ afforded
by early recombination scenarios.
In strong contrast, \Planck{} combined with DESI yields $h \approx 0.704 \pm 0.011$,
apparently a mere $2 \sigma$ discrepancy with Pantheon+ and SH0ES.
We reiterate that a substantial contribution to DESI's preference for larger $h$ is driven by its
LRG data, which within \LCDM{} are in slight tension with SDSS's~\cite{DESI:2024mwx}.
Finally, because $h$ varies rather rapidly with the electron mass at fixed $\theta_s$, the
discrepancy in measurements of $m_{e, i}$ in \cref{fig:me-scalar-all-low-z-compare-h-me-amL} is less
severe but still impedes a quantitative identification of a consensus constraint from cosmological
datasets on $m_{e, i}$.

\begin{figure}[t!]
    \centering
    \includegraphics[width=0.6\textwidth]{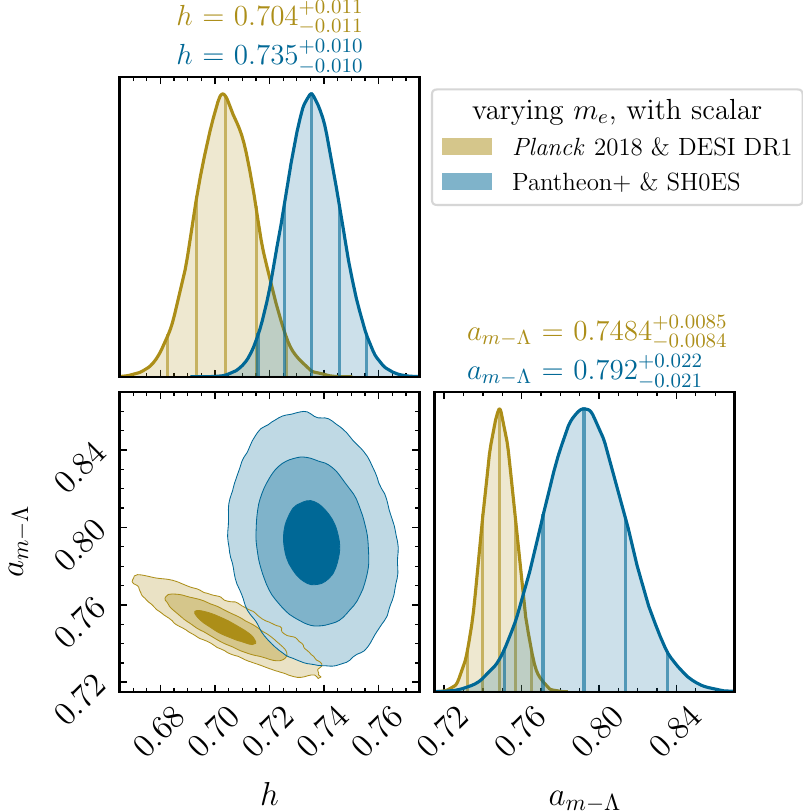}
    \caption{
        Posterior distribution over $h$ and $\amL$ from \Planck{} 2018 and DESI DR1 BAO data for the
        electron-coupled scalar model (gold) and from Pantheon+ with SH0ES-calibrated Cepheids
        (blue).
        The two datasets are in tension in both dimensions, which is best illustrated by the offset
        of their posterior in the $h$-$\amL$ plane; these results are hardly changed for the
        standard varying-$m_e$ model (without a scalar) and are virtually identical when instead
        varying both $m_{e, i}$ and the neutrino mass sum $\sum m_\nu$.
        Combining \Planck{} with SNe or BAO datasets other than DESI shifts the posterior to larger
        $\amL$ but lower $h$ (see \cref{fig:me-scalar-all-low-z-compare-h-me-amL}).
        The vertical lines in the diagonal panels depict the median and $2.5$th, $16$th, $84$th,
        and $97.5$th percentiles of the marginalized posterior distributions.
        These one-dimensional posteriors are kernel density estimates normalized relative to their
        peak value to facilitate comparison.
        The median and corresponding $\pm 1 \sigma$ uncertainties for each parameter are reported
        above the diagonal panels.
        The lower panels display the the 1, 2, and $3 \sigma$ contours (i.e., the $39.3\%$,
        $86.5\%$, and $98.9\%$ mass levels) of the marginalized, two-dimensional posterior.
    }
    \label{fig:h-amL-tension}
\end{figure}

Setting aside the more discrepant datasets (\cref{fig:me-scalar-all-low-z-compare-h-me-amL}), even
the improved agreement on $h$ in early recombination scenarios between the \Planck{} and DESI
combination and the SH0ES-calibrated distance ladder does not fully capture the degree of the
tension.
These two dataset combinations are also in a $2 \sigma$ tension in their (marginalized) posteriors
over $\amL$, as seen in \cref{fig:me-scalar-all-low-z-compare-h-me-amL}.
Moreover, $h$ and $\amL$ are anticorrelated in early recombination models as constrained by
\Planck{}, such that decreasing the discrepancy in one of $h$ or $\amL$ only exacerbates that in the
other.
Concordance must be assessed in the full two-dimensional parameter space that characterizes the
late-Universe expansion history; as shown in \cref{fig:h-amL-tension}, the posteriors in the
$h$-$\amL$ plane hardly overlap even in their $3 \sigma$ mass levels.
While a late dark matter contribution could increase $h$ without reducing $\amL$ in principle
(\cref{sec:scalar-impact-on-cosmological-background}), within the hyperlight scalar field model
\Planck{} does not allow abundances large enough to do so.
Even in a scenario where this additional degeneracy is realized, the discrepant inference of $\amL$
by BAO and SNe data (which trace the same interval of redshift) cannot be fully reconciled---a
problem for any model in which the late-time expansion history is well described by flat \LCDM{}.

For these reasons, we do not analyze the combination of \Planck{} 2018 data with the
SH0ES-calibrated distance ladder.
Posteriors for these combined datasets would localize at parameters between their individual
preferences, obfuscating the fact that the individual likelihoods are degraded in this resulting
parameter space.
Prior work has in addition admonished the usage of a SH0ES-derived prior directly on $H_0$,
emphasizing the importance of including the full magnitude-redshift sample from Pantheon SNe,
calibrated by SH0ES Cepheids, to provide genuine constraints on modifications to late-time
cosmology~\cite{Efstathiou:2021ocp, Camarena:2021jlr}.
Our results demonstrate the importance of doing so more generally, because modifications to
early-time cosmology can correlate to changes in the late-time expansion history.
In early recombination models, the disproportionate increase in the matter and dark-energy densities
cause the late-time expansion history to depart (in shape, or $\amL$) from that directly measured by
current SN datasets, underscoring the potential hazards in interpreting analyses that combine the
Cepheid-calibrated distance ladder with CMB (and BAO) data.

As anticipated in \cref{sec:degeneracies}, our results bear a close resemblance to those for the
(counterfactual) cosmologies in which the present-day CMB temperature $T_0$ is treated as an
otherwise unconstrained parameter~\cite{Ivanov:2020mfr, Wen:2020txi}.
The correlated impact on the late-time expansion history is common to both scenarios, as is the role
of CMB lensing and the ISW effect~\cite{Ivanov:2020mfr, Carron:2022eum}.
The preceding considerations also apply (to a reasonable extent) to
early recombination mechanized by small-scale inhomogeneities in the baryon
distribution~\cite{Jedamzik:2020krr, Rashkovetskyi:2021rwg, Thiele:2021okz, Lee:2021bmn} mentioned
in \cref{sec:degeneracies}.
These scenarios only prefer a larger Hubble constant when combining \Planck{} data with datasets
like SH0ES~\cite{Jedamzik:2020krr}, indicating that (superficially) alleviating the tension
comes at the cost of degrading the \Planck{} likelihoods.

Having assessed the joint effect of varying constants and hyperlight scalars on the background
evolution in the $h$-$\amL$ space, we now turn to the model's impact on matter clustering and the
Hubble constant.
In \LCDM{}, low-redshift probes prefer larger $h$ but smaller $\sigma_8$ (or
$S_8 \equiv \sqrt{\Omega_m / 0.3})$, but a common feature of proposed \LCDM{} extensions is a
positive correlation between, e.g., $h$ and $\sigma_8$~\cite{Planck:2015lwi, SPT:2016izt}.
This expectation may be understood on general grounds via the arguments of
\cref{sec:varying-constants}: the radiation driving and early ISW effects constrain the CDM
abundance at recombination, requiring a larger matter abundance today if recombination occurs at
higher density (whether because it occurs early or in the presence of additional new degrees of
freedom).
Absent other effects, early-time modifications to cosmology that lead to inference of larger $h$
typically also yield a larger amplitude of matter clustering.

\begin{figure}[t!]
    \centering
    \includegraphics[width=0.495\textwidth]{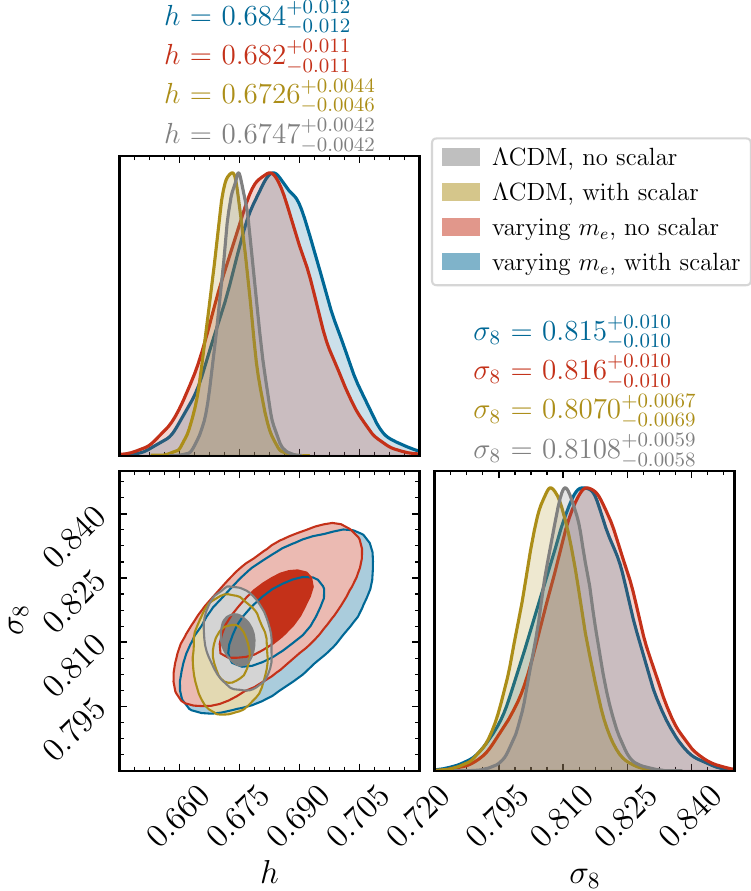}
    \includegraphics[width=0.495\textwidth]{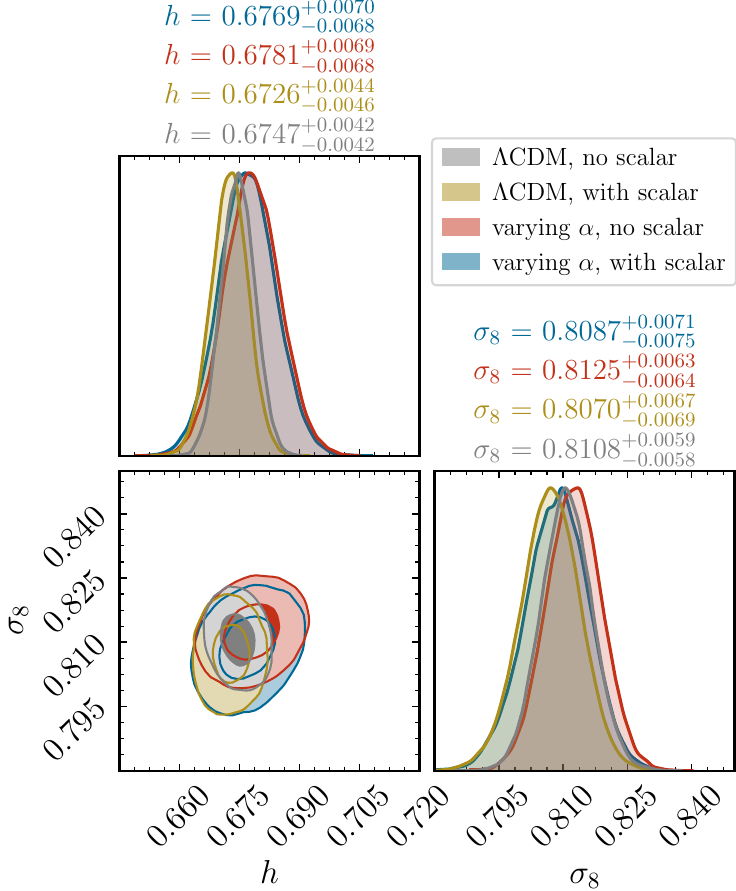}
    \caption{
        Posterior distribution over $h$ and $\sigma_8$ (the latter a derived quantity) in scenarios
        vary $m_{e, i} / m_{e, 0}$ (left) or $\alpha_i / \alpha_0$ (right).
        Both cases also compare results with \LCDM{} with and without a scalar.
        Scenarios with a scalar field take its mass to be $10^{-30}~\eV$, and all cases use
        \Planck{} 2018 likelihoods and the SDSS+ BAO datasets enumerated in the text.
    }
    \label{fig:h-sigma8}
\end{figure}

On the other hand, per \cref{sec:scalar-impact-on-cosmological-perturbations} (and, e.g.,
Ref.~\cite{Rogers:2023ezo}), a hyperlight scalar subcomponent of dark matter in general suppresses
matter clustering.
The correlation between the abundance of the scalar ($\fphi$) and the early-time shift in the electron
mass ($m_{e, i} / m_{e, 0} - 1$) outlined in \cref{sec:scalar-impact-on-cosmological-background}
would achieve an anticorrelation between $h$ and $\sigma_8$ if it had borne out in the full data
analysis.
Indeed, \cref{fig:h-sigma8} evinces a broadening of the posteriors (projected into the space of $h$
and $\sigma_8$) in this direction for varying-$m_e$ cosmologies that include a scalar compared to
those that do not.
However, \Planck{} data itself ultimately curtail the extent of this effect in parameter space; the
results here ultimately do not depart dramatically from those of Ref.~\cite{Rogers:2023ezo}, which
considered hyperlight scalars with only gravitational interactions.
The degree of the effect is comparable to the extent hyperlight scalars may reduce $\sigma_8$ in
standard \LCDM{}, except that (as explained in \cref{sec:results-scalars}) $\fphi$ and $h$ are
themselves anticorrelated in this case.
Coupled hyperlight scalars therefore provide at the least an example model that can accommodate CMB
data at larger $h$ but smaller $\sigma_8$.

To summarize, in this section we confront the models of \cref{sec:varying-constants} and predictions
of \cref{sec:cosmology-hyperlight-scalars} with data, carefully considering the role of individual
probes and trends in the most recent datasets.
In \cref{sec:results-varying-constants}, we confirm the degeneracy direction introduced by varying
the electron mass and quantify the extent to which a large early-time electron mass (early
recombination) is disfavored by secondary effects in the CMB data; deviations in the fine structure
constant, on the other hand, are severely constrained by changes to the damping tail.
Late-time datasets prove crucial, breaking degeneracies to yield much stronger constraints on the
phenomenological varying-$m_e$ scenario.
Our results in \cref{sec:results-scalars} affirm the expectations of
\cref{sec:cosmology-hyperlight-scalars} that the gravitational impact of a late dark matter
component (for instance, a hyperlight scalar or massive neutrinos) opens a new degeneracy direction
with early recombination.
However, \cref{sec:results-varying-constants-planck} shows that \Planck{} does not allow large
enough scalar field abundances to fully explore this degeneracy and qualitatively alter the
resulting parameter constraints.
In \cref{sec:concordance} we find that, like many proposed \LCDM{} extensions, a hyperlight scalar
coupled to the electron can relax but never fully resolve tensions in $h$ and $\sigma_8$ between
early and late-time datasets.
More importantly, we show that this conclusion is heavily dataset dependent: recent BAO and SNe
datasets exhibit increasingly discrepant preferences for the late-time expansion history (in its
shape, i.e., independent of the Hubble tension) that, per
\cref{fig:me-scalar-all-low-z-compare-h-me-amL}, have a striking effect on cosmological parameter
inference in modified recombination scenarios.

%% file: scalar-implementation.tex
\section{Numerical implementation of cosmological scalar fields}\label{app:scalar-implementation}

The gravitational effects of cosmological scalar fields have been studied in contexts including
ultralight dark matter~\cite{Hu:1998kj, Hu:2000ke, Hwang:2009js, Marsh:2010wq, Hlozek:2014lca,
Hlozek:2016lzm, Hlozek:2017zzf, Lague:2021frh, Rogers:2023ezo, Winch:2023qzl}, scalar field dark
energy~\cite{Bean:2003fb,Ballesteros:2010ks}, and early dark energy~\cite{Poulin:2018cxd,
Poulin:2018dzj, Smith:2019ihp, Poulin:2023lkg}.
A scalar field's potential typically introduces a new timescale of cosmological importance, which in
the simplest case of potentials with minima is the scalar's oscillation period.
Scalars with quadratic potentials generically oscillate with a fixed period much shorter than the
age of the Universe (unless their mass is close to the Hubble constant $H_0$, a regime to which we
do not restrict ourselves).

To mitigate the onerous computational cost required to accurately solve for scalar fields' dynamics,
a variety of effective fluid treatments for scalar fields have been utilized in prior
literature~\cite{Hu:1998kj, Hu:2000ke, Hwang:2009js, Hlozek:2014lca, Urena-Lopez:2015gur,
Poulin:2018dzj, Cookmeyer:2019rna, Smith:2019ihp}.
We follow Ref.~\cite{Passaglia:2022bcr}, which systematically developed a fluid approximation for
massive scalar fields that enabled semianalytically and empirically calibrating the effective
equation of state and sound speed describing the fluid.
The scheme devised by Ref.~\cite{Passaglia:2022bcr} entails solving the scalar's equation of motion
(the Klein-Gordon equation) directly for its first few oscillations (when fluid approximations are
largely insufficient) before switching to the effective fluid approximation.
The matching procedure of Ref.~\cite{Passaglia:2022bcr} also improves substantially upon prior
methods.
We extend \textsf{CLASS}~\cite{Blas:2011rf, Lesgourgues:2011re} by implementing the scheme proposed
by Ref.~\cite{Passaglia:2022bcr} to enable an efficient and accurate treatment of scalar fields.
(The implementation is available at Ref.~\cite{class-uls}.)
For completeness, we briefly review the formalism and definitions necessary to specify the scheme,
referring to Ref.~\cite{Passaglia:2022bcr} for full details and comparison with previous fluid
treatments.

\subsection{Dynamics}\label{app:scalar-dynamics}

In a general spacetime, the action for a minimally coupled scalar field is
$S_\phi = \int \ud^4 x \, \sqrt{-g} \mathcal{L}_\phi$ in terms of its Lagrangian density
\begin{align}
    \mathcal{L}_\phi
    &= - \frac{1}{2} \partial_\mu \phi \partial^\mu \phi
        - V(\phi).
    \label{eqn:free-scalar-lagrangian}
\end{align}
The scalar's Euler-Lagrange equation is
\begin{align}
    \label{eqn:scalar-ele}
    \nabla_\mu \nabla^\mu \phi
    &= \frac{\ud V}{\ud \phi},
\end{align}
where $\nabla_\mu$ is the covariant derivative compatible with the metric $g$.
We parametrize a general, perturbed, conformal-time FLRW metric with
\begin{align}
    g_{\mu \nu}
    \equiv a(\tau)^2 \left( \eta_{\mu\nu} + h_{\mu\nu} \right)
\end{align}
where $\eta_{\mu\nu}$ is the Minkowski metric with the mostly positive signature and $h_{\mu \nu}$ a
small perturbation we decompose as
\begin{subequations}\label{eqn:metric-svt-decomposition}
\begin{align}
    \label{eqn:metric-svt-decomposition-h00}
    h_{00}
    &= - E
    \\
    \label{eqn:metric-svt-decomposition-h0i}
    h_{i 0}
    &= \partial_i F
    \\
    \label{eqn:metric-svt-decomposition-hij}
    h_{i j}
    &= A \delta_{i j}
        + \partial_i \partial_j B
    .
\end{align}
\end{subequations}
Because neither primordial vectors nor tensors have been detected, and because we introduce no new
physics that would source them, \cref{eqn:metric-svt-decomposition} includes only scalar
perturbations $A$, $B$, $E$, and $F$.
This parametrization is general but redundant due to gauge invariance; we discuss concrete gauge
choices below.

Expanding to linear order with $\phi(t, \three{x}) = \bar{\phi}(t) + \delta \phi(t, \three{x})$,
the equation of motion \cref{eqn:scalar-ele} becomes
\begin{subequations}\label{eqn:scalar-eom-decomposed}
\begin{align}
    0
    &= \bar{\phi}''
        + 2 \mathcal{H} \bar{\phi}'
        + a^2 \frac{\ud V}{\ud \phi}
    \label{eqn:bar-phi-eom}
    \\
\begin{split}
    \label{eqn:delta-phi-eom}
    0
    &= \delta \phi''
        + 2 \mathcal{H} \delta \phi'
        - \partial_i \partial_i \delta \phi
        + a^2 \frac{\ud^2 V}{\ud \phi^2} \delta \phi
        + \frac{\bar{\phi}'}{2} \left(
            \partial_i \partial_i \left[ B' - 2 F \right]
            + 3 A'
            - E'
        \right)
        + a^2 \frac{\ud V}{\ud \phi} E.
\end{split}
\end{align}
\end{subequations}
Here primes denote derivatives with respect to conformal time $\tau$.
The conformal Newtonian gauge has $E = 2 \Psi$, $A = - 2 \Phi$, and both $F$
and $B$ zero, in which case
\begin{align}
    \label{eqn:delta-KG-equation-newtonian}
        0
        &= \delta \phi''
            + 2 \mathcal{H} \delta \phi'
            - \partial_i \partial_i \delta \phi
            + a^2 \frac{\ud^2 V}{\ud \phi^2} \delta \phi
            - \left( 3 \Phi' + \Psi' \right) \bar{\phi}'
            + 2 a^2 \frac{\ud V}{\ud \phi} \Psi
        .
\end{align}
The synchronous gauge used in Ref.~\cite{Ma:1995ey} sets
$A = - 2 \eta$ and $\partial_i \partial_i B = h + 6 \eta$ with $E$ and $F$ zero;
\cref{eqn:delta-phi-eom} then reads
\begin{align}
    \label{eqn:delta-KG-equation-ma-bertschinger}
    0
    &= \delta \phi''
        + 2 \mathcal{H} \delta \phi'
        - \partial_i \partial_i \delta \phi
        + a^2 \frac{\ud^2 V}{\ud \phi^2} \delta \phi
        + \frac{1}{2} \bar{\phi}' h'
        .
\end{align}

The scalar field's contribution to the stress-energy tensor is
\begin{align}
    \label{eqn:scalar-stress-energy}
    \left( T_{\mu\nu} \right)^{\phi}
    = - 2 \frac{\partial \mathcal{L}_\phi}{\partial g^{\mu \nu}}
        + g_{\mu\nu} \mathcal{L}_\phi
    &= \partial_\mu \phi \partial_\nu \phi
        + g_{\mu\nu} \left(
            - \frac{1}{2} \partial_\alpha \phi \partial^\alpha \phi
            - V(\phi)
        \right).
\end{align}
At the background level, the effective energy density
$\bar{\rho}_\phi = - \left( \bar{T}_0^{\hphantom{0} 0} \right)^{\phi}$
and pressure $\bar{P}_\phi = \left( \bar{T}_i^{\hphantom{i} i} \right)^{\phi} / 3$ are
\begin{subequations}
\begin{align}
    \bar{\rho}_\phi
    &= \frac{\left( \bar{\phi}' \right)^2}{2 a^2} + V(\bar{\phi})
    \label{eqn:rho-phi-bg}
    \\
    \bar{P}_\phi
    &= \frac{\left( \bar{\phi}' \right)^2}{2 a^2} - V(\bar{\phi}).
    \label{eqn:P-phi-bg}
\end{align}
\end{subequations}
We parametrize the scalar perturbations to the stress-energy tensor in terms of density, pressure,
and velocity perturbations $\delta \rho$, $\delta P$, and $\delta u$, as well as anisotropic stress $\pi^S$:
\begin{subequations}
\label{eqn:stress-tensor-decomp-mixed}
\begin{align}
    \label{eqn:stress-tensor-decomp-mixed-T00}
    \delta T^{0}_{\hphantom{0}0}
    &= - \delta \rho
    \\
    \label{eqn:stress-tensor-decomp-mixed-T0i}
    \delta T^{0}_{\hphantom{0}i}
    &= \left( \bar{\rho} + \bar{P} \right)
        \partial_i \delta u
    \\
    \label{eqn:stress-tensor-decomp-mixed-Tij}
    \delta T^{i}_{\hphantom{i}j}
    &= \delta_{ij} \delta P
        + \left( \partial_i \partial_j - \frac{1}{3} \delta_{ij} \partial_k \partial_k \right) \pi^S
    .
\end{align}
\end{subequations}
Often the velocity perturbation is written in terms of
$\theta = \partial_i \partial_i \delta u$.
The corresponding contributions from \cref{eqn:scalar-stress-energy} (at linear order) are
\begin{subequations}
\begin{align}
    \delta \rho_\phi
    &= - \frac{E}{2 a^2} \left( \bar{\phi}' \right)^2
        + \frac{\bar{\phi}' \delta \phi'}{a^2}
        + \frac{\ud V}{\ud \phi} \delta \phi
    \\
    \left( \bar{\rho}_\phi + \bar{P}_\phi \right) \delta u_\phi
    &= - \frac{\bar{\phi}' \delta \phi}{a^2}
    \\
    \delta P_\phi
    &= - \frac{E}{2 a^2} \left( \bar{\phi}' \right)^2
        + \frac{\bar{\phi}' \delta \phi'}{a^2}
        - \frac{\ud V}{\ud \phi} \delta \phi
    .
\end{align}
\end{subequations}
Anisotropic stress from a scalar field is zero to leading order in perturbation theory.
Recall again that $E = 2 \Psi$ in Newtonian gauge and $0$ in synchronous gauge.

\subsection{Effective fluid treatment}\label{app:effective-fluid}

We now turn to effective fluid approximations for scalars with purely quadratic potentials,
$V(\phi) = m_\phi^2 \phi^2 / 2$.
Following Ref.~\cite{Passaglia:2022bcr}, we fix synchronous gauge for the remainder of the
discussion and switch to cosmic time $t$ defined by $\ud t = a \ud \tau$.
With these choices,
\begin{align}
    0
    &= \ddot{\bar{\phi}}
        + 3 H \dot{\bar{\phi}}
        + m_\phi^2 \bar{\phi}
    \label{eqn:bar-phi-eom-cosmic}
    \\
\begin{split}
    \label{eqn:delta-phi-eom-cosmic-synchronous}
    0
    &= \delta \ddot{\phi}
        + 3 H \delta \dot{\phi}
        + \frac{k^2}{a^2} \delta \phi
        + m_\phi^2 \delta \phi
        + \frac{1}{2} \dot{\bar{\phi}} \dot{h},
\end{split}
\end{align}
writing the perturbation equation in Fourier space.
In the absence of metric perturbations, these equations both exhibit oscillatory solutions with a
decaying envelope.

\subsubsection{Background}

Beginning at the background level Ref.~\cite{Passaglia:2022bcr} decomposes the solution to
\cref{eqn:bar-phi-eom-cosmic} onto its two oscillatory modes as
\begin{align}\label{eqn:def-phi-c-s}
    \bar{\phi}(t)
    &= \bar{\varphi}_c(t) \cos \left( m_\phi t - m_\phi t_\star \right)
        + \bar{\varphi}_s(t) \sin \left( m_\phi t - m_\phi t_\star \right),
\end{align}
where the amplitudes $\bar{\varphi}_c$ and $\bar{\varphi}_s$ evolve slowly (i.e., on timescales
order $1/H$).
\Cref{eqn:bar-phi-eom-cosmic} expands to
\begin{align}
\begin{split}\label{eqn:kg-equation-phi-c-s}
    0
    &= \left[
            \ddot{\bar{\varphi}}_c
            + 2 m_\phi \dot{\bar{\varphi}}_s
            + 3 H \left(
                \dot{\bar{\varphi}}_c
                + m_\phi \bar{\varphi}_s
            \right)
        \right]
        \cos \left( m_\phi t - m_\phi t_\star \right)
    \\ &\hphantom{ {}={} }
        + \left[
            \ddot{\bar{\varphi}}_s
            - 2 m_\phi \dot{\bar{\varphi}}_c
            + 3 H \left(
                \dot{\bar{\varphi}}_s
                - m_\phi \bar{\varphi}_c
            \right)
        \right]
        \sin \left( m_\phi t - m_\phi t_\star \right),
\end{split}
\end{align}
which may be solved by solving the individual expressions in brackets as a set of two coupled
differential equations.
In brief, Ref.~\cite{Passaglia:2022bcr} defines an effective energy density and pressure
\begin{subequations}\label{eqn:rho-P-phi-ef}
\begin{align}
    \bar{\rho}_{\phi, \mathrm{ef}}
    &\equiv \frac{1}{2} \left(
            \frac{1}{2} \left[
                \dot{\bar{\varphi}}_c^2
                + \dot{\bar{\varphi}}_s^2
            \right]
            + m_\phi \left[
                \dot{\bar{\varphi}}_c \bar{\varphi}_s
                - \dot{\bar{\varphi}}_s \bar{\varphi}_c
            \right]
            + m_\phi^2 \left[
                \bar{\varphi}_s^2
                + \bar{\varphi}_c^2
            \right]
        \right)
    \\
    \bar{P}_{\phi, \mathrm{ef}}
    &\equiv \frac{1}{2} \left(
            \frac{1}{2} \left[
                \dot{\bar{\varphi}}_c^2
                + \dot{\bar{\varphi}}_s^2
            \right]
            + m_\phi \left[
                \dot{\bar{\varphi}}_c \bar{\varphi}_s
                - \dot{\bar{\varphi}}_s \bar{\varphi}_c
            \right]
        \right)
\end{align}
\end{subequations}
that, via the Klein-Gordon equation \cref{eqn:kg-equation-phi-c-s}, exactly satisfy the standard fluid conservation law
\begin{align}\label{eqn:ef-fluid-eom}
    \dot{\bar{\rho}}_{\phi, \mathrm{ef}}
    &= - 3 H \left( \bar{\rho}_{\phi, \mathrm{ef}} + \bar{P}_{\phi, \mathrm{ef}} \right).
\end{align}
In the scheme of Ref.~\cite{Passaglia:2022bcr}, one solves the full Klein-Gordon equation
[\cref{eqn:bar-phi-eom-cosmic}] until some transition time $t_\star$ and matches onto the amplitudes
$\bar{\varphi}_c$ and $\bar{\varphi}_s$ in order to evaluate the effective fluid energy density
$\bar{\rho}_{\phi, \mathrm{ef}}$.
Matching error (due to oscillations of $\rho_\phi$ about its time-averaged $1/a^3$ redshifting) is
minimized by choosing
\begin{align}
    \left. \frac{\ddot{\bar{\varphi}}_{I}}{\dot{\bar{\varphi}}_I} \right\vert_{t=t_\star}
    &= - \frac{\left\langle H \right\rangle}{2}
        \left(
            3
            - \frac{2 \ud \left\langle H \right\rangle / \ud \tau}{a \left\langle H \right\rangle^2}
        \right)
    \equiv \mathcal{M}.
    \label{eqn:matching-M}
\end{align}
for $I = c$, $s$.
Here $\left\langle H \right\rangle$ denotes the Hubble parameter averaged over the oscillations
induced by the scalar's contribution to FLRW expansion; we may neglect such effects in practice and
simply substitute $H$.
Inserting the decomposed Klein-Gordon equation into \cref{eqn:matching-M} and solving yields
\begin{align}\label{eqn:match-varphi-cs-dot}
    \begin{pmatrix}
        \dot{\bar{\varphi}}_c
        \\
        \dot{\bar{\varphi}}_s
    \end{pmatrix}
    &= - \frac{3 H}{\mathcal{M}^2 + 3 H \mathcal{M} + 4 m_\phi^2}
        \begin{pmatrix}
            2 m_\phi & 3 H + \mathcal{M}
            \\
            - \mathcal{M} & 2 m_\phi
        \end{pmatrix}
        \begin{pmatrix}
            m_\phi \bar{\phi}
            \\
            \dot{\bar{\phi}}
        \end{pmatrix}
\end{align}
and
\begin{subequations}
\begin{align}\label{eqn:match-varphi-cs}
    \bar{\varphi}_c
    &= \bar{\phi}
    \\
    \bar{\varphi}_s
    &= \dot{\bar{\phi}} / m_\phi
        - \dot{\bar{\varphi}}_c / m_\phi.
\end{align}
\end{subequations}
Reference~\cite{Passaglia:2022bcr} finds these choices reduce the matching error between
$\bar{\rho}_\phi$ and $\bar{\rho}_{\phi, \mathrm{ef}}$ to $\mathcal{O}([H_\star / m_\phi]^3)$,
where $H_\star = H(t_\star)$.

After matching, one solves the dynamics for the effective fluid approximation (efa)
to the energy density $\bar{\rho}_{\phi, \mathrm{efa}}$,
\begin{align}\label{eqn:efa-fluid-eom}
    \dot{\bar{\rho}}_{\phi, \mathrm{efa}}
    &= - 3 H \left( 1 + w_{\phi, \mathrm{efa}} \right) \bar{\rho}_{\phi, \mathrm{efa}} .
\end{align}
At late times, $\bar{\rho}_\phi$ redshifts like $a^{-3}$ to good precision, motivating the choice
$w_{\phi, \mathrm{efa}} = 0$.
However, optimal computational efficiency switches at as large of $H_\star / m_\phi$ as possible, at
which point the (average) equation of state and rate of redshift deviate appreciably from that of
cold dark matter.
Reference~\cite{Passaglia:2022bcr} therefore calibrated a time-dependent ansatz for
$w_{\phi, \mathrm{efa}} = 3 (H / m_\phi)^2 / 2$, informed by computing \cref{eqn:rho-P-phi-ef} using
numerical solutions to the Klein-Gordon equation.
However, Ref.~\cite{Passaglia:2022bcr} focused on scalars that begin oscillating in the radiation
era ($m_\phi \gtrsim 10^{-28}~\eV$), whereas for the scenarios relevant here, the scalars
necessarily begin to oscillate in the matter era.
Using numerical solutions, Ref.~\cite{Passaglia:2022bcr} demonstrated that
$w_{\phi, \mathrm{ef}} \approx 3 (H / m_\phi)^2 / 2$ for $H \gg m_\phi$, noting that the result could also be obtained by the
analytic solution to the homogeneous Klein-Gordon equation in the radiation era.
Since we are interested in lighter scalars (for which matching will occur in the matter era or
during the onset of dark-energy domination), we extend this result to FLRW backgrounds with
arbitrary equations of state.\footnote{
    Generalization to lighter masses is also of interest for future cosmological constraints on
    scalar subcomponents of dark matter with such masses~\cite{Hlozek:2014lca, Hlozek:2016lzm,
    Hlozek:2017zzf, Lague:2021frh, Rogers:2023ezo}, given projected biases from fluid approximations
    for future CMB experiments from Ref.~\cite{Cookmeyer:2019rna} (which considered masses greater
    than $10^{-27}~\eV$).
    It would also be interesting to extend the formalism of Ref.~\cite{Passaglia:2022bcr} to
    potentials with minima steeper than quadratic in $\phi$, as relevant to early dark
    energy~\cite{Poulin:2018cxd, Poulin:2018dzj, Smith:2019ihp, Hill:2020osr, Ivanov:2020ril};
    however, the computational impetus to do so is likely not nearly so severe because scalars in
    steeper potentials oscillate with a frequency that decreases as the field amplitude redshifts
    (see, e.g., Ref.~\cite{Agrawal:2019lmo}).
}

The Hubble parameter in a Universe with constant equation of state $w$ is $H(t) = n / t$, where
$n = 2 / 3 (1 + w)$.
In such a background, the solution to the homogeneous Klein-Gordon equation (with constant field
value at early times $t \ll 1/m_\phi$) is $\phi(t) \propto J_\nu(m_\phi t) / t^\nu$ where
$\nu = (3 n - 1) / 2$.
Taking the asymptotic (i.e., $m_\phi t \gg 1$) expansion of the energy density and pressure for this
solution (at next-to-leading order, to correctly compute the $1/t^2$ term) yields
$w_{\phi, \mathrm{ef}} = 3 n / 4 (m_\phi t)^2 + \mathcal{O}[(m_\phi t)^{-4}]$.
In a multicomponent Universe (like $\Lambda$CDM), the instantaneous equation of state may be
obtained by combining the Friedmann equations to give
$w(t) = - 2 \dot{H}(t) / 3 H(t)^2 - 1$.
Given that the fluid approximation is meant to be used only when $m_\phi / H \gtrsim 10$,
approximating expansion as a power law at any given time should be wholly sufficient [barring exotic
cosmologies with $w(t)$ evolving on timescales much shorter than the Hubble scale].
As such, we may promote our single-component result above to a general Universe via
\begin{align}
    w_{\phi, \mathrm{ef}}
    &= - \frac{3}{4} \frac{\dot{H}}{m_\phi^2}
    \label{eqn:effective-eos}
\end{align}
In practice, we find \cref{eqn:effective-eos} works exceedingly well.
Since $\dot{H} = - 2 H^2$ deep in the radiation era, it reproduces the results of
Ref.~\cite{Passaglia:2022bcr} for transitions occurring well before matter-radiation equality.
Moreover, for transitions occurring in the matter era, we observe even more rapid convergence,
improving with $(H_\star / m_\phi)^{4}$ for $m_\phi \geq 10^{-28}~\eV$ compared to
cubic convergence for transitions in the radiation era.

\subsubsection{Perturbations}

In spite of the added complications that the equation of motion for $\delta \phi$ depends on wave
number $k$ and is sourced by metric perturbations, the treatment for spatial perturbations proceeds
similarly.
We briefly enumerate the corresponding definitions, referring to Ref.~\cite{Passaglia:2022bcr} for
further discussion and justification.
The perturbation is again decomposed onto oscillatory modes as
\begin{align}\label{eqn:def-delta-phi-c-s}
    \delta \phi(t)
    &= \delta \varphi_c(t) \cos \left( m_\phi t - m_\phi t_\star \right)
        + \delta \varphi_s(t) \sin \left( m_\phi t - m_\phi t_\star \right).
\end{align}
The effective fluid variables defined as
\begin{subequations}
\begin{align}
\begin{split}
    2 \delta \rho_{\phi, \mathrm{ef}}
    &= \dot{\bar{\varphi}}_c \delta \dot{\varphi}_c
        + \dot{\bar{\varphi}}_s \delta \dot{\varphi}_s
        + m_\phi \left(
            \bar{\varphi}_s \delta \dot{\varphi}_c
            - \bar{\varphi}_c \delta \dot{\varphi}_s
        \right)
        + m_\phi \left(
            \dot{\bar{\varphi}}_c \delta \varphi_s
            - \dot{\bar{\varphi}}_s \delta \varphi_c
        \right)
    \\ &\hphantom{ {}={} }
        + 2 m_\phi^2 \left[
            \bar{\varphi}_c \delta \varphi_c
            + \bar{\varphi}_s \delta \varphi_s
        \right]
\end{split}
    \\
    \left( \bar{\rho}_{\phi, \mathrm{ef}} + \bar{P}_{\phi, \mathrm{ef}} \right)
        \theta_{\phi, \mathrm{ef}}
    &= \frac{k^2}{2 a} \left[
            \left( \dot{\bar{\varphi}}_c + m_\phi \bar{\varphi}_s \right)
            \delta \varphi_c
            + \left( \dot{\bar{\varphi}}_s - m_\phi \bar{\varphi}_c \right)
            \delta \varphi_s
        \right]
    \\
    \delta P_{\phi, \mathrm{ef}}
    &= \delta \rho_{\phi, \mathrm{ef}}
        - m_\phi^2 \left[
            \bar{\varphi}_c \delta \varphi_c
            + \bar{\varphi}_s \delta \varphi_s
        \right]
\end{align}
\end{subequations}
exactly satisfy the perturbed fluid equations
\begin{subequations}
\begin{align}
    0
    &= \delta \rho_{\phi, \mathrm{ef}}'
        + 3 \mathcal{H}
        \left( \delta \rho_{\phi, \mathrm{ef}} + \delta P_{\phi, \mathrm{ef}} \right)
        + \left( \bar{\rho}_{\phi, \mathrm{ef}} + \bar{P}_{\phi, \mathrm{ef}} \right)
        \left(
            \theta_{\phi, \mathrm{ef}}
            + \frac{h'}{2}
        \right)
    \\
    0
    &= \partial_\tau \left[
            \left( \bar{\rho} + \bar{P} \right) \theta_{\phi, \mathrm{ef}}
        \right]
        + 4 \mathcal{H}
        \left( \bar{\rho}_{\phi, \mathrm{ef}} + \bar{P}_{\phi, \mathrm{ef}} \right)
        \theta_{\phi, \mathrm{ef}}
        - k^2 \delta P_{\phi, \mathrm{ef}}.
\end{align}
\end{subequations}
Matching proceeds identically to the background, i.e., using the analogous matching condition to
\cref{eqn:matching-M} and solving for $\delta \varphi_c$, $\delta \varphi_s$, and their time
derivatives in analogy to \cref{eqn:match-varphi-cs,eqn:match-varphi-cs-dot}.

After matching (and computing $\delta_\phi$ and $\theta_\phi$ at $t_\star$), one solves
the effective fluid equations~\cite{Hu:1998kj}
\begin{subequations}
\begin{align}
    \begin{split}
        \delta_\phi'
        &= - \left( 1 + w_{\phi, \mathrm{ef}} \right)
            \left( \theta_\phi + \frac{h'}{2} \right)
            - 3 \left( c_s^2 - w_{\phi, \mathrm{ef}} \right) \mathcal{H} \delta_\phi
            - 9 \mathcal{H}^2 \left( 1 + w_{\phi, \mathrm{ef}} \right) \left( c_s^2 - c_a^2 \right) \frac{\theta_\phi}{k^2}
    \end{split}
    \\
    \theta_\phi'
    &= - \left( 1 - 3 c_s^2 \right) \mathcal{H} \theta_\phi
        + \frac{c_s^2 k^2}{1 + w_{\phi, \mathrm{ef}}} \delta_\phi,
\end{align}
\end{subequations}
where $c_s^2$ is the rest-frame fluid sound speed and the ``adiabatic sound speed'' squared is
\begin{align}
    c_a^2
    &= w_{\phi, \mathrm{ef}} - \frac{w_{\phi, \mathrm{ef}}'}{3 \mathcal{H} \left( 1 + w_{\phi, \mathrm{ef}} \right)}.
\end{align}
Reference~\cite{Passaglia:2022bcr} used a Wentzel-Kramers-Brillouin approximation to compute an
effective sound speed $c_s^2 = (\sqrt{1 + \kappa^2} - 1) / \kappa$ (using the shorthand
$\kappa = k / a m_\phi$) in the limit that expansion is negligible.
Reference~\cite{Passaglia:2022bcr} also calibrated an $\mathcal{O}[(H / m_\phi)^2]$ correction to the
squared sound speed, finding (empirically from numerical solutions) the coefficient to be $5/4$,
different from the factor $3/2$ found for the equation of state that one might naively expect.
We observe in addition that a coefficient $9/8$ works well in the matter era.
To reproduce both regimes, we take the ansatz
\begin{align}
    c_s^2
    &= \frac{\sqrt{1 + (k / a m_\phi)^2} - 1}{k / a m_\phi}
        - \frac{1}{4} \frac{\dot{H}}{m_\phi^2}
        + \frac{3}{4} \left( \frac{H}{m_\phi} \right)^2,
    \label{eqn:effective-sound-speed}
\end{align}
which we empirically observe works well across a wide range of masses.\footnote{
    Curiously, the correction to the sound speed does coincide with that to equation of state
    [\cref{eqn:effective-eos}] in the matter era.
}
We observe roughly an order-of-magnitude improvement in precision compared to using the sound speed
from Ref.~\cite{Passaglia:2022bcr} (at the same $H_\star / m_\phi$) for scalars that begin
oscillating in the matter era.
We defer a more thorough investigation of convergence and possible analytic justification for
\cref{eqn:effective-sound-speed} to future work.

\subsection{Initial conditions}\label{app:initial-conditions}

We set initial conditions in the radiation era, when the scale factor evolves as
$a(\tau) = \sqrt{\Omega_r} H_0 \tau$.
The homogeneous equation of motion for the scalar has a slow-roll solution (i.e., where
$\ud V / \ud \phi$ is approximately constant) of
\begin{align}\label{eqn:slow-roll-sol}
    \bar{\phi}'(\tau)
    &= - \frac{\Omega_r H_0^2}{5} \frac{\ud V}{\ud \phi} \tau^3.
\end{align}
In the superhorizon limit, $h = C (k \tau)^2$.
While $\bar{\phi}$ is slowly rolling in the radiation era, the superhorizon limit of the equation of motion for $\delta \phi$
(i.e., at leading order in $k \tau \ll 1$) is solved by
\begin{align}
    \delta \phi
    &\approx C \frac{\Omega_r H_0^2}{210 k^4} \frac{\ud V}{\ud \phi}
        (k \tau)^{6}.
\end{align}
Plugging the slow-roll solution \cref{eqn:slow-roll-sol} and noting that
\begin{align}
    \bar{\rho}_\phi
    + \bar{P}_\phi
    &= \frac{\left( \bar{\phi}' \right)^2}{a^2},
\end{align}
the corresponding initial conditions for the fluid variables are
\begin{align}
    \frac{\delta_\phi}{C}
    &\equiv \frac{\delta \rho_\phi}{C \bar{\rho}_\phi}
    = - \frac{\Omega_r H_0^2}{1050 k^4 V(\bar{\phi})}
        \left( \frac{\ud V}{\ud \phi} \right)^2
        (k \tau)^{6}
    \\
    \frac{\theta_\phi}{C}
    &\equiv \frac{-k^2 \delta u_\phi}{C}
    = \frac{k^2 \delta \phi}{C \bar{\phi}'}
    = - \frac{k (k \tau)^{3}}{42},
\end{align}
consistent with those in Ref.~\cite{Smith:2019ihp}.

\subsection{Implementation details}

Finally, we mention a few nontrivial aspects of the implementation of the above formalism in
\textsf{CLASS}, particularly related to switching solving schemes.
To begin with, switching between solving the Klein-Gordon equation and effective fluid equations
changes the system of equations (and degrees of freedom) that are solved for at some intermediate
time, based on a condition that itself depends on the instantaneous value of a dynamical quantity
(namely, $H / m_\phi$).
Aside from the fact that the exact time this transition occurs cannot be known in advanced (i.e.,
not without integrating the equations in question), performing such transition while using adaptive
ordinary differential equation (ODE) integrations is tricky.
In particular, the transition condition may be reached in the middle of an integration interval;
changing the system of equations midstep introduces discontinuities that violate the smoothness
conditions required by ODE solvers.
One must therefore execute the transition between integration steps.
(ODE solvers evaluate the system of equations at multiple subintervals of each step.)
However, when using ``black box'' integrators (i.e., where one passes a function that evaluates the
differential equations to a solver routine that performs the entire integration from an initial
condition until some end point, as implemented in \textsf{CLASS}) it is nontrivial to determine
whether a function evaluation is the first evaluation of an integration step or some intermediate
evaluation.
When integrating the background cosmology, we therefore implement the switching logic in a callback
method that \textsf{CLASS}'s integrator executes between integration step (i.e., the
\texttt{background\_sources} method, in which various derived quantities are tabulated).
Still, we find that \textsf{CLASS}'s stiff (implicit) ODE solver still does not work with the above
procedure, requiring that we use its explicit Runge-Kutta method instead.
This could be due to the resulting change in the Jacobian, which cannot be accounted for because the
user cannot decide when the Jacobian is recomputed (and it is not recomputed every integration
step).
For the solution of perturbations, \textsf{CLASS} does have a built-in system for scheduling and
executing approximation scheme transitions (used extensively for other sectors, e.g., tight-coupling
approximations for the photon-baryon plasma).
We therefore simply add another such approximation scheme to switch on the effective fluid treatment
for scalar perturbations.

Several issues also arise due to \textsf{CLASS}'s method for tabulating and interpolating the
background cosmology.
A standard feature of ODE solvers is so-called dense output, which combines function evaluations
used for integration to additionally produce interpolating functions over solver steps.
The interpolants are designed to be accurate to some high order in the step size (comparable to the
accuracy of the integration itself).
\textsf{CLASS} instead simply tabulates the results at fixed intervals in $\ln a$ and constructs a cubic Hermite interpolant.
This interpolation is therefore sensitive to the transition between the Klein-Gordon and effective
fluid solutions, which can induce spurious errors in the interpolant that have a non-negligible effect on later stages in \textsf{CLASS}'s execution (discussed below).
However, it is not tractable to simply specify that the background be tabulated with a smaller
$\ln a$ interval: \textsf{CLASS} restricts integration step sizes to be no larger than this
interval, drastically increasing the runtime of the background solution.
(Standard ODE solver libraries instead take as large of steps as the adaptive method permits and
evaluates the dense output interpolants.)
Decreasing the tabulation interval by just two orders of magnitude over the
default [$\Delta \ln a = \ln(a_0 / a_i) / N \approx 8 \times 10^{-4}$, where $a_i / a_0 = 10^{-14}$
and $N = 40,000$ by default] makes the background runtime comparable to that of the perturbations
solution.
Because intervals in physical time increase as $\Delta t = \Delta \ln a / H$, this tabulation
becomes insufficient to accurately capture the scalar's oscillatory behavior when
$m_\phi \Delta t = \Delta \ln a \cdot m_\phi / H \gtrsim 1/10$, which for the default parameters occurs when $m_\phi / H \sim 100$.
Because of the drastic effect it has on subsequent calculations, we increase the default number of
subintervals to $N = 10^5$, but even this is insufficient to efficiently and rigorously test the
accuracy of the fluid approximation by comparing choices of $m_\phi / H_\star$ as large as, say,
$10^4$.
This choice is, however, sufficient for a switching time $m_\phi / H_\star = 10$ as we use in all
results.
The relative difference in, e.g., CMB spectra compared to solutions with larger $m_\phi / H_\star$ is better than the $10^{-3}$ level.

The errors from the inadequate tabulation of the background solution naturally propagate to the
evolution of scalar perturbations and can qualitatively alter the dynamics if unmitigated.
They also can trigger \textsf{CLASS} to crash when determining the sampling points in time at which
to evaluate the CMB source functions (in the \texttt{perturbations\_timesampling\_for\_sources}
routine).
To determine the time sampling required to capture the late ISW effect, \textsf{CLASS} estimates the metric growth rate, which factors in
$2 a'' / a - (a' / a)^2 = - a^2 \bar{P} / \Mpl^2$.
In the matter era, the \emph{only} contribution to $\bar{P}$ is that from the scalar, which
oscillates between $\bar{\rho}$ and $-\bar{\rho}$.
This metric is therefore strongly sensitive to any interpolation errors, which over the course of
parameter sampling often triggered \textsf{CLASS} to crash due to invalid time sampling determined
by this routine.
To mitigate these issues, we implemented a smooth transition between the Klein-Gordon and effective
fluid solutions.
Namely, over some number of oscillations (specified by a parameter) after $H$ drops below the
specified threshold $H_\star$, we solve both the Klein-Gordon and effective fluid equations.
In this intermediate interval, when computing the scalar's energy density and pressure, we average
the results of the two methods with weights determined by a $\mathrm{tanh}$-based window function in
time.
In effect, this procedure (exponentially) smoothly damps the effect of oscillations in the true
(Klein-Gordon) solution before fully switching to solving the effective fluid equations alone.
Since the perturbation matching requires knowing $\bar{\varphi}_c$, $\bar{\varphi}_s$, their time
derivatives, and $\bar{\rho}_{\phi, \mathrm{ef}}$ as well, we record these at the midpoint of this
transition interval for use when matching perturbations.
Note that we use still $\bar{\rho}_{\phi, \mathrm{ef}}$ and $\bar{P}_{\phi, \mathrm{ef}}$ when
matching to effective fluid perturbations, \emph{not} the weighted energy density and pressure,
which are only used in the Friedmann equations.

Finally, a minor complication arises from the effective equation of state $w_{\phi, \mathrm{ef}}$
[\cref{eqn:effective-eos}] depending on $\dot{H}$.
The Friedmann equations set $\dot{H} = - (\bar{\rho} + \bar{P}) / 2 \Mpl^2$, but one does not know
the pressure contribution from the scalar $\bar{P}_{\phi, \mathrm{ef}}$ without knowing the current
value of $w_{\phi, \mathrm{ef}}$.
Since we only consider scenarios where the scalar contributes a subfraction of the total matter
density, we simply iteratively compute $\dot{H}$ (beginning by neglecting the scalar field's
pressure contribution) and $w_{\phi, \mathrm{ef}}$ until the values converge to a relative precision
of $10^{-12}$.

%% file: parameter-inference.tex
\section{Parameter inference}\label{app:parameter-inference}

\subsection{Likelihoods}\label{app:likelihoods}

We use the implementation of the standard 2018 \Planck{}
likelihoods~\cite{Planck:2018vyg,Planck:2019nip} (PR3) in \textsf{clik}~\cite{clik} with supporting
data obtained from the \Planck{} Legacy Archive~\cite{pla}.
For the high-$\ell$ likelihoods (covering $30 \leq \ell < 2500$ in temperature and
$30 \leq \ell < 2000$ for $E$-mode polarization and temperature-polarization cross correlation) we
use the \texttt{Plik\_lite} (\texttt{plik\_lite\_v22\_TTTEEE}) variant that is marginalized over the
parameters of the foreground models.
For the low-$\ell$ likelihoods ($2 \leq \ell \leq 30$) we use \texttt{Commander}
(\texttt{commander\_dx12\_v3\_2\_29}) for temperature and \texttt{SimAll}
(\texttt{simall\_100x143\_offlike5\_EE\_Aplanck\_B}) for $E$-mode polarization.
Finally, we use \Planck{}'s lensing autopower spectrum likelihood
(\texttt{smicadx12\_Dec5\_ftl\_mv2\_ndclpp\_p\_teb\_consext8}) over multipoles $8 \leq L \leq 400$.

We use BAO measurements from a variety of surveys.
The 6dFGS~\cite{Beutler:2011hx} measured $\theta_\mathrm{BAO} = 0.327 \pm 0.0142$ at an effective
sample resdhift $z = 0.106$.
Reference~\cite{Beutler:2011hx} used fitting functions from Ref.~\cite{Eisenstein:1997ik} to compute
$r_\mathrm{d}$, which for their fiducial cosmology with $h = 0.7$, $\omega_b = 0.02227$, and
$\Omega_m = 0.27$ gives $\mathrm{d} = 154.06~\mathrm{Mpc}$, compared to $149.94~\mathrm{Mpc}$ from
\textsf{CLASS}.
The 6dFGS measurement of $\theta_\mathrm{BAO} = 0.336 \pm 0.015$ (both the mean and standard
deviation) is thus rescaled by $149.94 / 154.06$.
The SDSS Main Galaxy Sample DR7 provides a non-Gaussian
likelihood for $1/\theta_\mathrm{BAO}$ evaluated at $z = 0.15$ with tabulated values (specified
relative to a fiducial value $D_V / r_\mathrm{d} = 638.95 / 148.69$) available in supplementary material to
Ref.~\cite{Ross:2014qpa}.
The BAO measurements from the BOSS DR12 galaxies (at $z = 0.38$ and $0.51$) are a multivariate Gaussian likelihood
for $\theta_{\mathrm{BAO}, \perp}$ and $\theta_{\mathrm{BAO}, \parallel}$ at
each redshift~\cite{BOSS:2016wmc}.
Likewise, LRGs from the eBOSS DR16~\cite{eBOSS:2020lta, eBOSS:2020hur} provide a multivariate Gaussian likelihood
for $1/\theta_{\mathrm{BAO}, \perp}$ and $1/\theta_{\mathrm{BAO}, \parallel}$ at $z = 0.70$.
The latter are available on the SDSS-IV SVN Software Repository~\cite{sdss-data}.

In addition to the former ``standard'' set of BAO measurements, we also implement Gaussian
likelihoods for the BAO measurements from DESI DR1~\cite{DESI:2024mwx, DESI:2024lzq, DESI:2024uvr}
as tabulated in Table 1 of Ref.~\cite{DESI:2024mwx} (which has not yet been published at the time of
writing).
Reference~\cite{DESI:2024mwx} notes that each each redshift bin are effectively uncorrelated and the
results from individual tracers are reasonably concordant, justifying their combination.
Notably, rounding errors in Table 1 of Ref.~\cite{DESI:2024mwx} have a nonnegligible effect on the
likelihoods.
In particular, the effective redshift for the bright galaxy sample measurement of
$\theta_\mathrm{BAO}$ is reported as $0.30$, rounded from $0.295$.
The impact of the rounding error is notable, shifting the full DESI DR1 dataset's preferred $h
r_\mathrm{d}$ upward by about half a percent and $\amL$ downward by nearly a percent.
To our knowledge, at the time of writing the actual data are only available at
Ref.~\cite{desi-data}.

Finally, we use the Pantheon~\cite{Pan-STARRS1:2017jku}, Pantheon+~\cite{Brout:2022vxf,
Scolnic:2021amr}, and DES 5YR~\cite{DES:2024tys} measurements of the apparent magnitude
[\cref{eqn:apparent-magnitude}] of type Ia supernovae.
We also use a recent joint analysis of supernovae from numerous datasets,
Union3~\cite{Rubin:2023ovl}.
Per Ref.~\cite{Brout:2022vxf}, SN Ia at redshifts less than $10^{-2}$, which are more sensitive to
bias due to peculiar velocities and other effects, are excluded for Pantheon+.
The Pantheon+ dataset also includes Cepheid host distances from SH0ES~\cite{Riess:2021jrx} that can
optionally be used as calibrators.
Likelihoods are Gaussian in the distance moduli $\mu = m - M_B$, with measured values and covariance
provided online~\cite{pantheon-github,pantheon-plus-github,des-github,union3-data}.

\subsection{Sampling methods}\label{app:sampling}

To generate posterior samples, we use MCMC methods from the Python
package \textsf{emcee}~\cite{Foreman-Mackey:2012any,Hogg:2017akh,Foreman-Mackey:2019}.
We utilize the ensemble move proposal based on kernel density
estimation~\cite{Farr:2013tia,kombine}, which (for the posteriors presented here) we find yields
MCMC chains with shortest autocorrelation times of \textsf{emcee}'s methods.
We sample with 80 walkers, chosen so that each step, which \textsf{emcee} performs in two batches,
runs (unparallelized) likelihood evaluations on each of the 40 cores on the CPUs we use.
Autocorrelation times (for sample parameters) span from $\sim 7$ (for the best-constrained
posteriors) to $\sim 30$ steps (for, e.g., the varying-constant scenarios with a scalar, since $\fphi$
is poorly constrained, or for varying-$m_e$ when using only \Planck{} data, for which the posteriors
are broad).
We sample for 5,000 steps or for 10,000 steps for those cases with more challenging posteriors,
discard the first ten autocorrelation times' worth of steps, and thin by one autocorrelation time.
These choices ensure that reported posteriors include at least $\sim 20,000$-$30,000$ independent
samples, enough to draw robust posterior mass contours even in the presence of extremely strong
correlations (e.g., \cref{fig:vary-me-corner-wb-wc-h-me}) or parameters whose constraints only place
upper limits (e.g., $\fphi$).
These sample sizes are entirely generous for one-dimensional marginalized posteriors and summary
statistics thereof, which are all consistent among smaller subsamples of the full posterior.

Finally, we note that we checked the robustness of our results with \textsf{CLASS}'s default
precision settings by comparing likelihood evaluations with increased precision over $1000$ samples
of the posteriors for each scenario we consider.
In all cases, for $95\%$ of samples the logarithm of the high-$\ell$ Planck likelihood differs by
$\lesssim 0.15$ when increasing precision, and only $0.3$ at most; the others deviate negligibly.
Namely, the distribution of errors is no worse than that for \LCDM{} at the default precision
settings.